\documentclass{cornell}
\usepackage{amsmath,amssymb,rotating}
\usepackage{graphicx}

\title{A high-precision measurement \\ of the di-electron widths \\ of the
Upsilon(1S), Upsilon(2S), \\ and Upsilon(3S) mesons \\ at CLEO-III}

\abstracttitle{A high-precision measurement of the di-electron widths of the
Upsilon(1S), Upsilon(2S), and Upsilon(3S) mesons \\ at CLEO-III}

\newlength{\Mheight}
\newlength{\cwidth}
\newcommand{\emc}{\settoheight{\Mheight}{M}\settowidth{\cwidth}{c}M\parbox[b][\Mheight][t]{\cwidth}{c}}

\author{James \emc Cann Pivarski}

\begin{document}

\newcommand{\subs}[1]{{\mbox{\scriptsize #1}}}
\newcommand{\inv}{$^{-1}$}
\newcommand{\PM}{$\pm$}
\newcommand{\ups}{$\Upsilon$}
\newcommand{\gee}{$\Gamma_{ee}$}
\newcommand{\us}{$\Upsilon(1S)$}
\newcommand{\uss}{$\Upsilon(2S)$}
\newcommand{\usss}{$\Upsilon(3S)$}
\newcommand{\es}{$\epsilon_{1S}$}
\newcommand{\ess}{$\epsilon_{2S}$}
\newcommand{\esss}{$\epsilon_{3S}$}
\newcommand{\ee}{$e^+e^-$}
\newcommand{\mumu}{$\mu^+\mu^-$}
\newcommand{\tautau}{$\tau^+\tau^-$}
\newcommand{\gamgam}{$\gamma\gamma$}
\newcommand{\gggamma}{$gg\gamma$}
\newcommand{\qqbar}{$q\bar{q}$}
\newcommand{\bee}{${\mathcal B}_{ee}$}
\newcommand{\bmm}{${\mathcal B}_{\mu\mu}$}
\newcommand{\btt}{${\mathcal B}_{\tau\tau}$}
\newcommand{\bcas}{${\mathcal B}_\subs{cas}$}
\newcommand{\geehadtot}{$\Gamma_{ee}\Gamma_\subs{had}/\Gamma_\subs{tot}$}
\newcommand{\twotoone}{$\Upsilon(2S) \to \pi^+\pi^- \Upsilon(1S)$}
\newcommand{\pipi}{$\pi^+\pi^-$}
\newcommand{\evis}{$\epsilon_\subs{vis}$}
\newcommand{\ecuts}{$\epsilon_\subs{cuts}$}
\newcommand{\ebeam}{$E_\subs{beam}$}
\newcommand{\ecm}{$E_\subs{CM}$}
\newcommand{\pmax}{$|\vec{p}_\subs{max}|$}
\newcommand{\visen}{$E_\subs{vis}$}
\newcommand{\dxy}{$d_\subs{XY}$}
\newcommand{\dz}{$d_\subs{Z}$}
\newcommand{\vtd}{$V_{td}$}
\newcommand{\twotrack}{{\tt two-track}}
\newcommand{\hadron}{{\tt hadron}}
\newcommand{\radtau}{{\tt rad-tau}}
\newcommand{\eltrack}{{\tt $e^\pm$-track}}
\newcommand{\barrelbhabha}{{\tt barrel-bhabha}}
\newcommand{\axial}{{\tt AXIAL}}
\newcommand{\stereo}{{\tt STEREO}}
\newcommand{\cblo}{{\tt CBLO}}
\newcommand{\cbmd}{{\tt CBMD}}
\newcommand{\cbhi}{{\tt CBHI}}

\maketitle
\makecopyright
\begin{abstract}
  The di-electron width of an Upsilon meson is the decay rate of the
  Upsilon into an electron-positron pair, expressed in units of
  energy.  We measure the di-electron width of the Upsilon(1S) meson
  to be 1.354 $\pm$ 0.004 $\pm$ 0.020 keV (the first uncertainty is
  statistical and the second is systematic), the di-electron width of
  the Upsilon(2S) to be 0.619 $\pm$ 0.004 $\pm$ 0.010 keV and that of
  the Upsilon(3S) to be 0.446 $\pm$ 0.004 $\pm$ 0.007 keV.  We
  determine these values with better than 2\% precision by integrating
  the Upsilon production cross-section from electron-positron
  collisions over their collision energy.  Our incident electrons and
  positrons were accelerated and collided in the Cornell Electron
  Storage Ring, and the Upsilon decay products were observed by the
  CLEO-III detector.  The di-electron widths probe the wavefunctions
  of the Strongly-interacting bottom quarks that constitute the three
  Upsilon mesons, information which is especially interesting to check
  high-precision Lattice QCD calculations of the nuclear Strong force.
\end{abstract}
\begin{biosketch}
  James Adam \emc Cann was induced to be born on Friday, July
  2, 1976, as that weekend was one that even delivery surgeons wanted
  to have off.  His loving parents are Tom \emc Cann and
  Donna Scott.  James later married Melanie Pivarski, and in an effort
  to balance an overwhelming cultural practice, took her last name.
  Thus, he is now known as James \emc Cann Pivarski, and may
  be the only man on earth whose middle initial is an
  ``\emc.''

  James (Jim) was educated in public schools in Westfield, MA, later
  studied physics with a minor in mathematics at Carnegie Mellon
  University in Pittsburgh, PA, and finally experimental particle
  physics at Cornell University in Ithaca, NY.  With this document, he
  completes his twenty-fifth year of schooling.  Soon Jim and Melanie
  will move to College Station, TX, to do research in physics and
  mathematics (respectively) at Texas~A\&M University.
  
  Jim was originally interested in physics as a means of mystifying
  his understanding of the world, rather than making it clearer, which
  some of the conclusions of modern physics provide for.  His
  assumption that the universe was not what it seemed to be blended
  with his affinity for the existentialism of Albert Camus and his
  former skepticism of all experience, including subjective
  experience.  Trying to take these views seriously, he suffered a
  philosophical breakdown and reconsidered his world-view at a
  fundamental level.  Today, he begins by assuming that things are as
  they appear to be, modifying that assumption as observed
  complications arise.  This is why Jim is now an experimentalist, and
  it is also related to his conversion to Catholicism (when applied to
  subjective experience, rather than objective).  Before college,
  Jim's aspiration was to create special effects for movies, to
  convince the senses of something which isn't true.  Now he does the
  opposite.
\end{biosketch}
\begin{dedication}
  \includegraphics[width=\linewidth]{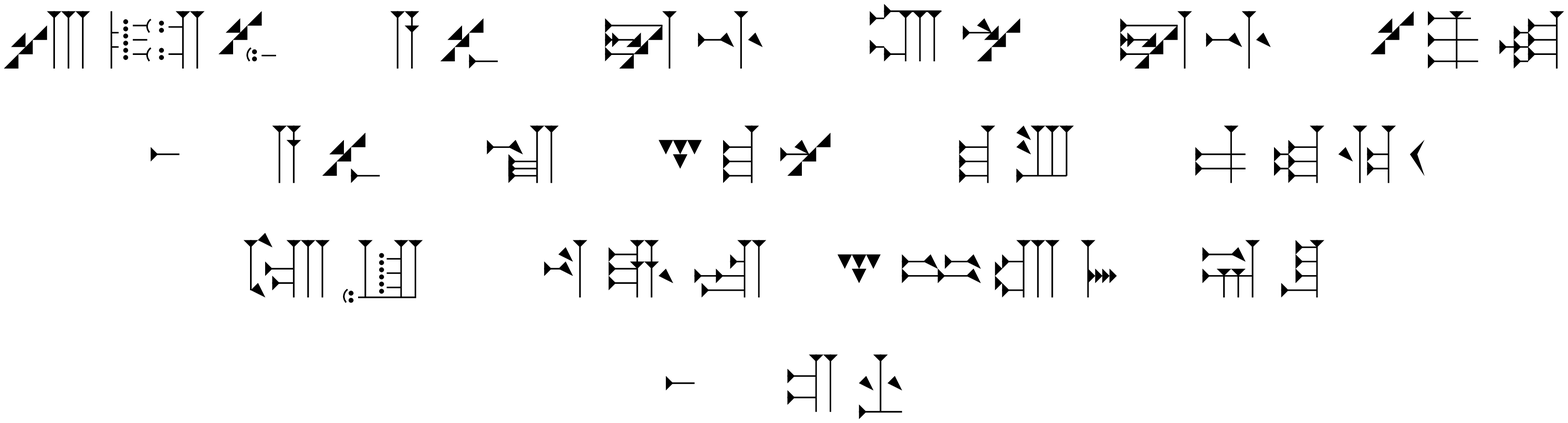}
\end{dedication}
\begin{acknowledgements}
  The following is by no means purely my own work.  The most input
  came (naturally) from Ritchie Patterson, my advisor, who taught me
  how to follow my nose by pointing which way to go.  Also very
  influential were Karl Berkelman and Rich Galik.  It was Karl's idea
  to measure hadronic efficiency with \twotoone\ transitions, and he
  was very involved in radiative corrections and resonance-continuum
  interference: in fact, he wrote the routine that we use to compute
  them both.  Rich organized this project at its earliest stages and
  chaired the committee that oversaw the development of a publication.
  Brian Heltsley, Istvan Danko, and Surik Mehrabyan also gave a great
  deal of input.  I should also note that Brian and Surik were the
  primary persons involved in determining integrated luminosity from
  Bhabha counts (Chapter~\ref{chp:luminosity}).  This document quotes
  their work (CLEO internal note CBX 05-17).

  Widening the circle, this project couldn't even begin without an
  operational collider and detector (Chapter~\ref{chp:hardware}), so
  the entire staff of the Cornell Electron Storage Ring and the CLEO
  collaboration in a real way helped to make this analysis possible.
  In particular, Stu Peck tuned the beam to optimize running
  conditions for every \ecm\ increment, and he was very patient with
  our special requests.  Also, Mike Billing took the time to teach me
  how to use the beam simulation when I was concerned about changes in
  beam energy spread.

  It was also helpful (and stimulating) to learn about the Lattice QCD
  calculation which motivated our experient.  For this, I thank Peter
  Lepage and Christine Davies for their detailed explainations.

  This work was supported by the A.P.~Sloan Foundation, the National
  Science Foundation, and the U.S. Department of Energy.
\end{acknowledgements}
\contentspage
\tablelistpage
\figurelistpage

\normalspacing
\setcounter{page}{1}
\pagenumbering{arabic}
\pagestyle{cornellc}

\chapter{Introduction and Motivation}
\label{chp:introduction}

\section{The \boldmath \ups\ Di-electron Width and Why it is Important}

An Upsilon (\ups) meson is a composite particle consisting of a bottom quark
($b$) and an anti-bottom quark ($\bar{b}$) bound with their spins
aligned in a $J=1$ quantum mechanical wavefunction, where $J$ is the
total angular momentum.  This meson is a nuclear analogy of
ortho-positronium in atomic physics.  The di-electron width is the
rate of \ups\ decay into an electron/positron pair, and measuring it
provides unique experimental access to the physical size of the
$b\bar{b}$ wavefunction and its total decay rate--- the average
extension of the \ups\ meson in both space and time.

The $b\bar{b}$ system, also known as bottomonium, is the most
non-relativistic system of quarks bound by the nuclear Strong force.
This is because the bottom quark is the heaviest quark that can
participate in the Strong force, the top decaying immediately into
bottom by the Electroweak force.  Unlike much more abundant protons
and neutrons, whose masses consist almost entirely of the kinetic
energy of the constituent quarks and gluons, 94\% of the mass of the
lightest \ups\ consists of the mass of its two bottom quarks.  This
simplifies the dynamics of bottomonium and even permits description in
terms of a potential, making it a good testing ground for Strong force
calculations.

Quantum Chromodynamics (QCD) has long been accepted as the correct
description of the nuclear Strong force (with possible modifications
only at TeV energies and above) because of its success in predicting
scattering interactions above one GeV and its qualitative explanation
of low-energy phenomena like quark confinement.  Today, the Lattice
QCD technique, which simulates QCD on a computer, is yielding
few-percent calculations of low-energy phenomena from first
principles--- in particular, \ups\ properties such as the di-electron
widths.  Precise experimental knowledge of the \ups\ di-electron
widths will test the new Lattice QCD techniques that made this advance
possible.

The di-electron widths check Lattice QCD in a way that is key for
Electroweak physics.  The CP violation parameters \vtd\ and $V_{ub}$,
fundamental constants in the Standard Model, could be extracted from
existing hadronic measurements much more precisely if the strength of
the force between quarks were better known.  Lattice QCD can help, but
precise Lattice results will only be trusted if similar calculations
can be experimentally verified.  The \ups\ di-electron width closely
resembles the factor that limits our knowledge of \vtd, and thus will
provide a cross-check that will either lend credence to or cast doubt
on the \vtd\ extraction.

Di-electron widths of the \ups\ resonances have been measured before,
but not with the precision that is now being demanded by Lattice QCD.
This document represents a comprehensive study of the \us, \uss, and
\usss\ di-electron widths, with 50 times the data of any previous
measurement.  We present di-electron width measurements of the \us,
\uss, and \usss\ with 1.5\%, 1.8\%, and 1.8\% total uncertainty,
respectively.  This is the second-ever measurement of the \usss\
di-electron width, improving its precision by a factor of five.
Furthermore, measuring all three resonances in the same study permits
us to derive very precise ratios of di-electron widths, where the
tightest constraint on theory is likely to be.  Without this
measurement, comparisons with Lattice QCD would probably be limited by
experiment.

\section{The Bottomonium Potential and Mass Eigenstates}

\begin{figure}[p]
  \begin{center}
    \includegraphics[width=\linewidth]{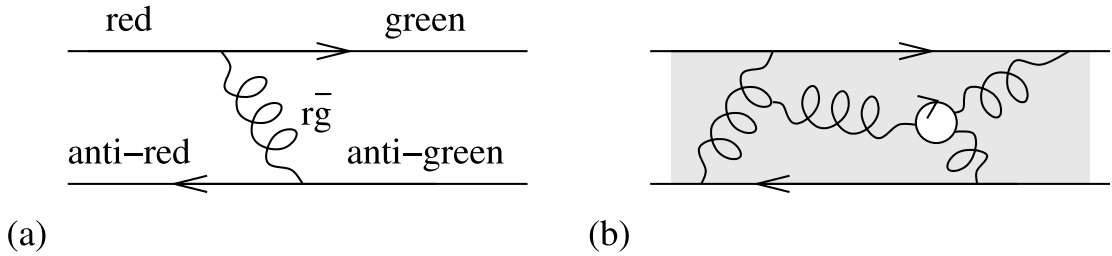}
  \end{center}
  \caption[Color exchange and the complexity of QCD]{\label{gluejunk} (a)
  An example of a gluon as a force propagator and a carrier of color
  charge.  Directed lines are quarks and springs are gluons.  (b) An
  exchange between two quarks involving a complicated network of
  gluons and a light quark/anti-quark pair.  Grey shading indicates
  the sum of all such amplitudes.}
\end{figure}

The $b$-quark and the $\bar{b}$-quark in bottomonium attract each
other by the nuclear Strong force, which in QCD is mediated by gluons,
the nuclear analogy of virtual photons.  The two quarks are charged
with opposite ``colors,'' in a quantum mechanical superposition of
red/anti-red, blue/anti-blue, and green/anti-green states, which are
constantly traded for each other by the doubly-colored gluons.  (The
exchange of a red/anti-green gluon will turn a red/anti-red $b\bar{b}$
system green/anti-green, for instance.  See Figure~\ref{gluejunk}(a).)
Because the gluons carry color charge, they can interact with other
gluons and spawn complicated networks of interactions between the two
quarks (Figure~\ref{gluejunk}(b)), which increases the interaction
strength with distance.  Bottom quarks are usually separated by a
femtometer, and at this distance scale, the coupling constant of QCD
is of order unity.  Feynman diagrams with many vertices are not
suppressed relative to simple diagrams, and therefore a calculation of
the force between the two quarks does not submit to a perturbative
expansion.

\begin{figure}[p]
  \begin{center}
    \includegraphics[width=0.8\linewidth]{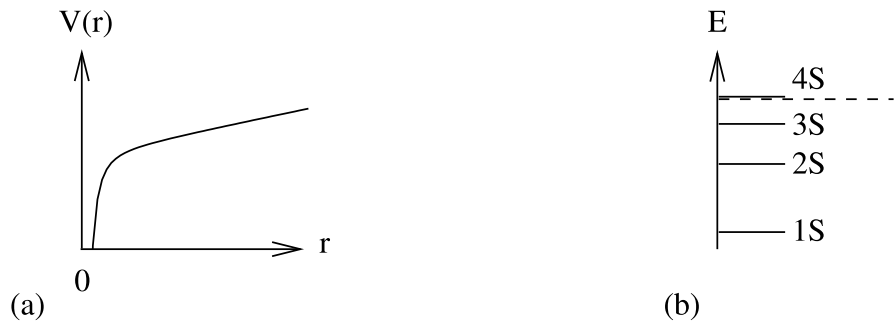}
  \end{center}
  \caption[Bottomonium potential and energy
  eigenvalues]{\label{cornellpotential} (a) Schematic of the potential
  energy between two quarks as a function of separation.  (b)
  Quantitative level diagram of $J=1$ $b\bar{b}$ solutions (\ups).
  The $\Upsilon(4S)$ lies just above the string-breaking threshold
  (dashed line).}
\end{figure}

A very successful model of the force between $b$ and $\bar{b}$
consists of a Coulomb-like potential at short distances (though about
50 times stronger) and a linear potential at large distances (which
limits to a constant force of about 14 tons), as illustrated in
Figure~\ref{cornellpotential}(a) \cite{cornellpotential}.  At large
distances, a string of self-interacting gluons, stretched between the
two quarks, is responsible for the linear component.  This string will
generate a real quark/anti-quark pair and snap if stretched with
sufficient energy.  There are three $J=1$ solutions to the
Schr\"odinger equation below this threshold: they are labeled \us,
\uss, and \usss.  States above this threshold have a very different
pattern of decay modes and are beyond the scope of this study.

These three mass eigenstates are the bottomonium equivalent of atomic
energy levels--- discrete lines of allowed mass-energies.  But, just
as in atomic spectra, the short lifetimes of these states imply a
broadening of their spectral lines: they are not perfect
time-independent solutions.  The full-width of an \ups\ resonance at
half-maximum, $\Gamma$, is equal to its decay rate, in analogy with
excited atomic resonances.  If we partition the \ups\ decays into
distinguishable modes, one being $\Upsilon \to e^+e^-$, the total
decay rate is a sum of those modes.  Hence, $\Gamma_{ee} = \Gamma \,
{\mathcal B}_{ee}$, where \gee\ is the di-electron width and \bee\ is
the fraction of \ups\ mesons that decay to \ee, that is, the branching
fraction to \ee.
  
\begin{figure}[p]
  \begin{center}
    \includegraphics[width=0.5\linewidth]{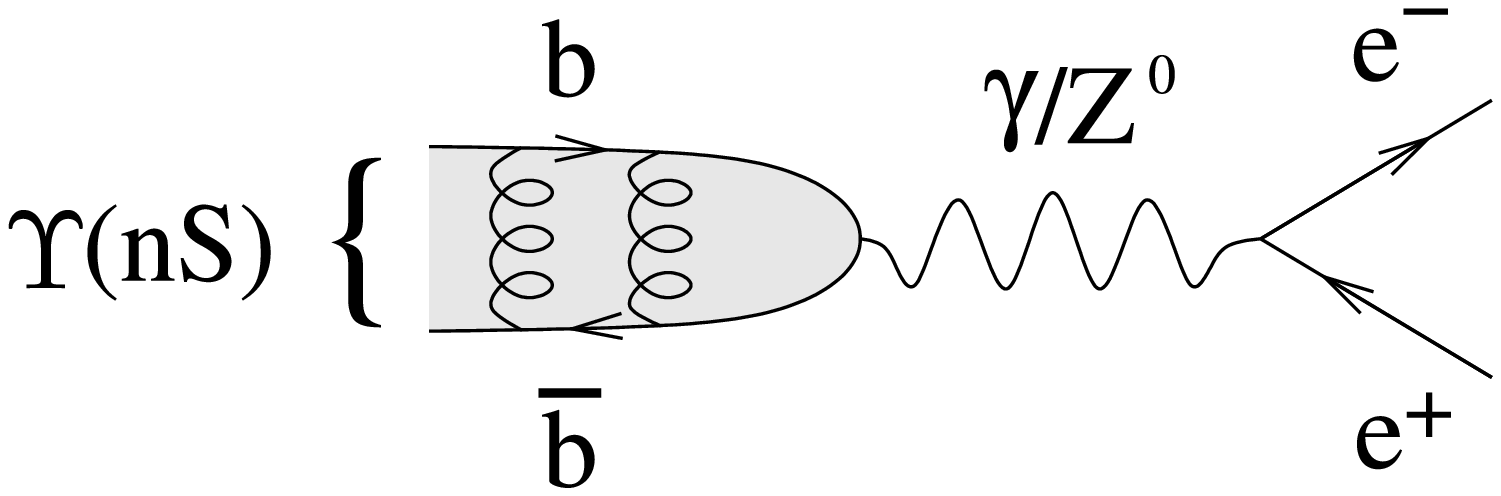}
  \end{center}
  \caption[Decay diagram of $\Upsilon(nS) \to e^+e^-$]{\label{diagramgee}
  Decay diagram of $\Upsilon(nS) \to e^+e^-$.  The $Z^0$ contribution is
  0.25\% of the total rate.}
\end{figure}

The \ups\ meson decays into \ee\ by $b\bar{b}$ annihilation
(Figure~\ref{diagramgee}), which is a point-like interaction.  The $b$
and the $\bar{b}$ must fluctuate to the same point in space for the
reaction to proceed.  This probability, which is the square of the
$b\bar{b}$ spatial wavefunction evaluated at the origin
($|\psi(0,0,0)|^2$), is therefore a factor in \gee.
\begin{equation}
  \Gamma_{ee} = 3 {Q_b}^2 \, \frac{16 \pi \alpha^2}{3} \, \frac{|\psi(0,0,0)|^2}{{M_\Upsilon}^2}
  \label{eqn:waveatorigin}
\end{equation}
where $Q_b=1/3$, the $b$-quark electric charge, $\alpha$ is the
Electromagnetic fine structure constant and $M_\Upsilon$ is the \ups\
mass \cite{ps}.  This is a non-relativistic approximation:
relativistic corrections replace the wavefunction at the origin with
an integral of values very close to the origin.  Because of this
dependence on knowledge of the $b\bar{b}$ wavefunction, and therefore
the potential, a first-principles calculation \gee\ will require
non-perturbative QCD.

\section{Lattice QCD}

\begin{figure}[p]
  \begin{center}
    \includegraphics[width=\linewidth]{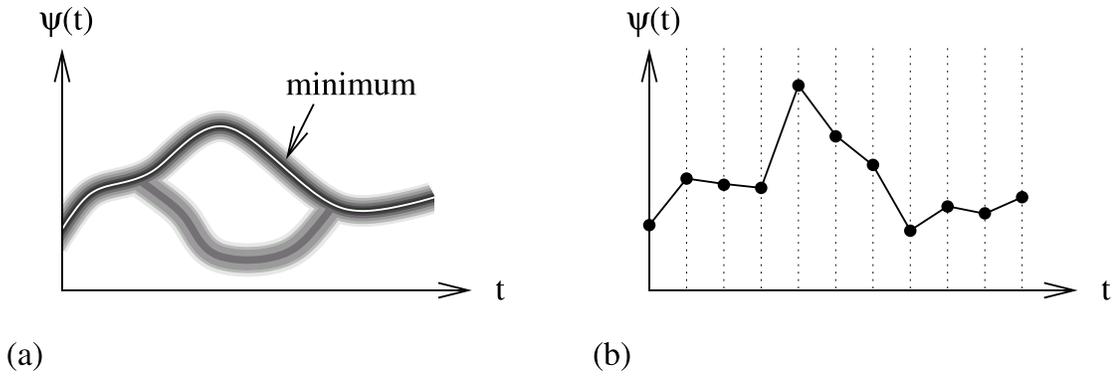}
  \end{center}
  \caption[Path integrals and path integrals in discretized
  space-time]{\label{pathintegrals} (a) Paths in field strength
  ($\psi$) versus time.  The minimum (white) is the classical solution
  and the path which contributes the most to the quantum amplitude;
  shades of grey represent other quantum paths with smaller
  contributions to the total amplitude.  (b) A path approximated by a
  discrete time sequence.}
\end{figure}

Feynman path integrals provide a general approach to quantum field
theory that don't rely on a perturbative expansion.  In this
formalism, the amplitude of a process is calculated as a weighted sum
of all possible ways it can proceed.  The value of every field at
every point in space in the initial state is allowed to vary as an
arbitrary function of time to the final state, and these paths are
weighted by their action.  This is a generalization of Lagrange's
method in classical physics, in which the true path is the one which
minimizes action.  In quantum physics, all paths which nearly minimize
action contribute to the amplitude of a process (Figure~\ref{pathintegrals}(a)).

To calculate a sum over a set of arbitrary paths, one must discretize
space-time into time slices and space cubes.  The path of a field
value in a space cube from the initial state to the final state is a
sequence of values for each time slice (Figure~\ref{pathintegrals}(b)).
A sequence of $N$ values is a vector in an $N$-dimensional vector
space: the weighting factor is integrated over these vector spaces.
To obtain a realistic result, one must afterward limit the
discretization scale to zero.

In general, realistic problems like QCD, the integral will not be
analytic and must be solved by numerical integration.  The integral
will have a fixed number of dimensions, which implies a fixed
discretization of space-time that can only be lifted by extrapolating
several calculations toward zero lattice size.  This discretization is
the lattice of Lattice QCD.  Quark field values are represented on a
four-dimensional lattice of space-time points with gluon field values
on the edges connecting them.

This is a very computationally intensive problem, since the number of
dimensions in the integral scales with the number of grid points, and
one must maximize the number of grid points to extrapolate to the
continuum limit.  Over the past 30 years, theorists have improved the
algorithms of Lattice QCD and sought approximations to make realistic
calculations tractable.

\begin{figure}[t]
  \begin{center}
    \includegraphics[width=0.35\linewidth]{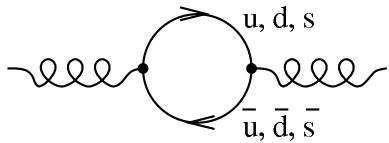}
  \end{center}
  \caption[Vacuum polarization by light
  quarks]{\label{vacuumpolarization} Vacuum polarization by light
  quarks.}
\end{figure}

The most time-consuming part of most Lattice QCD calculations is the
polarization of the vacuum by light quarks.  In terms of Feynman
diagrams, these are interruptions of a gluon propagator by loops of
$u\bar{u}$, $d\bar{d}$ and $s\bar{s}$ pairs, which can be ignored or
suppressed by assuming infinite or very heavy up, down, and strange
quark masses (Figure \ref{vacuumpolarization}).  This approximation is
known as the quenched approximation, and it permits calculations with
10--20\% systematic uncertainties.

\begin{figure}[p]
  \begin{center}
    \includegraphics[width=\linewidth]{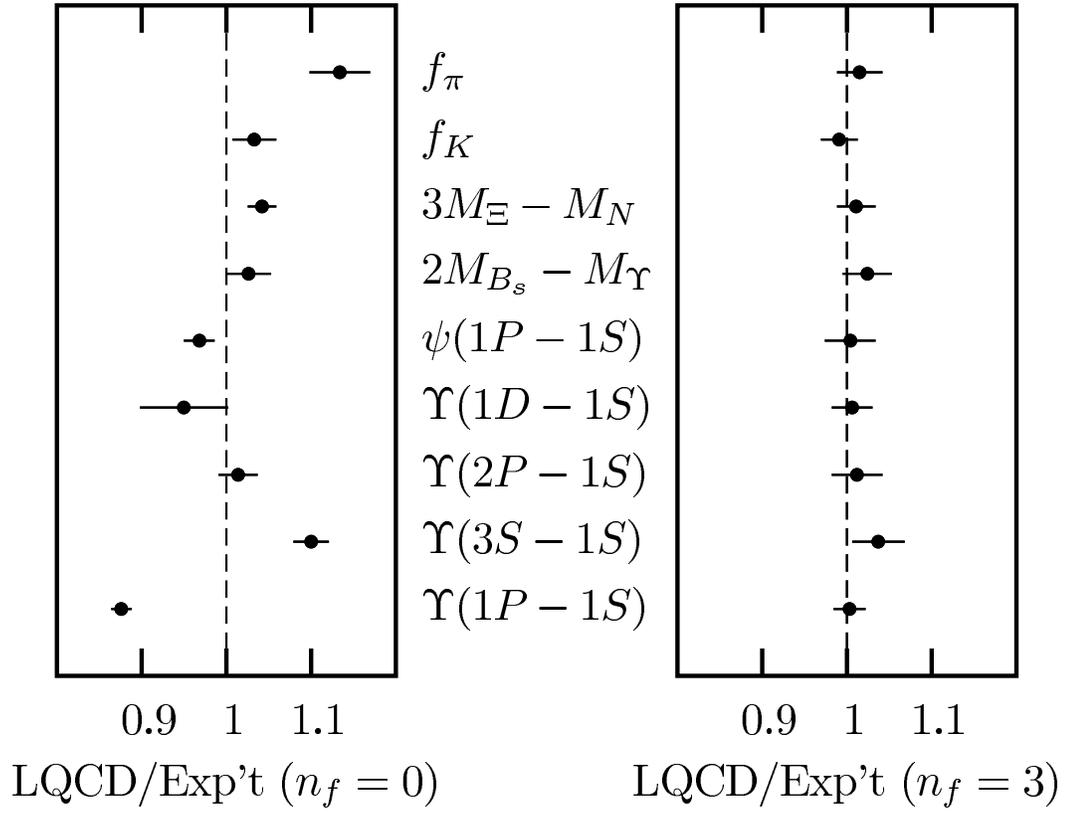}
  \end{center}
  \caption[Lattice QCD results with and without vacuum
  polarization]{\label{latticevictory} Lattice QCD calculations
  divided by experimental measurements for nine quantities, without
  and with quark vacuum polarization (left and right panels,
  respectively).  (The notation $\Upsilon(n - 1S)$ refers to the mass
  difference between $\Upsilon(n)$ and \us.)}
\end{figure}

This situation was dramatically improved in the late 1990's by the
development of new algorithms based on the Symanzik-improved
staggered-quark formalism \cite{confronts}.  These algorithms are by
far the most efficient known, and the formalism features an exact
chiral symmetry that particularly benefits simulations with small
light quark masses.  Realistic up and down quark masses are still out
of reach, but simulations using masses three times too large can be
accurately extrapolated with chiral perturbation theory.  Thus,
``unquenched'' calculations are now possible, resulting in the
accurate prediction of many masses and decay rates, as demonstrated in
Figure~\ref{latticevictory}.

This algorithmic speed comes at a conceptual price: the
staggered-quark formalism introduces four equivalent species of each
quark field, called ``tastes.''  These are artifacts of the formalism
and are unrelated to quark flavor.  Each of these tastes contributes
to the vacuum polarization, resulting in loop contributions which are
four times too large.  To correct for this, the quark determinant in
the action is replaced by its fourth root, a procedure which is
rigorous in the free-field theory and in perturbative QCD, but
introduces violations of Lorentz symmetry at short distances in the
lattice simulation.  Although these non-physical effects can be
removed by interpolating between the lattice points with perturbative
QCD, this issue makes the new algorithms controversial, and it is one
aspect that will be tested by confrontations with experiment.

The di-electron width may be determined from \ups\ simulations by
extracting the $b\bar{b}$ wavefunction at the origin and applying
Equation~\ref{eqn:waveatorigin}.  In a path integral context, the
wavefunction is the quark field amplitude.  These simulations employ a
non-relativistic QCD action with relativistic corrections (NRQCD),
because the de Broglie wavelength of a massive $b$ quark would require
impractically narrow time slices.

\begin{figure}[p]
  \begin{center}
    \includegraphics[width=\linewidth]{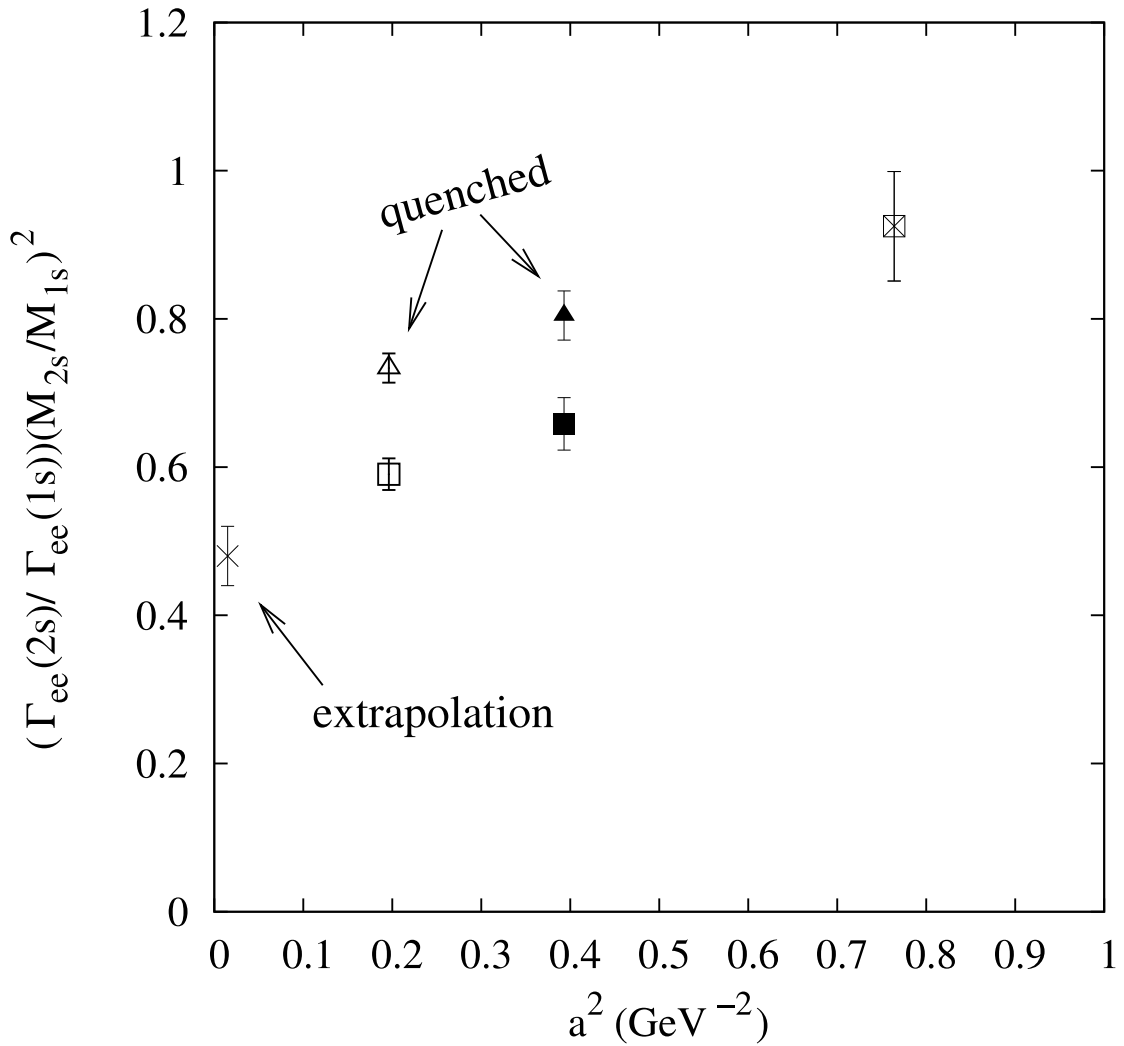}
  \end{center}
  \caption[HPQCD results for
  $\Gamma_{ee}(2S)/\Gamma_{ee}(1S)$]{\label{latticespacing} HPQCD
  calculations of $\Gamma_{ee}(2S)/\Gamma_{ee}(1S)$ times the ratio of
  masses squared as a function of lattice grid size $a$.  Square data
  points represent calculations with light quark masses close to their
  natural values, and triangular points are quenched (infinite light
  quark masses).}
\end{figure}

Simulations of the \ups\ mesons have been generated by the HPQCD
collaboration, but the determination of \gee\ from them is not yet
complete \cite{ukqcd}.  To properly calculate \gee, one needs to
correct the lattice wavefunction for discretization effects through a
constant, $Z_\subs{match}^\subs{vector}$, that matches the lattice
approximations of the virtual photon current to a continuum
renormalization scheme.  The leading term in
$Z_\subs{match}^\subs{vector}$ is on the order of the Strong coupling
constant $\alpha_s$, about 20\%.  This calculation is in progress.
However, $Z_\subs{match}^\subs{vector}$ largely cancels in ratios of
\gee: for instance, \gee\ from the \uss\ divided by \gee\ from the
\us\ can already be extracted with a 10\% uncertainty.
\begin{equation}
  \left. \frac{\Gamma_{ee}(2S)}{\Gamma_{ee}(1S)}
  \right|_\subs{HPQCD} = 0.43 \pm 0.04 \mbox{.}
  \label{eqn:latticeresult}
\end{equation}
This uncertainty is primarily due to residual discretization errors,
evident from the steep dependence of the result on lattice spacing
size (Figure~\ref{latticespacing}).  When the discretization
correction has been calculated, the uncertainty in this ratio should
be only a few percent, while absolute \gee\ values should have
uncertainties on the 10\%-level.  This is why experimentally precise
ratios of \gee\ are also valuable.  The ratios test the NRQCD
treatment of the $b$ quarks with high precision, though they are less
sensitive to the $Z_\subs{match}^\subs{vector}$ corrections.

\section{Relationship to Electroweak Parameter Extraction}

\begin{figure}[t]
  \begin{center}
    \includegraphics[width=0.4\linewidth]{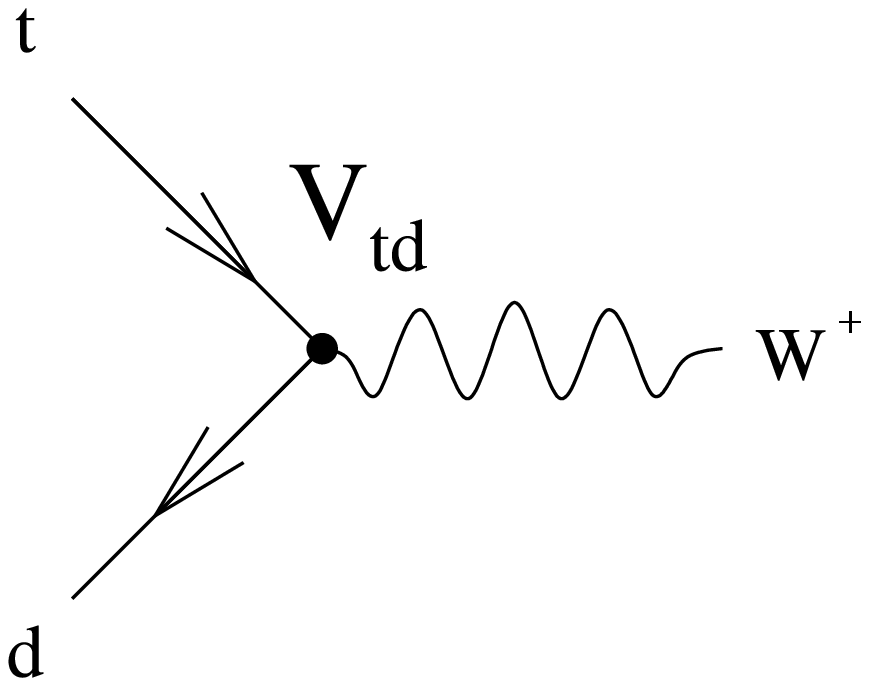}
  \end{center}
  \caption[Vertex joining top, down, and $W^\pm$]{\label{vtdvertex}
  The vertex joining top, down, and $W^\pm$.}
\end{figure}

Lattice verification of \gee\ is particularly significant for an
application of the technique to Electroweak physics.  Vertices in
Feynman diagrams that join a top quark, a down quark, and a $W$ boson
contribute an a priori unknown factor, \vtd, to the amplitude
(Figure~\ref{vtdvertex}).  This parameter is a fundamental constant in
the Standard Model and is essential for violation of charge-parity
(CP) symmetry: if \vtd\ were zero, the Standard Model would be CP
symmetric (exchanging particles for antiparticles and
mirror-transforming space would preserve all observables).  To
determine \vtd, one must resort to measurements of bound quark
systems, because bare quarks do not exist in nature.  The transition
rates for these systems depend both on \vtd\ and on QCD factors
related to the structure of the bound state.  Lattice QCD can
calculate these factors and thereby extract \vtd.

\begin{figure}[p]
  \begin{center}
    \includegraphics[width=0.7\linewidth]{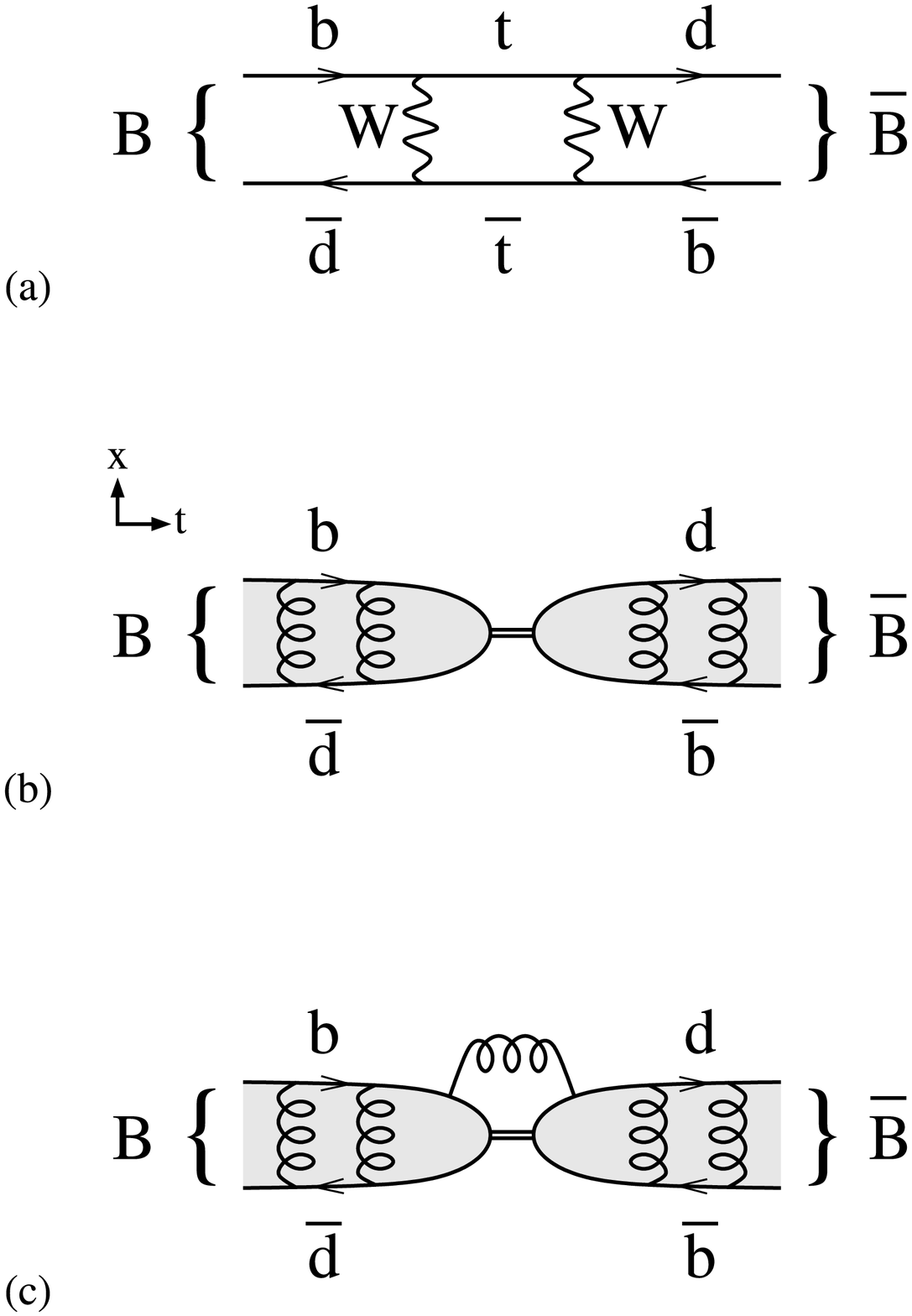}
  \end{center}
  \caption[$B^0$-$\bar{B^0}$ mixing diagrams and the meaning of
  $f_B$]{\label{bmixing} (a) One of the two diagrams dominating
  $B^0$-$\bar{B^0}$ mixing (the other exchanges $t \leftrightarrow
  W$).  (b) The same diagram emphasizing the Strong force between the
  quarks and the relative size of the $t$-$W$ loop.  (c) Diagrams that
  contribute to the Bag parameter, which is not a part of $f_B$.}
\end{figure}

The most sensitive probe of \vtd\ is $B^0$-$\bar{B^0}$ mixing.  A
$B^0$ meson is a bound state of $d$ and $\bar{b}$ quarks, and a
$\bar{B^0}$ meson is its charge conjugate, $b\bar{d}$.  These two
mesons can mix, that is, a $B^0$ can transform into a $\bar{B^0}$ and
vice-versa, through the diagram illustrated in
Figure~\ref{bmixing}(a).  The heavy top quark dominates in this loop
and provides a factor of \vtd\ for each vertex with a down quark.  The
rate of this process is extremely well-known: 0.509 \PM\ 0.004~ps\inv,
a 1\% measurement \cite{hfag}.

Despite this precision, the \vtd\ extraction has 20\% uncertainties
from Strong interactions.  To illustrate the influence of the Strong
interaction on $B^0$-$\bar{B^0}$ mixing, we re-draw the Feynman
diagram as a space-time diagram in Figure~\ref{bmixing}(b).  The
$W$-$t$ loop is a very short-range process ($\sim$0.001~fm).  By
comparison, the average distance between the $b$ and $\bar{d}$ quarks
is set by the QCD potential ($\sim$fm), just as it is for $b\bar{b}$
in the \ups\ meson.  Just as in \gee, the rate of $B^0$-$\bar{B^0}$
mixing is determined by the probability that the two quarks will
fluctuate to the same point in space, and this probability is
characterized by the $B$ meson decay constant $f_B$.
\begin{equation}
  \Gamma(B \to \bar{B}) = \mbox{(known factors)} \times \bigg|
  \underbrace{{f_B}^2 B_B}_\subs{QCD} \times {V_{td}}^2 \bigg|^2
  \mbox{.}
\end{equation}
The $B^0$-$\bar{B^0}$ mixing amplitude depends on two factors of
$f_B$, one from the $B^0$ wavefunction and the other from $\bar{B^0}$
(see Figure~\ref{bmixing}(b)).  Another factor, known as the Bag
parameter $B_B$, corrects for gluons connecting the $B^0$ and
$\bar{B^0}$ (Figure~\ref{bmixing}(c)).  Its uncertainty is more easily
controlled.  Our knowledge of \vtd\ is therefore dominated by the
uncertainty in~$f_B$.

\begin{figure}[p]
  \begin{center}
    \includegraphics[width=0.5\linewidth]{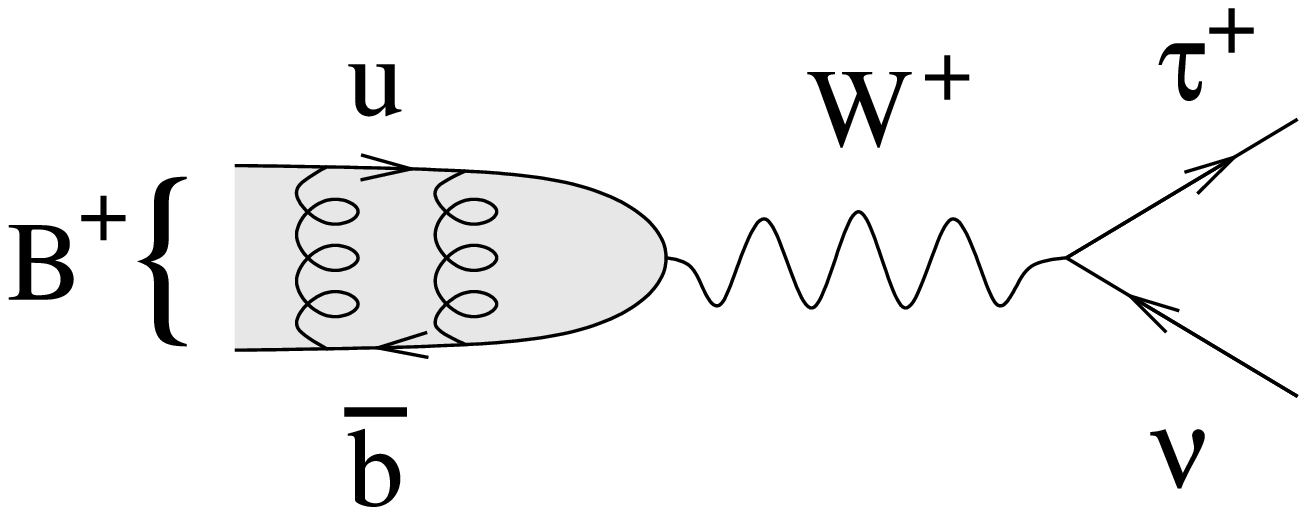}
  \end{center}
  \caption[Decay diagram of $B^+ \to \tau^+ \nu$]{\label{btomunu}
  Decay diagram of $B^+ \to \tau^+ \nu$.}
\end{figure}

In principle, one can measure $f_B$ experimentally through $B^+ \to
\mu^+ \nu$ or $\tau^+ \nu$, illustrated in Figure~\ref{btomunu}.  The
charged $B^+$ has different quark content from the neutral $B^0$, but
its decay rate depends on $f_B$ because up and down quark masses are
both much smaller than the bottom quark mass, and flavors do not enter
the QCD calculation.  Unfortunately, this process is suppressed by
$V_{ub}$, to the extent that it has yet to be observed in 88.9~million
$B^\pm$ decays at BaBar \cite{btaunu}.  Given the low rate of this
decay and the challenge of reconstruction, it will take a long time to
accumulate enough data to make a statistically significant measurement
of~$f_B$.

\begin{figure}[p]
  \begin{center}
    \includegraphics[width=\linewidth]{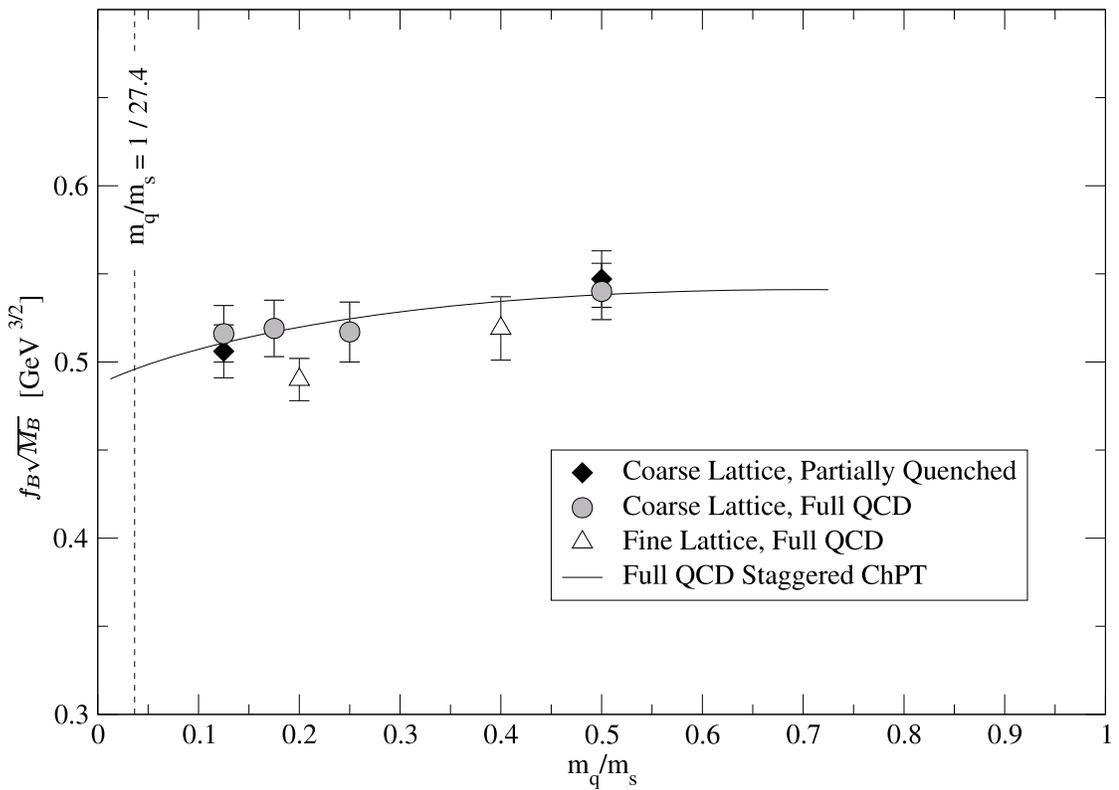}
  \end{center}
  \caption[Extrapolation of Lattice QCD results to physical light
  quark masses]{\label{fbresults} HPQCD calculations of $f_B \sqrt{M_B}$ as
  a function of the light quark mass used in the simulations.  The
  solid curve extrapolates from the simulations to the natural
  $m_q/m_s$ of 1/27.4.}
\end{figure}

The $B$ decay constant may instead be extracted from Lattice QCD
simulations of $B$ mesons in much the same way as \gee\ is from \ups\
simulations: by sampling the wavefunction at the origin.  The HPQCD
collaboration has found $f_B$ to be 216 \PM\ 22~MeV (see
Figure~\ref{fbresults}) \cite{fb}.  Like the \ups, the $B$ meson is
modeled with NRQCD, and the discretization issues and corrections to
this calculation are analogous to \gee.  The largest uncertainty in
$f_B$ is in the $Z_\subs{match}^\subs{axial}$ constant that
matches lattice approximations of the virtual $W^\pm$ current to a
continuum renormalization scheme.  This $Z_\subs{match}^\subs{axial
vector}$ has been calculated to first order in $\alpha_s$, but
uncertainty in the ${\alpha_s}^2$ term imposes a 9\% uncertainty on
$f_B$, which dominates the 10\% uncertainty cited above.  Finer
calculations of $f_B$ are in progress.

Lattice calculations of $f_B$ would be viewed with suspicion if
calculations of \gee\ do not match experiment at a comparable level of
precision.  From the lattice's perspective, the only difference
between these two calculations is the mass of one of the two quarks: a
bottom quark is replaced by a light quark.  This is not a trivial
distinction: it is also worthwhile checking the lattice calculation of
the $D$ meson decay constant, in which a charm quark and a light quark
annihilate, with experimental results from CLEO-c that are now
becoming available \cite{fd}.  The $D$ meson is a heavy/light quark
combination, just like the $B$ meson, so $f_D$ is physically more
analogous to $f_B$ than \gee\ is.  However, the $D$ meson is a more
relativistic system, the charm quark being four times lighter than
bottom, so instead of simulating charm quarks with NRQCD, the $D$
meson simulations use a relativistic approximation known as the
FermiLab action \cite{confronts}.  Thus, \gee\ tests the treatment of
heavy quarks in the $f_B$ calculation and $f_D$ tests the heavy-light
simulation and matching the virtual $W^\pm$ current to the continuum
with $Z_\subs{match}^\subs{axial}$.  Ratios of \gee\ are
particularly applicable to this test, since they will be more
sensitive to the treatment of heavy quarks than to
$Z_\subs{match}^\subs{vector}$ in \gee.  Experimental verification of
\gee\ and \gee\ ratio calculations are therefore key to our confidence
in $f_B$ and the extraction of \vtd.

\chapter{Measurement Technique}
\label{chp:technique}

\section{Scans of \boldmath \ups\ Resonance Production}

\begin{figure}[p]
  \begin{center}
    \includegraphics[width=0.6\linewidth]{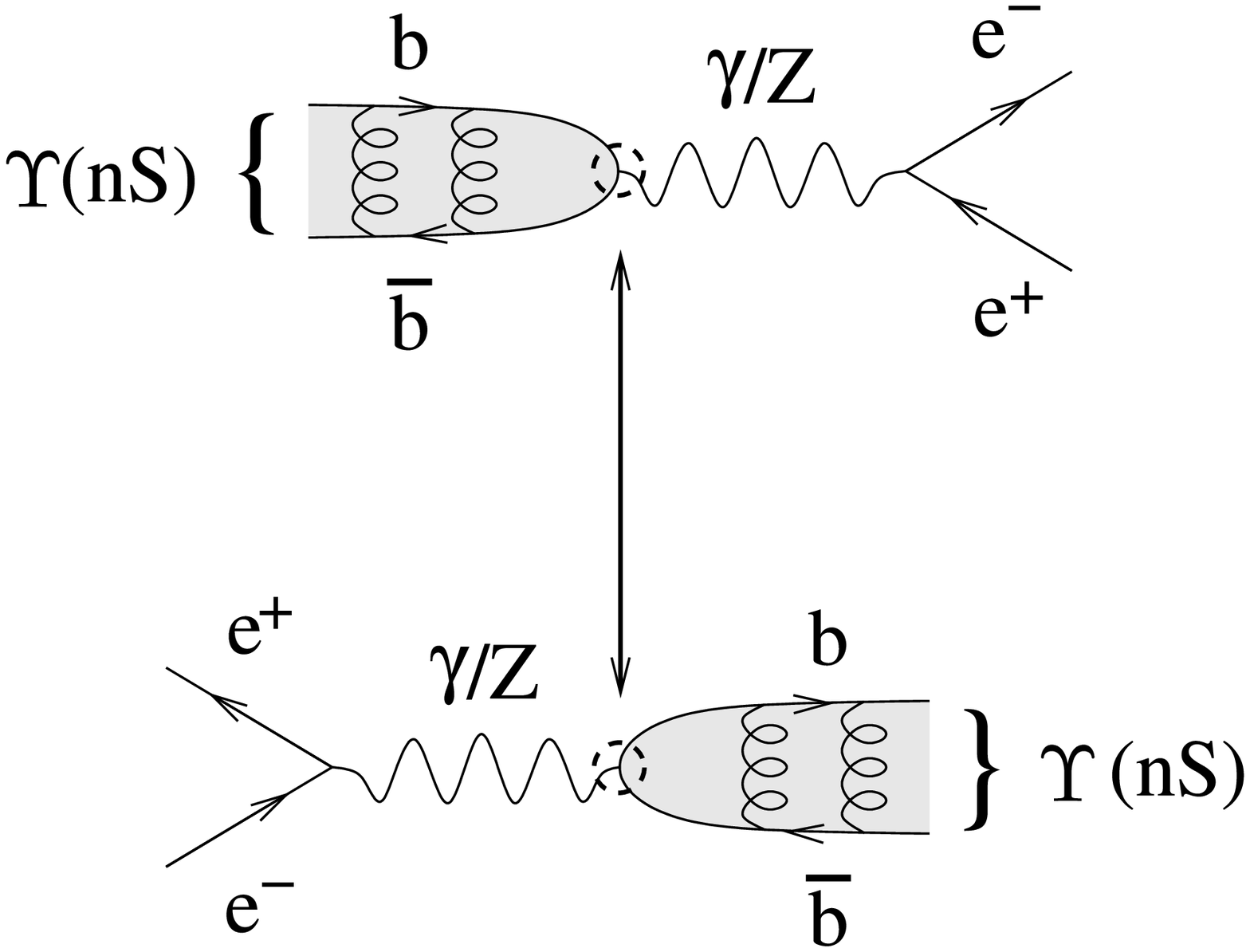}
  \end{center}
  \caption[Diagrams of $\Upsilon \to e^+e^-$ and $e^+e^- \to
  \Upsilon$]{\label{timereversed} Diagrams of $\Upsilon \to e^+e^-$
  and $e^+e^- \to \Upsilon$.  Both feature the same
  $b$-$\bar{b}$-$\gamma/Z^0$ vertex whose rate is set by
  $|\psi(0,0,0)|^2$.}
\end{figure}

To determine the decay rate of $\Upsilon \to e^+e^-$, we use a special
strategy available to \ee\ colliders: we measure the total
cross-section of $e^+e^- \to \Upsilon$, the time-reversed process.
This cross-section is related to the $b\bar{b}$ wavefunction at the
origin for the same reason as \gee\ (Figure~\ref{timereversed}, an
application of crossing symmetry).
\begin{equation}
  \int \sigma(e^+e^- \to \Upsilon) \, dE_\subs{CM} = 3{Q_b}^2 \, 64
  \pi^3 \alpha^2 \frac{|\psi(0,0,0)|^2}{{M_\Upsilon}^4}
  \label{eqn:crosssecwaveorigin}
\end{equation}
where the $dE_\subs{CM}$ integration is performed over \ee\
center-of-mass energies \cite{ps}.  To obtain \gee\ in terms of $\int
\sigma(e^+e^- \to \Upsilon) \, dE_\subs{CM}$, we combine the above
with Equation~\ref{eqn:waveatorigin}.
\begin{equation}
  \Gamma_{ee} = \frac{{M_\Upsilon}^2}{6\pi^2} \int \sigma(e^+e^- \to
  \Upsilon) \, dE_\subs{CM} \mbox{.}
  \label{eqn:gee}
\end{equation}
This is more general than Equations \ref{eqn:waveatorigin} and
\ref{eqn:crosssecwaveorigin}, which only apply in the non-relativistic
limit.  In the fully relativistic case, $|\psi(0,0,0)|^2$ must be
replaced with an integral of wavefunction values near the origin,
which cancels in Equation~\ref{eqn:gee}.

This may seem like a very indirect way of measuring \gee.  Why do we
not count $\Upsilon \to e^+e^-$ decays relative to the number of \ups\
mesons produced, for instance?  The reason is because such a fraction
would be \bee, rather than the decay rate.  We would need to multiply
by $\Gamma$, the rest mass distribution of the \ups\ meson, to
determine \gee, and this is experimentally inaccessible: $\Gamma$ is
on the order of 50 keV, which is about a hundred times narrower than
the beam energy spread of an \ee\ collider and a thousand times
narrower than detector resolution.  Equation~\ref{eqn:gee} provides
direct access to \gee, which, as a collateral benefit, can be combined
with \bee\ to obtain $\Gamma$.

To evaluate $\int \sigma(e^+e^- \to \Upsilon) \, dE_\subs{CM}$, we fit
a curve to the \ups\ lineshape, that is, the \ups\ production
cross-section as a function of \ee\ collision energy.  We then
integrate this curve analytically.  To construct our fit function, we
begin with the natural lineshape of the \us, \uss, and \usss\
resonances, a Breit-Wigner:
\begin{equation}
  \sigma(e^+e^- \to \Upsilon)(E_\subs{CM}) = \left(
  \frac{6\pi^2}{{M_\Upsilon}^2} \, \Gamma_{ee} \right)
  \frac{\Gamma/2\pi^2}{(E_\subs{CM} - M_\Upsilon)^2 + (\Gamma/2)^2} \mbox{.}
  \label{eqn:breitwigner}
\end{equation}
The observed spectrum is smeared by a unit Gaussian spread in incident
beam energies ($\sim$4 MeV), as discussed above.  We represent this
smearing by a convolution, but the integral is unchanged.

\begin{figure}[p]
  \begin{center}
    \includegraphics[width=0.7\linewidth]{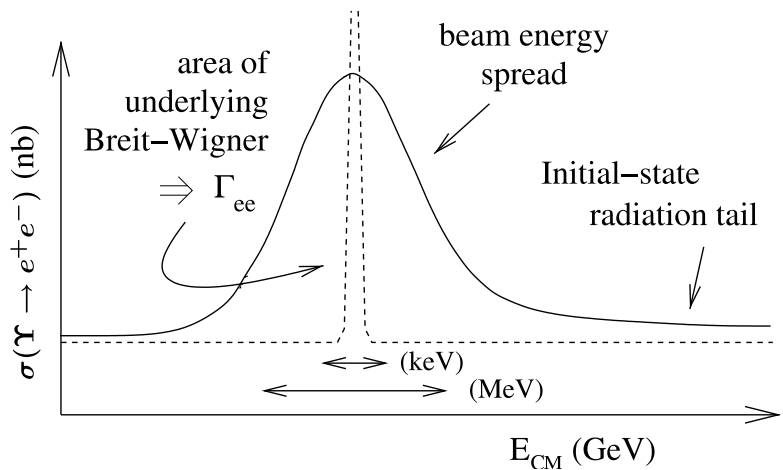}
  \end{center}
  \caption[Anatomy of an \ups\ lineshape scan]{\label{cartoon} The
  anatomy of an \ups\ lineshape scan (cross-section versus
  center-of-mass energy), including the natural lineshape (dashed
  peak), beam energy spread and ISR tail (solid), and backgrounds
  (vertical offset).}
\end{figure}

The high-energy side of the lineshape is also distorted by
initial-state radiation (ISR): $e^+e^- \to \Upsilon$ events are hard
to distinguish from $e^+e^- \to \gamma \Upsilon$ for sufficiently
small photon energies $E_\gamma$.  These events add to the apparent
cross-section, and contribute a high-energy tail that falls of as
$1/E_\gamma$, causing the integral to diverge.  We could introduce an
artificial cut-off, but then the \gee\ we report would be a function
of that cut-off.  Instead, we include the ISR distortion in our fit
function to match the data, but report the integral with no ISR
contribution, a procedure depicted in Figure~\ref{cartoon}.  This
means that $\sigma(e^+e^- \to \Upsilon)$ in Equation~\ref{eqn:gee}
represents the \ups\ cross-section with no ISR photons at all, and the
\gee\ we derive is devoid of final-state radiation ($\Upsilon \to
\gamma e^+e^-$).

\begin{figure}[p]
  \begin{center}
    \includegraphics[width=\linewidth]{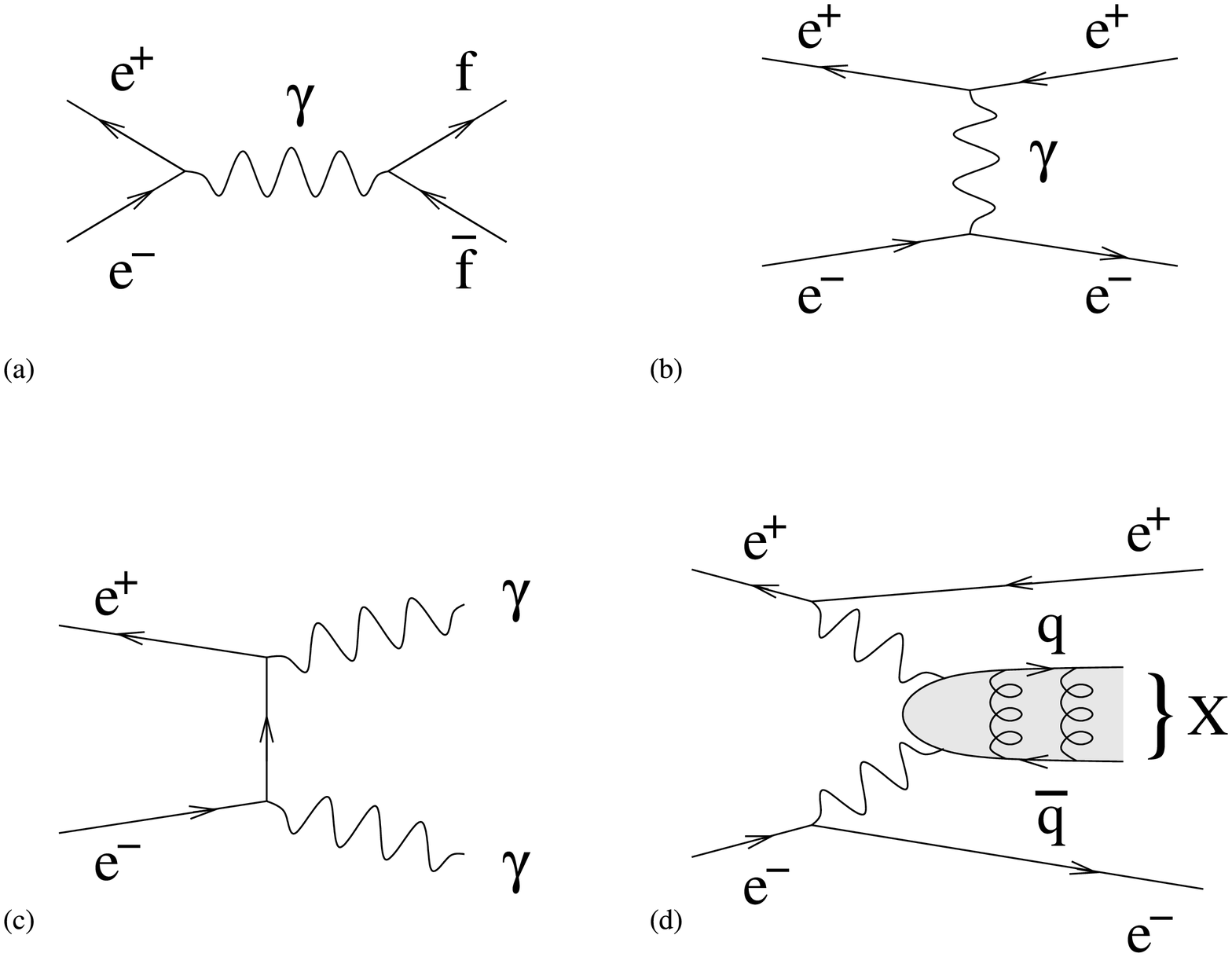}
  \end{center}
  \caption[Survey of continuum backgrounds]{\label{continuum} Survey
  of continuum backgrounds: (a)~$s$-channel fermion pair production,
  $f\bar{f}$ may be \qqbar, \mumu, or \tautau, (b)~$t$-channel
  exchange which dominates Bhabha (\ee) scattering, (c)~\ee\
  annihilation into two real photons (the other, undrawn diagram
  simply exchanges the identity of the two outgoing photons),
  (d)~fusion of two virtual photons into a low-momentum hadronic state
  accompanied by high-energy \ee.}
\end{figure}

In addition to $e^+e^- \to \Upsilon$ and $\gamma \Upsilon$,
electron-positron collisions in the 9.4--10.4~GeV range can also
undergo the following continuum processes, which also contribute to
the observed cross-section:  \renewcommand{\labelenumi}{\alph{enumi}.}
\begin{enumerate}

  \item \qqbar, \mumu, and \tautau\ purely through annihilation
    ($s$-channel, Figure~\ref{continuum}(a)),

  \item Bhabha \ee\ through annihilation ($s$-channel) and Coulomb
    scattering \\ (\mbox{$t$-channel,} Figure~\ref{continuum}(b)),

  \item \gamgam\ through annihilation (Figure~\ref{continuum}(c)), and

  \item $e^+e^- X$ via the fusion of two virtual photons from a
    grazing collision (Figure~\ref{continuum}(d)).

\end{enumerate}
At these energies, muon- and tau-pair cross-sections are 1~nb, and
\qqbar\ are 4~nb, decreasing with center-of-mass energy as $1/s$ ($s =
{E_\subs{CM}}^2$).  Bhabha events are by far the most abundant; in
fact, the Bhabha cross-section diverges if glancing-angle scatters are
included.  The \gamgam\ cross-section diverges also, but less rapidly
as a function of angle.  Bhabha and \gamgam\ cross-sections both
decrease as $1/s$.  The last process, two-photon fusion, generates
low-momentum hadronic particles $X$ and two electrons (\ee), at least
one of which grazes the incident beam-line.  The two-photon fusion
cross-section increases with center-of-mass energy, but very slowly,
as $\log s$.  Some of these non-\ups\ processes can be hard to
distinguish or are indistinguishable from \ups\ events, and therefore
can be confused with signal.  Fortunately, the continuum cross-section
is a much smoother function of \ecm\ than the \ups\ cross-section, so the
\ups\ peak appears to stand on a flat continuum plateau, also depicted
in Figure~\ref{cartoon}.  We add terms to the fit function to
accommodate these effects as well.

When a continuum final state is truly indistinguishable from an \ups\
decay, as is the case for $e^+e^- \to q\bar{q}$ and $e^+e^- \to
\Upsilon \to q\bar{q}$, the cross-sections don't simply add.  The
complex amplitudes for these processes add, the square of which is
proportional to cross-section: \label{sec:earlyinterference}
\newcommand{\res}{{\mbox{\scriptsize res}}}
\newcommand{\cont}{{\mbox{\scriptsize cont}}}
\newcommand{\inter}{{\mbox{\scriptsize int}}}
\newcommand{\aye}{{\mathcal A}}
\begin{equation}
  \sigma_{\res+\cont}(E_\subs{CM}) \propto \big| \aye_\res + \aye_\cont \big|^2 =
  \big| \aye_\res \big|^2 + \big| \aye_\cont \big|^2 + 2 {\mathcal
  Re}({\aye_\res}^* \aye_\cont) \mbox{,}
\end{equation}
so we can re-write $\sigma_{\res+\cont}$ as 
\begin{equation}
  \sigma_{\res+\cont}(E_\subs{CM}) =
  \sigma_\res(E_\subs{CM}) + \sigma_\cont(E_\subs{CM}) + \tilde{\sigma}_\inter(E_\subs{CM})
  \label{eqn:rescontinter}
\end{equation}
where $\sigma$ denotes cross-section and $\aye$ amplitude, ``res'' for
the resonant (\ups) contribution and ``cont'' for the continuum.  The
interference term, $\tilde{\sigma}_\inter$, is a function of \ecm, like the
familiar $\sigma_\res$ and $\sigma_\cont$, but it can be negative and
sometimes larger than $\sigma_\res$.  That is, introducing another way
for \ee\ to produce \qqbar\ can actually decrease the \qqbar\
cross-section!  We calculate this interference term from a
Breit-Wigner amplitude (Feynman propagator) of the form
\begin{equation}
  \aye_\res \propto \frac{1}{E_\subs{CM} - M_\Upsilon + i\Gamma/2}
  \label{eqn:bwasprop}
\end{equation}
and a constant continuum with constant phase $\phi_0$ (resonance phase
minus continuum phase at $E_\subs{CM} \ll M_\Upsilon$).  We find
\begin{eqnarray}
  \tilde{\sigma}_\inter(E_\subs{CM}) &=& \alpha_\subs{int} \, \sigma_\subs{res}(E_\subs{CM}) \,
  \left(2 \frac{E_\subs{CM} - M_\Upsilon}{\Gamma} \cos \phi_0 + \sin
  \phi_0 \right) \\
  \mbox{where } \alpha_\subs{int} &=& \sqrt{\sigma_\cont \, ({M_\Upsilon}^2 / 3 \Gamma_{ee}) \, \Gamma_f} \mbox{.}
  \label{eqn:yint}
\end{eqnarray}
The magnitude of each interference correction is characterized by
$\alpha_\subs{int}$, which is a constant derived from the continuum
cross-section, the resonance magnitude \gee, and the decay rate
$\Gamma_f$ to the given final state (in this case \qqbar).  Given the
continuum \qqbar\ cross-section (through $R = \sigma(e^+e^- \to
q\bar{q})/\sigma(e^+e^- \to \mu^+\mu^-)$ \cite{novor}) and the
resonance \qqbar\ branching ratio $\Gamma_{q\bar{q}}/\Gamma_{ee}$
(assuming \bee\ = \bmm, this is $R$), we calculate $\alpha_\subs{int}$
to be 0.0186, 0.0179, and 0.0182 for the \us, \uss, and \usss,
respectively, with 3\% uncertainties.  Note that if $\phi_0 = 0$,
$\tilde{\sigma}_\inter < 0$ below $M_\Upsilon$ and
$\tilde{\sigma}_\inter > 0$ above $M_\Upsilon$.  If, however, $\phi_0
= \pm\pi/2$, $\tilde{\sigma}_\inter$ will have exactly the same energy
dependence as $\sigma_\res$, and thus be indistinguishable from the
\ups\ cross-section itself.

\begin{figure}
  \begin{center}
    \includegraphics[width=0.9\linewidth]{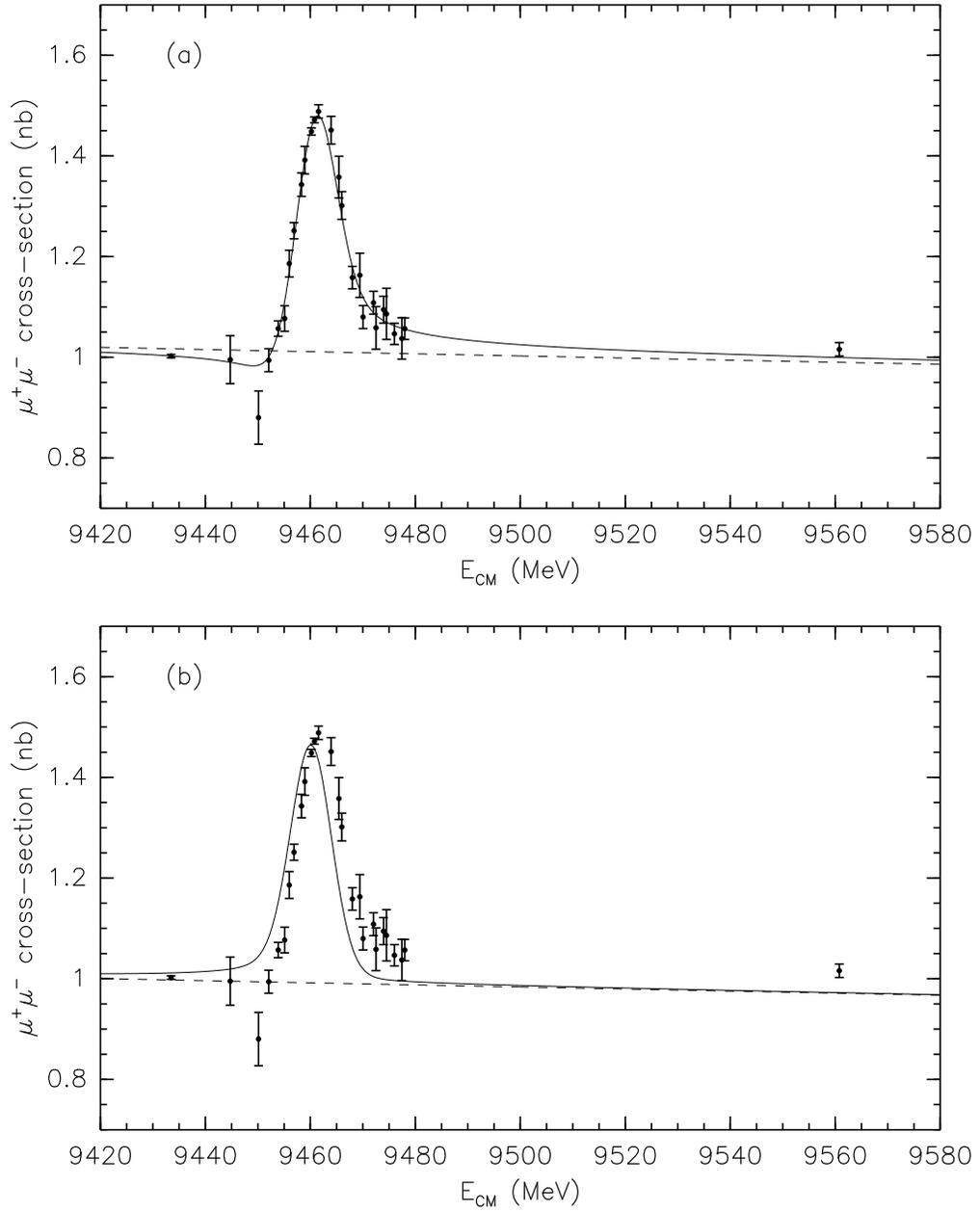}
  \end{center}
  \caption[Interference in the \mumu\ channel]{\label{newmumu}
  Efficiency-corrected, background-subtracted \mumu\ cross-section
  versus \ecm, fitted with (a) resonance, continuum, and interference
  normalization allowed to float independently, and (b) interference
  fixed with the opposite phase.  The fit prefers $\phi_0=0$ over
  $\phi_0=\pi$ by 18.5 standard deviations.}
\end{figure}

Continuum Bhabhas, \mumu, \tautau, and \qqbar\ all interfere with
$e^+e^- \to \Upsilon \to f\bar{f}$ with phase angle $\phi_0 = 0$.  We
can see this by considering that all of these processes are purely QED
except for the formation, propagation, and disintegration of the \ups\
resonance.  The tree-level QED amplitudes are real because they all
feature an even number of photon vertices (each of which contributes a
factor of $i$).  While \ups\ production introduces a factor of the
conjugated wavefunction at the origin, $\psi(0,0,0)^*$, the \ups\
decay part of the diagram multiplies it by $\psi(0,0,0)$.  The \ups\
propagator (Equation \ref{eqn:bwasprop}) is real for \ecm\ $\ll$
$M_\Upsilon$.  Therefore, $\phi_0=0$ or $\phi_0=\pi$.  We see in a
scan of $\Upsilon(1S) \to \mu^+\mu^-$ (Figure~\ref{newmumu}) that
interference is destructive below resonance and constructive above,
which indicates $\phi_0=0$.

\section{\boldmath \ups\ Final States and Hadronic Cross-section}

In our \ee\ collisions, \ups\ mesons are produced nearly at rest and
immediately decay.  We only ever observe the \ups\ decay products.  An
\ups\ may decay into
\renewcommand{\labelenumi}{\alph{enumi}.}
\begin{enumerate}

  \item leptonic final states: \ee, \mumu, and \tautau, through an
    $s$-channel virtual photon (or $Z^0$, with 1.5\% contribution to
    the rate),

  \item hadronic final states via the hadronization of \qqbar, $ggg$,
    or \gggamma,

  \item lower-mass $b\bar{b}$ states, accompanied by pions or photons,

  \item neutrino pairs exclusively through $Z^0$, and

  \item possibly other, exotic, modes.

\end{enumerate}

The \mumu\ branching fractions, \bmm, have been measured with 2--5\%
precision for the \us, \uss, and \usss\ \cite{istvan}, and the \ee\
and \tautau\ decays are expected to have the same amplitudes as \mumu.
Thus, the branching fractions, \bee, \bmm, and \btt, are nearly equal,
with only a tiny correction from lepton mass, which is 0.05\% for the
heavy $\tau$ lepton.  This assumption is called Lepton Universality.

States containing bare quarks or gluons (partons), like \qqbar, $ggg$,
and \gggamma, must hadronize before propagating to the detector.  That
is, strings of self-interacting gluons, stretched between the partons,
will generate new quark/anti-quark pairs when stretched sufficiently
far.  These new quarks will clothe the original partons, such that a
macroscopic detector will only ever observe mesons ($q\bar{q}$ bound
states) and baryons ($qqq$ or $\bar{q}\bar{q}\bar{q}$).  Hadronization
is a random process, leading to a broad spectrum of event topologies,
with as many as twenty particles in the final state.  Most \ups\
mesons decay hadronically.

\begin{sidewaysfigure}[p]
  \begin{center}
    \includegraphics[width=0.9\linewidth]{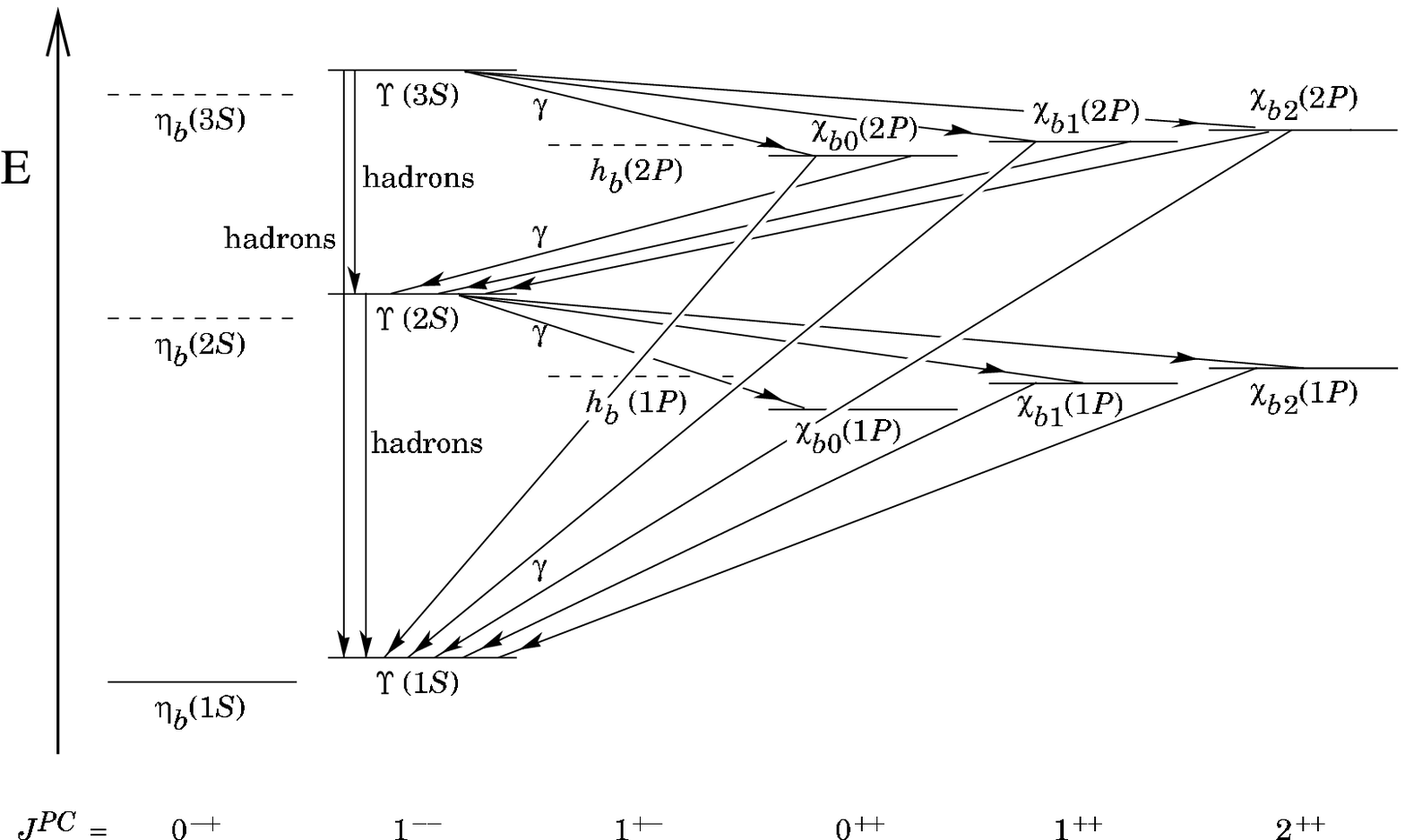}
  \end{center}
  \caption[States and transitions in the $b\bar{b}$
  system]{\label{bbtransitions} States and transitions in the
  $b\bar{b}$ system.  Only $J=1$ \ups\ mesons are produced directly by
  \ee, and of those, only \uss\ and \usss\ decay significantly into
  lower $b\bar{b}$ states.}
\end{sidewaysfigure}

Only the \uss\ and the \usss\ have appreciable decay rates to other
$b\bar{b}$ states.  (The $\Upsilon(1S) \to \eta_b(1S)$ branching
fraction is expected to be less than $10^{-4}$ \cite{ustoetab}.)
These decays are the bottomonium equivalent of atomic transitions, but
in addition to emitting monoenergetic photons in $\Delta J = 1$
decays, bottomonium can emit monoenergetic $\pi\pi$ (charged or
neutral) and $\gamma\gamma$ when decaying with $\Delta J = 0$.
Figure~\ref{bbtransitions} plots the energy levels of the most
well-known $b\bar{b}$ states and the transitions between them.

The $Z^0$ boson at \ups\ masses is 80~GeV off-shell, while the photon
is only 10~GeV off shell, so the $Z^0$ contributes to 1.5\% of the
\ups\ meson's Electroweak decays (\ee, \mumu, \tautau, and \qqbar).
The Electroweak decays account for $(3+R){\mathcal B}_{\mu\mu}$ = 16\%
of all \ups\ decays, where $R$, the branching ratio of \qqbar\ to
\mumu, has a value of 3.58 \PM\ 0.14 \cite{novor}.  The $\Upsilon \to
Z^0$ branching fraction is therefore 0.25\%, and $\Upsilon \to Z^0 \to
\nu\bar{\nu}$ is 0.05\%, which is negligible at our level of
precision.

Finally, we do not exclude the possibility that unknown \ups\ decays
exist, or that their branching fractions are larger than a few
percent.  These modes may resemble hadronic decays, or have exotic
signatures that have been overlooked.

For the sake of this analysis, we will classify \ups\ decay modes as
``leptonic,'' meaning \ee, \mumu, and \tautau\ exclusively (no
$\nu\bar{\nu}$), or ``hadronic,'' meaning everything else.  By this
designation, there are ``hadronic'' final states that contain no
hadrons at all, such as the $\Upsilon(2S) \to \gamma \chi_{b1}(1P) \to
\gamma\gamma \Upsilon(1S) \to \gamma\gamma e^+e^-$ chain, $\Upsilon
\to \nu\bar{\nu}$, and $\Upsilon \to \mbox{\sc wimp }
\overline{\mbox{\sc wimp}}$ (where {\sc wimp}s are
cosmologically-motivated invisible particles).  This classification is
convenient and has been used by previous \gee\ analyses.

Experimentally, the \ups\ cross-section is the number of $e^+e^- \to
\Upsilon$ events that occurred divided by the time-integrated
luminosity of the \ee\ collisions.  To identify $e^+e^- \to \Upsilon$
events, we select events that look like hadronic \ups\ decays, because
the leptonic final states are hard to distinguish from continuum
backgrounds and account for only 6--7.5\% of the \ups\ decays.  Thus,
we count $e^+e^- \to \Upsilon \to \mbox{hadronic}$ events and our
cross-section is $\sigma(e^+e^- \to \Upsilon \to \mbox{hadronic})$.
This hadronic cross-section is a constant multiple of the total
cross-section
\begin{equation}
  \sigma(e^+e^- \to \Upsilon \to \mbox{hadronic}) = \sigma(e^+e^- \to
  \Upsilon) \times \Gamma_\subs{had} / \Gamma_\subs{tot}
\end{equation}
for all \ecm.  We fit our lineshape function to hadronic cross-section
versus \ecm\ data and thereby derive \geehadtot.  To obtain \gee, we
divide by ${\mathcal B}_\subs{had} =
\Gamma_\subs{had}/\Gamma_\subs{tot}$, which is $(1 - {\mathcal B}_{ee}
- {\mathcal B}_{\mu\mu} - {\mathcal B}_{\tau\tau})$ by definition.
Applying Lepton Universality, we use
\begin{equation}
  \Gamma_{ee} = \frac{\Gamma_{ee}\Gamma_\subs{had}/\Gamma_\subs{tot}}{1
  - 3 {\mathcal B}_{\mu\mu}}
  \label{eqn:geehadtot:to:gee}
\end{equation}
to take advantage of the well-measured \bmm.  With \gee, we again
assume ${\mathcal B}_{ee} = {\mathcal B}_{\mu\mu}$ to calculate
$\Gamma = \Gamma_{ee} / {\mathcal B}_{\mu\mu}$.

The upcoming chapters will each present one aspect of the \geehadtot\
measurement.
\begin{description}

  \item[Chapter \ref{chp:hardware}] will describe the \ee\ collider
    and particle detector that were used to generate and count $e^+e^-
    \to \Upsilon \to \mbox{hadronic}$ events.

  \item[Chapter \ref{chp:backgrounds}] will define the $e^+e^- \to
    \Upsilon \to \mbox{hadronic}$ selection criteria and explain how
    background events are removed from that count.

  \item[Chapter \ref{chp:efficiency}] will derive the correction
    for hadronic \ups\ events missing from the sample, that is, the
    efficiency of the selection criteria.

  \item[Chapter \ref{chp:luminosity}] will explain how we measure
    integrated luminosity, thereby converting our hadronic event
    counts into hadronic cross-sections.

  \item[Chapter \ref{chp:beamenergy}] will show how we use
    cross-section data to put an upper limit on the uncertainty in
    beam energy measurements.

  \item[Chapter \ref{chp:fitting}] will describe the fit function and
    fit results in detail, and

  \item[Chapter \ref{chp:results}] will present all \geehadtot, \gee,
  $\Gamma$, and $|\psi(0,0,0)|^2$ results.

\end{description}

\chapter{Collider and Detector}
\label{chp:hardware}

In this Chapter, we present the apparatus we used to collide electrons
and positrons and collect \ups\ decay products.

\section{Cornell Electron Storage Ring}

\begin{figure}[p]
  \begin{center}
    \includegraphics[width=\linewidth]{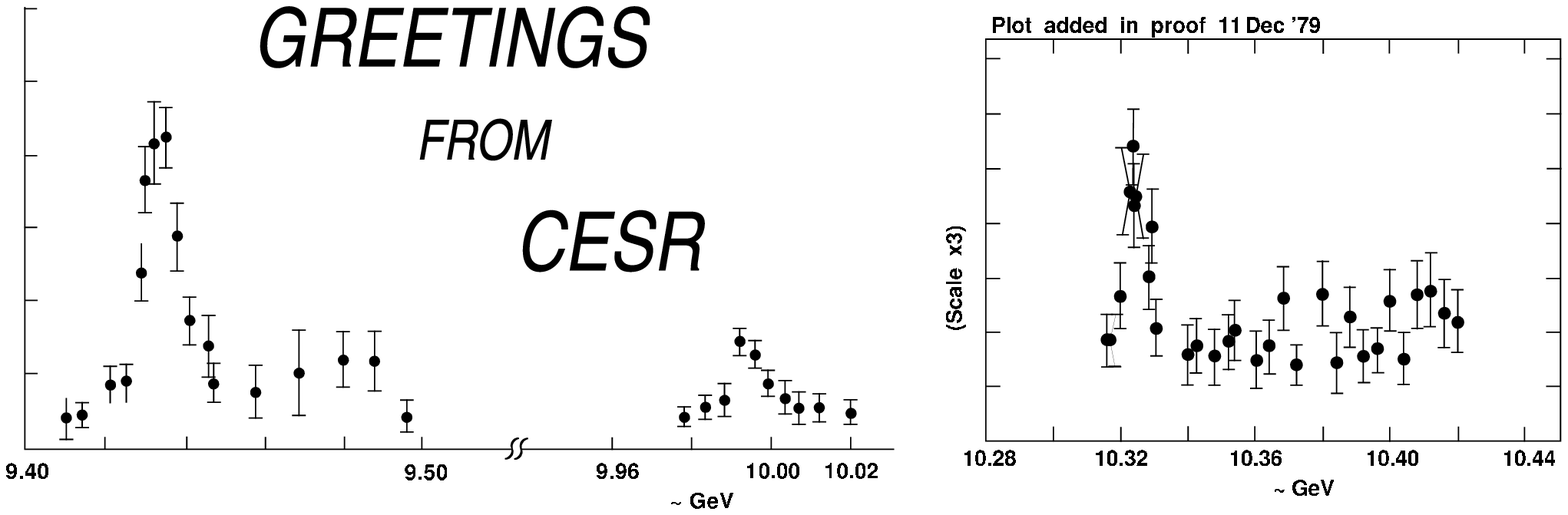}
  \end{center}
  \caption[First scan of \ups\ resonances by CESR in
  1979]{\label{greetingsfromcesr} Greeting card from CESR in 1979,
  demonstrating its success in colliding \ee\ at 10~GeV by scanning
  the lineshapes of the \us, \uss, and \usss\ (left to right).}
\end{figure}

Our electron-positron pairs collided in the Cornell Electron
Storage Ring (CESR), a 768~m-circumference, symmetric storage ring
and collider in Ithaca,~NY.  This collider covers a very
broad range of \ee\ energies, from the charmonium region at
$E_\subs{beam} = E_\subs{CM}/2 =$ 1.8~GeV through the excited \ups\
states at 5.5~GeV.  The \us, \uss, and \usss\ masses, between 4.7 and
5.2~GeV, lie in CESR's optimal range.  In fact, scans of the \us\
through \usss\ are among the first data taken by CESR in 1979
(Figure~\ref{greetingsfromcesr}).

Copper dipole magnets confine the beams to their orbits, alternating
with quadrapole and sextapole magnets for focusing.  Superconducting
quadruples provide the final focusing of the beams, only 30~cm from
the interaction point, allowing the collisions to reach instantaneous
luminosities of $10^{33}$~cm$^{-2}$~s\inv.  Like all synchrotrons, the
beam is pulsed to permit acceleration: beam bunches are timed to enter
radio frequency (RF) standing waves just when the electric field is
maximal.  In CESR, nine trains of five bunches circulate in the ring
at once, with 1.15$\times 10^{10}$ particles per bunch.  Both the
electron beam and the positron beam are enclosed in the same
beam-pipe, so they need to be separated electrostatically.  The
resulting orbit is called a ``pretzel orbit'' because the beams twine
around each other like twisted pair cable.

We determine the beam energy by measuring the magnetic field in two
dipole magnets, outside the ring but otherwise identical to the
others.  The current supplied to these two test magnets is in series
with the beam magnets to assure the same current, and the magnetic
field is measured with nuclear magnetic resonance (NMR) probes.  The
na\"ive beam energy,
\begin{equation}
  E_\subs{beam} = \mbox{electron charge} \times \mbox{magnetic field}
  \times \mbox{CESR radius}
\end{equation}
is then corrected for shifts in the RF frequency, steering and
focusing magnet currents, and the voltage of the electrostatic
separators.  This full reckoning misses the true beam energy by
0.172\%, which is 18~MeV in \ecm\ near the \ups\ masses, but it is
very stable with time and tracks beam energy differences with the same
precision.  Such a beam energy measurement will not improve the world
knowledge of \ups\ masses, but the scale uncertainty is small enough
to have negligible impact on the width, and therefore the area, of the
resonance scans.

Distributions of electron and positron energies in the CESR beam are
0.057\% wide at the \us\ (this is the ratio of the standard deviation
to the mean) and this width scales linearly with beam energy.  The
beam energy distribution is Gaussian.  Our lineshape data, which is
the world's most sensitive test of \ee\ beam energy distributions near
10~GeV, show no deviations from a pure Gaussian distribution.

\begin{figure}[p]
  \begin{center}
    \includegraphics[width=\linewidth]{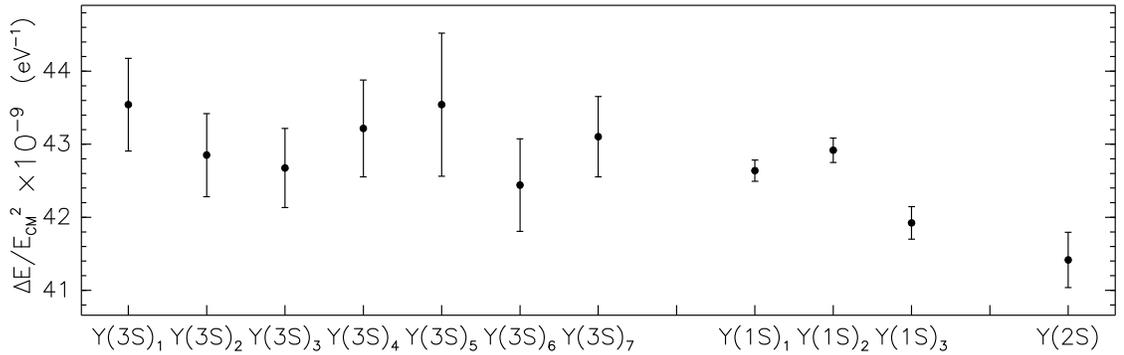}
  \end{center}
  \caption[Beam energy spread as a function of
  date]{\label{beamenergyspread} Beam energy spread fit results in the
  center-of-mass ($\Delta E = \sqrt{2}$ single beam energy spread)
  divided by \ecm$^2$, which is nearly constant, as expected.  The
  third \us\ measurement (April 2002) is 2.4\% lower than the second
  (March 2002).}
\end{figure}

The beam energy spread can vary by as much as 1--2\% a month, due to
perturbations in the beams' orbits from environmental conditions.  We
observed such a shift (Figure~\ref{beamenergyspread}), coincident with
large corrections to the horizontal steering magnets to compensate for
the new orbit.  We use records of these changes to track potential
shifts in the beam energy spread, and allow shifts in the widths of
the lineshapes by adding floating parameters to the
fit. \label{pag:beamenergyspread}

The beam-beam interaction region is a ribbon 0.18~mm tall (out of the
ring plane), 0.34~mm wide (in the ring plane but perpendicular to the
beam axis), and 1.8~cm long (in the ring, along the beam axis).
Electrons and positrons may collide anywhere within this constrained
distribution, and its center drifts by about 4~mm along the beam-line
and 1--2~mm perpendicular to it on a monthly timescale.  We can easily
track these changes.

In addition to \ee\ collisions, beam particles can interact with gas
nuclei inside the beam-pipe and with the wall of the beam-pipe itself
(2.1~cm in radius).  To minimize the number of beam-gas events, the
beam-pipe is kept evacuated at 2--4$\times 10^{-8}$~torr.  Beam-wall
events are minimized by focusing.  The electron and positron currents
can vary independently, so electron-induced beam-gas and beam-wall and
positron-induced beam-gas and beam-wall rates are not identical.  In
fact, we find that positron-induced rates are typically twice
electron-induced rates, suggesting a difference in cross-sections.

We collected data in eleven dedicated scans of the \us\ and one
high-energy point (100~MeV above the \us\ mass), totalling
0.09~fb\inv, and added to this 18~fb\inv\ of subsequent on-resonance
peak data (adjacent in time and limited to 48 hours after the
beginning of the scan).  We obtained six \uss\ scans with a
high-energy point (60~MeV above the \uss\ mass), totalling
0.05~fb\inv\, and added 0.03~fb\inv\ of subsequent peak data, and
seven \usss\ scans with a high-energy point (45~MeV above the \usss\
mass), totalling 0.08~fb\inv, and added 14~fb\inv\ of subsequent peak
data.  We present all the individual scans in Table~\ref{tab:scansa}.
In addition to scan data, we used the full 0.19~fb\inv, 0.41~fb\inv,
and 0.14~fb\inv\ off-resonance datasets, 20~MeV below the \us, \uss,
and \usss\ masses, to subtract continuum backgrounds.

\begin{table}
  \caption[Dates and integrated luminosity of \us\ scans]{\label{tab:scansa} Scans of the \us, including associated
  on-resonance peak data.  ``Spread'' indicates groups of scans which have
  the same beam energy spread (same labels as in
  Figure~\ref{beamenergyspread}).  The last entry is a point taken
  100~MeV above the \us\ mass.}
  \begin{center}
    \begin{tabular}{c c c l c}
    \hline\hline Scan & Int.\ Lum.\ (pb\inv) & Run Range & \hspace{1 cm} Dates \hfill & Spread \\\hline
      jan16 & 6.7  & 123164--123178 & Jan.\ 15--16, \hfill 2002 & $\Upsilon(1S)_1$ \\
      jan30 & 52.7 & 123596--123645 & Jan.\ 30--Feb.\  1, \hfill 2002 & $\Upsilon(1S)_1$ \\
      feb06 & 26.3 & 123781--123836 & Feb.\  6--8, \hfill 2002 & $\Upsilon(1S)_1$ \\
      feb13 & 7.8  & 124080--124092 & Feb.\ 19--20, \hfill 2002 & $\Upsilon(1S)_1$ \\
      feb20 & 21.0 & 124102--124159 & Feb.\ 20--22, \hfill 2002 & $\Upsilon(1S)_1$ \\
      feb27 & 23.9 & 124279--124338 & Feb.\ 27--Mar.\  1, \hfill 2002 & $\Upsilon(1S)_2$ \\
      mar06 & 19.6 & 124436--124495 & Mar.\  6--8, \hfill 2002 & $\Upsilon(1S)_2$ \\
      mar13 & 25.9 & 124625--124681 & Mar.\ 13--15, \hfill 2002 & $\Upsilon(1S)_2$ \\
      apr08 & 7.2  & 125254--125262 & Apr.\  8--9, \hfill 2002 & $\Upsilon(1S)_3$ \\
      apr09 & 5.6  & 125285--125295 & Apr.\  9--10, \hfill 2002 & $\Upsilon(1S)_3$ \\
      apr10 & 42.3 & 125303--125358 & Apr.\ 10--12, \hfill 2002 & $\Upsilon(1S)_3$ \\
   +100~MeV & 11.6 & 124960--124973 & Mar.\ 27--28, \hfill 2002 &  \\\hline\hline
    \end{tabular}
  \end{center}
\end{table}

\begin{table}
  \caption[Dates and integrated luminosity of \uss\ and \usss\
  scans]{\label{tab:scansb} Scans of the \uss\ and \usss, including
  associated on-resonance peak data.  ``Spread'' indicates groups of
  scans which have the same beam energy spread (same labels as in
  Figure~\ref{beamenergyspread}).  The ``+60~MeV'' and ``+45~MeV''
  entries are points taken 60 and 45~MeV above the \us\ and \usss\
  masses, respectively.  The dates of the ``+45~MeV'' data-taking
  overlap the ``dec26'' scan, but the integrated luminosity we quote
  do not.}
  \begin{center}
    \begin{tabular}{c c c l c}
    \hline\hline Scan & Int.\ Lum.\ (pb\inv) & Run Range & \hspace{0.75 cm} Dates & Spread \\\hline
     may29 & 14.6 & 126449--126508 & May.\ 29--31, \hfill 2002 & $\Upsilon(2S)$ \\
     jun11 & 9.9  & 126776--126783 & Jun.\ 11--12, \hfill 2002 & $\Upsilon(2S)$ \\
     jun12 & 23.6 & 126814--126871 & Jun.\ 12--14, \hfill 2002 & $\Upsilon(2S)$ \\
     jul10 & 18.8 & 127588--127615 & Jul.\ 10--11, \hfill 2002 & $\Upsilon(2S)$ \\
     jul24 & 5.8  & 127924--127933 & Jul.\ 23--24, \hfill 2002 & $\Upsilon(2S)$ \\
     aug07 & 9.3  & 128303--128316 & Aug.\  7--8, \hfill 2002 & $\Upsilon(2S)$ \\
   +60~MeV & 4.9  & 127206--127219 & Jun.\ 26--27, \hfill 2002 & \\\hline
     nov28 & 27.5 & 121884--121940 & Nov.\ 28--30, \hfill 2001 & $\Upsilon(3S)_1$ \\
     dec05 & 41.3 & 122069--122126 & Dec.\  6--8, \hfill 2001 & $\Upsilon(3S)_2$ \\
     dec12 & 41.3 & 122245--122298 & Dec.\ 12--14, \hfill 2001 & $\Upsilon(3S)_3$ \\
     dec19 & 24.2 & 122409--122452 & Dec.\ 19--22, \hfill 2001 & $\Upsilon(3S)_4$ \\
     dec26 & 27.7 & 122535--122579 & Dec.\ 25--26, \hfill 2001 & $\Upsilon(3S)_5$ \\
     jan02 & 27.7 & 122766--122821 & Jan.\  2--4, \hfill 2002 & $\Upsilon(3S)_6$ \\
     jan09 & 44.5 & 122993--123044 & Jan.\  9--11, \hfill 2002 & $\Upsilon(3S)_7$ \\
   +45~MeV & 10.8 & 122568--122575 & Dec.\ 26, \hfill 2001 & \\\hline\hline
    \end{tabular}
  \end{center}
\end{table}

\section{CLEO Detector}

\begin{sidewaysfigure}[p]
  \begin{center}
    \includegraphics[width=0.75\linewidth]{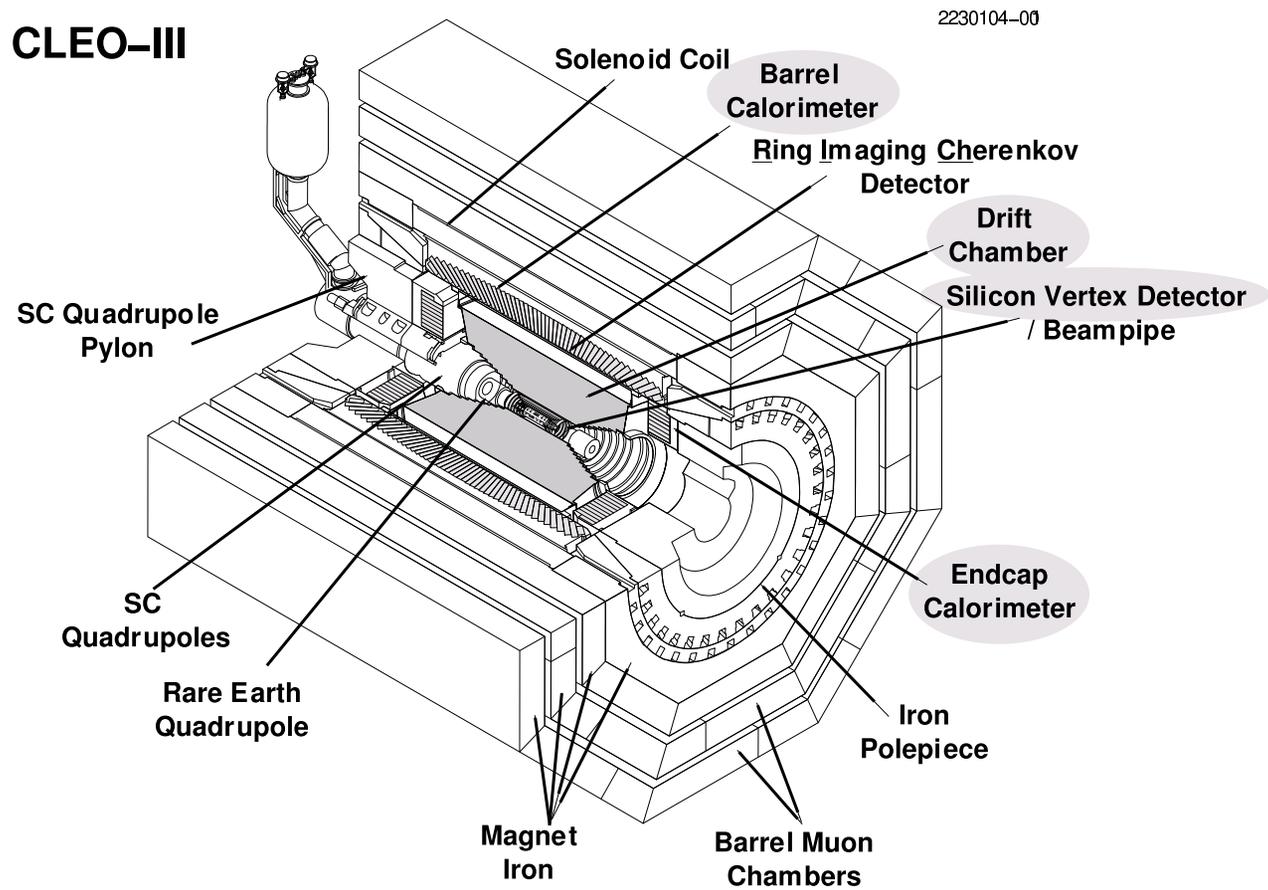}
  \end{center}
  \caption[Isometric view of the CLEO-III detector]{\label{cleoiii} An
  isometric view of the CLEO-III detector, highlighting the silicon
  vertex detector, the drift chamber, and the CsI crystal calorimeter
  in gray.}
\end{sidewaysfigure}

The CLEO detector is a general-purpose assembly of detectors built
concentrically around the CESR interaction point \cite{cleoiii}
\cite{driii}.  This analysis uses only three of CLEO's detectors: the
silicon vertex detector and central drift chamber for identifying
charged particles, and the CsI crystal calorimeter for measuring
electron and photon energies, and for simple particle identification.
The CLEO-III apparatus, which is the generation of CLEO in operation
in 2001--2002, is depicted in Figure~\ref{cleoiii}.

\begin{figure}[p]
  \begin{center}
    \includegraphics[width=\linewidth]{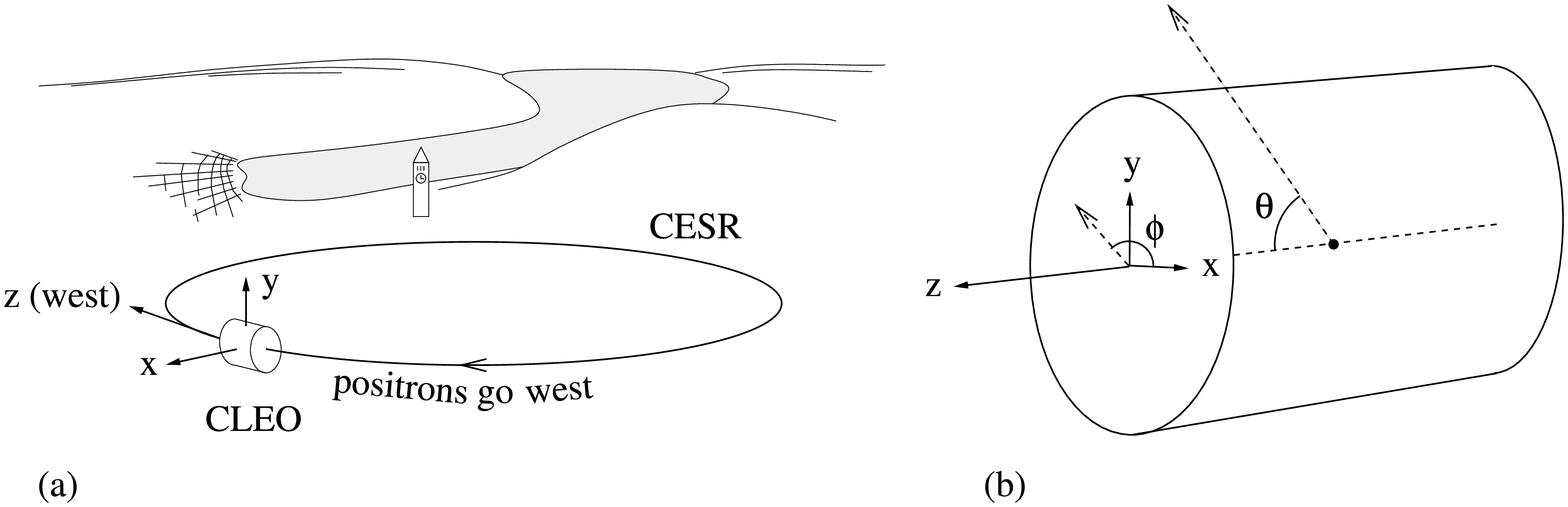}
  \end{center}
  \caption[Coordinate system of the CLEO
  detector]{\label{coordinatesystem} Coordinate system of the CLEO
  detector: (a) orientation with respect to the CESR ring and the
  Earth, (b) definition of $\theta$ and $\phi$.}
\end{figure}

We define the $z$ axis of our coordinate system to be parallel with
the beam-line, pointing in the direction of the incident positron
current (west).  Our coordinate system is right-handed, with $y$
pointing up and $x$ pointing away from the center of the CESR ring
(south).  The origin of the coordinate system is at the center of the
drift chamber, and lies within 1--2~mm of the beam-beam collision
point.  This coordinate system is illustrated in
Figure~\ref{coordinatesystem}.  The CLEO detector has an approximate
cylindrical symmetry around $z$, so we also define the polar angle
$\theta$ of a particle trajectory originating at the origin to be the
angle between the trajectory and the beam-line, or $\theta =
\tan^{-1}\left(\sqrt{x^2+y^2}/z\right)$.  We often use $\cos\theta$
and $\cot\theta$ to describe the polar angle.  The azimuthal angle
$\phi$ is the angle for which $\cos \phi = x/\sqrt{x^2+y^2}$ and $\sin
\phi = y/\sqrt{x^2+y^2}$.

\begin{figure}[p]
  \begin{center}
    \includegraphics[width=0.5\linewidth]{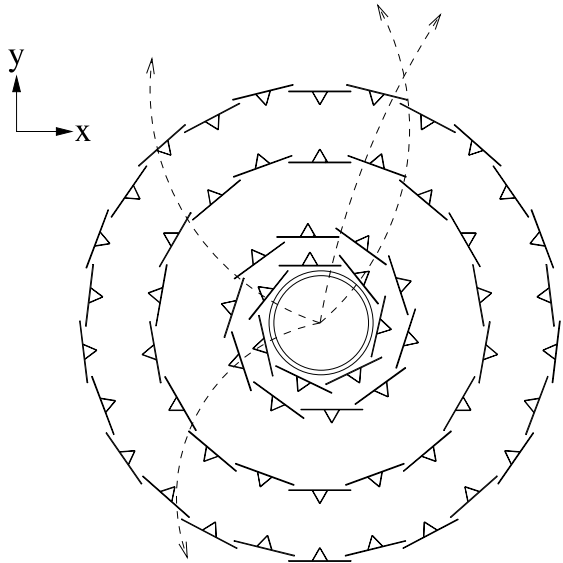}
  \end{center}
  \caption[Silicon vertex detector geometry]{\label{sidiagram} Silicon
  vertex detector geometry, projected onto the $x$-$y$ plane.  The
  trajectories of central ($\theta \approx \pi/2$) collision products
  (dashed lines) are roughly perpendicular to the wafers of silicon.
  Triangles represent diamond rods holding the wafers in place.}
\end{figure}

The silicon vertex detector and the drift chamber both detect tracks
left by charged particles by collecting charge left in the wake of
ionizing, high-energy particles.  In the vertex detector, the ionized
medium is silicon, cut into strips held perpendicular to the
trajectories of most particles (Figure~\ref{sidiagram}).  The charge
is conducted out of the detector for amplification along traces which
are parallel to the beam-line on one side of the strip and
perpendicular to it on the other, so that the two-dimensional point of
intersection may be reconstructed.

\begin{figure}[p]
  \begin{center}
    \includegraphics[width=0.75\linewidth]{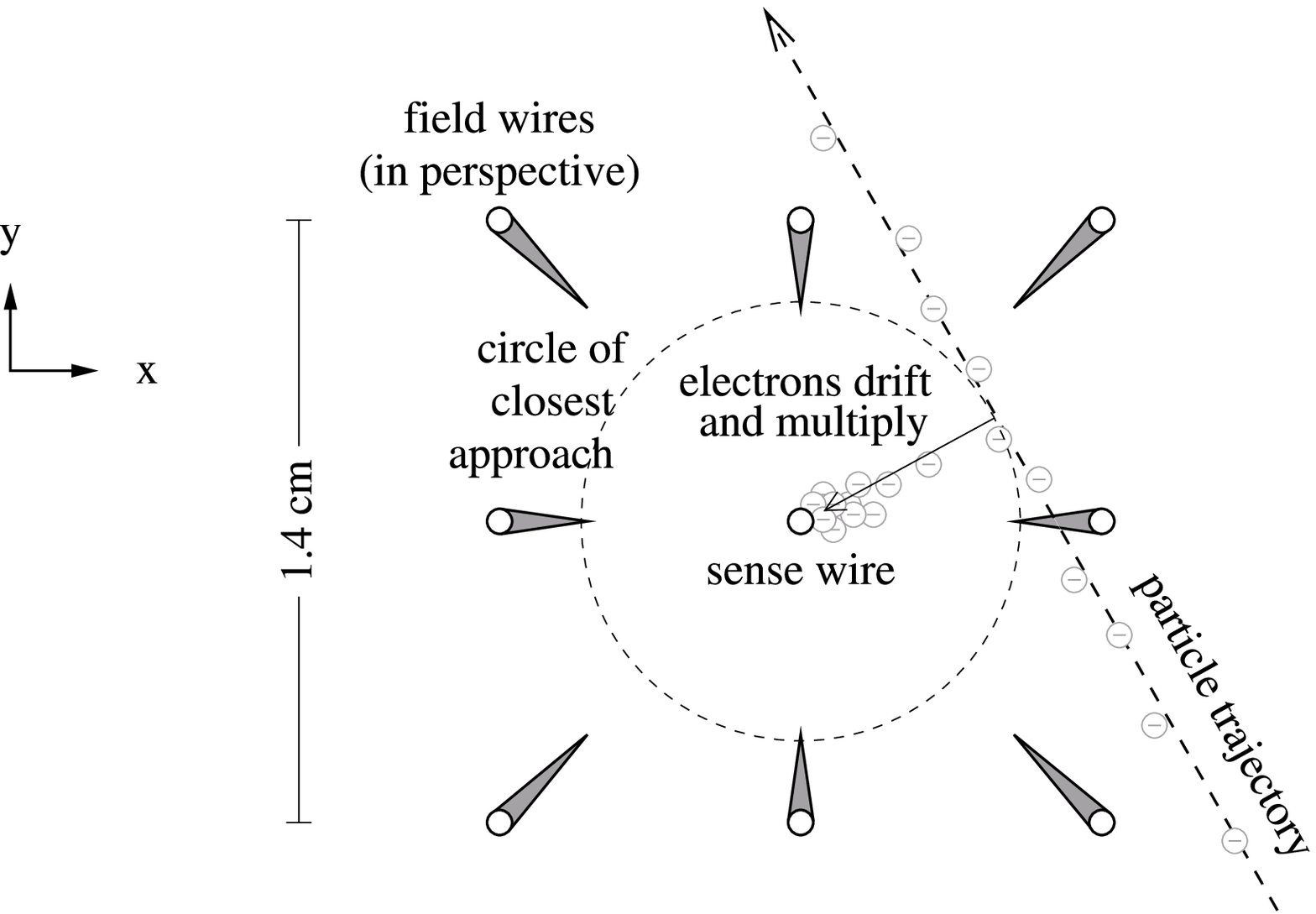}
    \mbox{\hspace{2 cm}}
  \end{center}
  \caption[Charge collection and multiplication in the drift
  chamber]{\label{driftcell} Charge collection and multiplication in
  the drift chamber.}
\end{figure}

\begin{figure}[p]
  \begin{center}
    \includegraphics[width=\linewidth, viewport=0 480 892 947]{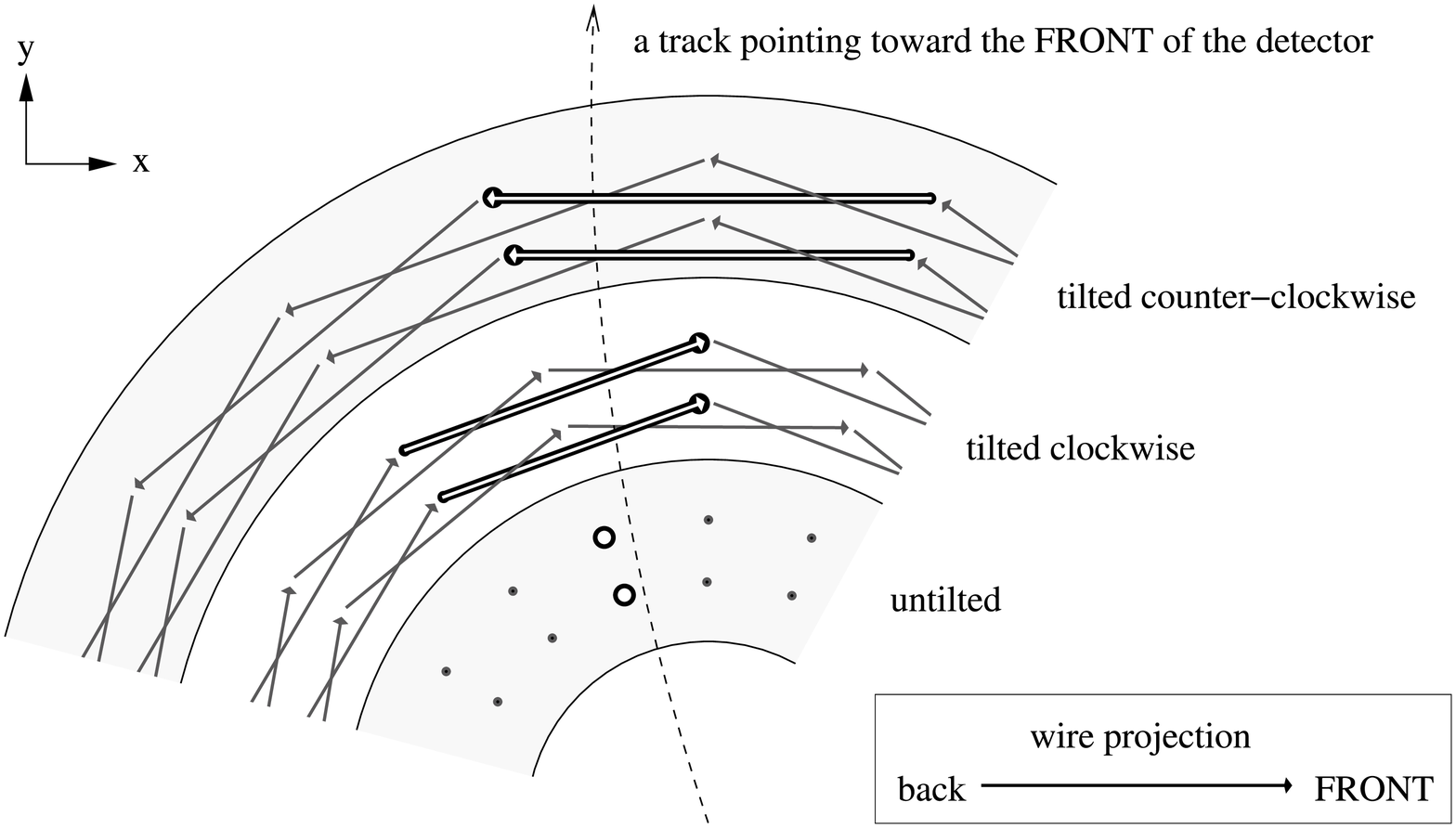}
  \end{center}
  \caption[Using tilted wires to obtain $z$ information in the drift
  chamber]{\label{stereotwist} Using tilted wires to obtain $z$
  information in the drift chamber.  Tilted wires extrude lines in the
  $x$-$y$ projection; position along a tilted wire indicates the $z$
  of the track helix near that wire.  The closest wires to the track
  (in three dimensions) are highlighted.}
\end{figure}

In the drift chamber, high-energy charged particles ionize a
helium-propane gas (60\% He, 40\% C$_3$H$_8$) in a strong electric
field generated by wires strung across the detector volume, parallel
with the $z$ axis.  One quarter of these wires, called sense wires,
are held at $+$2100~V, and the remaining three quarters, called field
wires, are held at ground.  The resulting field causes the freed
electrons to drift away from the field wires toward the sense wire,
which conducts the charge to amplifiers for analysis
(Figure~\ref{driftcell}).  As the electrons drift several millimeters
toward the sense wire, they ionize more atoms, causing an avalanche
that provides a 10$^7$ amplification.  We measure the time between the
first ionization (estimated from bunch collision times) and charge
collection on the sense wire to reconstruct the distance of closest
approach of the high-energy charged particle to the sense wire,
through the known electron drift speed of 28~$\mu$m/ns.  This
technique provides an average resolution of 88~$\mu$m in the $x$-$y$
plane.  Sensitivity to $z$ position is obtained by tilting the outer
wires, presented in more detail in Figure~\ref{stereotwist}.  The
outer 31~layers of wires, called the stereo section, are tilteded
21--28~mrad, yielding a $z$ position resolution of 3--4~mm at each
wire.  The inner 16~layers, called the axial section, are untilted.

Both tracking volumes are permeated by a 1.5~T magnetic field,
pointing along the $z$ axis.  Charged particle trajectories are
helical in this field: projections onto the $x$-$y$ plane are circles.
The polar angle $\theta$ of such a helical trajectory is a constant of
the motion, but not $\phi$.  We measure the charge $\times$ momentum
of particles through the radii of curvature of their tracks.  Only
electrons, muons, pions, kaons, protons, and deuterons are
sufficiently stable and abundantly produced to be observed as tracks,
and all of these particles have $\pm$1 units of charge, so the radius
of curvature provides access to momentum.  The momentum resolution,
dominated by drift chamber measurements, is 0.9\% for beam-energy
tracks, and position resolution near the interaction point, dominated
by silicon vertex measurements, is 40~$\mu$m in $x$-$y$ and 90~$\mu$m
in $z$.

\begin{sidewaysfigure}[p]
  \begin{center}
    \includegraphics[width=0.7\linewidth]{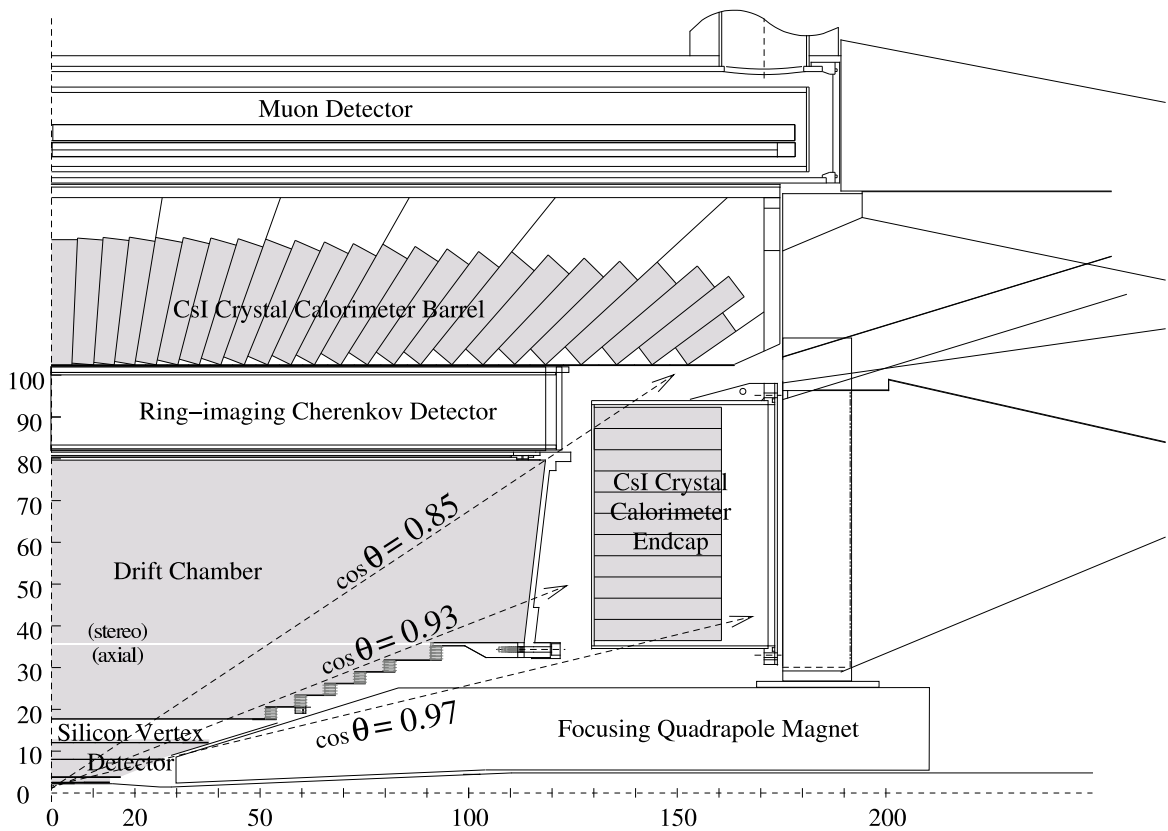}
  \end{center}
  \caption[Quarter-view of the CLEO detector]{\label{quarterview} A
  quarter of the CLEO detector, highlighting the silicon vertex
  detector, the drift chamber, and the CsI crystal calorimeter in
  gray.  Radial and $z$ distances are in centimeters; the angular
  reach of the calorimeter barrel, the drift chamber, and the
  calorimeter endcap are illustrated by dashed lines.}
\end{sidewaysfigure}

The outer radius of the drift chamber is 80~cm from the interaction
point, and the outer edges are $\pm$110~cm in $z$.  The drift
chamber's $z$ range is more limited for smaller radii to accommodate
the focusing quadrapole magnet, as shown in Figure~\ref{quarterview}.
It will later be to useful to know that charged particles with more than
60~MeV of $z$-momentum exit the detector before completing one
half-orbit in the magnetic field.  Such particles cannot generate
multiple tracks by spiralling inside the detector volume.
\label{pag:spiraling}

The CsI crystal calorimeter is sensitive to photons as well as charged
particles, by presenting a transparent, high-$Z$ material for them to
interact with Electromagnetically.  (Our thallium-doped CsI has a
radiation length of 1.83~cm.)  Incident electrons and photons are
destroyed by this interaction, and replaced by a shower of equal total
energy in less energetic photons, electrons, and positrons.  Other
particles deposit only a fraction of their energy.  Visible light from
the shower is collected on the back of the 30~cm-long crystals, from
which the incident energy is reconstructed.  Electrons and photons
with energies near the 5~GeV beam energy are fully reconstructed with
1.5\% resolution, but the energies of other particles is
underestimated.  Muons, for instance, deposit only 200~MeV in the
calorimeter, regardless of incident energy.  Combining the energy of
calorimeter showers with track momenta is sufficient to identify and
distinguish \ee, \mumu, and \gamgam\ events with negligible
backgrounds.

The calorimeter geometry is composed of three parts: a barrel
surrounding the tracking volume and two endcaps, beyond the tracking
volume in $z$ (see Figure~\ref{quarterview}).  The calorimeter barrel
covers polar angles with $|\cos\theta| < 0.85$ and the endcaps extend
this range to $|\cos\theta| < 0.97$.  The angular limits of the
tracking volume is between these two: $|\cos\theta| < 0.93$.

\begin{figure}[p]
  \begin{center}
    \includegraphics[width=\linewidth]{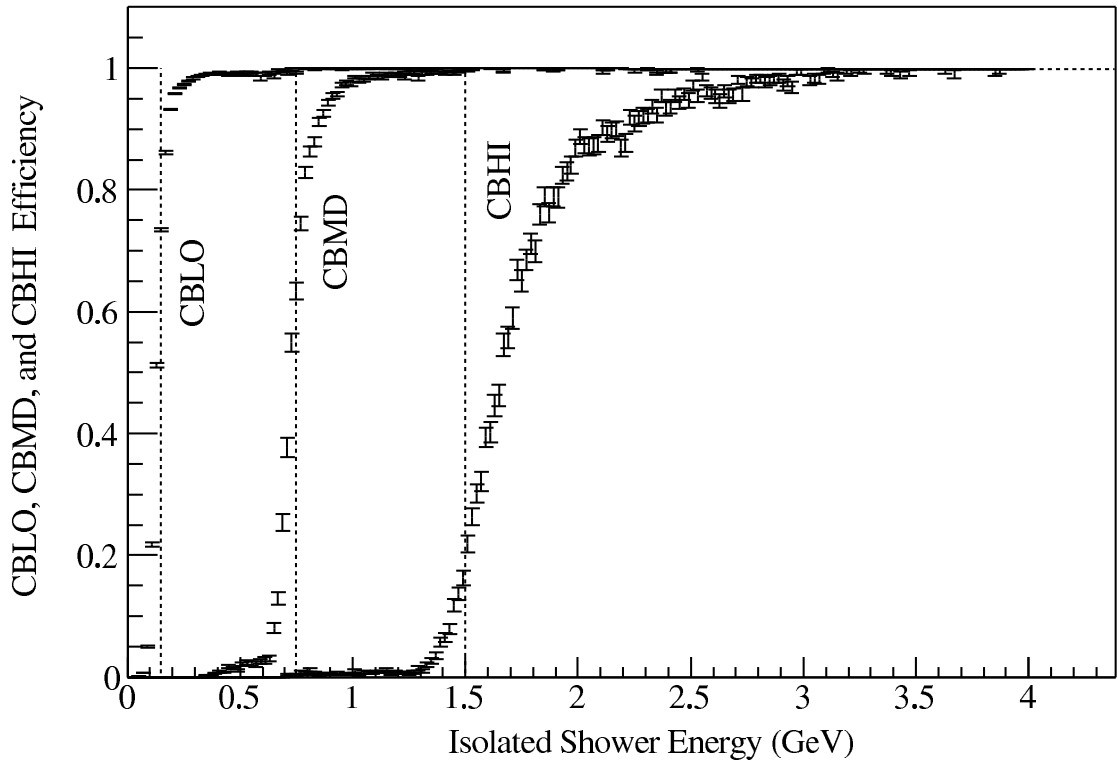}
  \end{center}
  \caption[Efficiencies of \cblo, \cbmd, and \cbhi\
  identification]{\label{topher} The efficiencies of \cblo, \cbmd, and
  \cbhi\ identification as a function of fully-reconstructed shower
  energies for isolated showers.  The efficiency curves are not
  symmetric around their thresholds because shower energy may be
  divided among tiles.}
\end{figure}

When a threshold amount of activity is observed in the drift chamber
and calorimeter, readout electronics are triggered to acquire a
snapshot of the detector and record all signals as an event.  This
activity is quantified in terms of the number of observed tracks and
the number of showers above given energy thresholds.  For speed in
triggering, tracks are counted using a lookup table of drift chamber
hits, trained by a simulation, and showers are approximated by summing
calorimeter barrel output over 2$\times$2 tiles, called clusters, and
counting the number that surpass a given threshold.  The number of
\axial\ tracks is the number of tracks reconstructed in the axial
section of the drift chamber, and \stereo\ is the number of tracks
which can be extended into the stereo section.  A \cblo\ cluster
exceeds 150~MeV, a \cbmd\ exceeds 750~MeV, and a \cbhi\ exceeds
1500~MeV.  Real showers can be distributed over as many as four tiles,
sometimes dividing their energy such that none of the clusters reach a
threshold.  This is a source of trigger inefficiency for final states
that rely on shower information (Figure~\ref{topher}).  After the data
have been recorded, we reconstruct tracks and showers with much finer
precision.

We use several triggers to accept events, all of which are
minimum-thresholds: an event is never rejected for having too many
tracks or clusters.  All of these triggers are active, and when an
event is recorded, it is tagged with the names of the triggers it
satisfied.  The trigger relevant for this analysis are
\begin{itemize}

  \item \twotrack, which requires $\ge$2 \axial\ tracks, prescaled by
  a factor of 19 (5.3\%~of the events satisfying this criterion are
  accepted),

  \item \hadron, which requires $\ge$3 \axial\ tracks and $\ge$1 \cblo, \label{pag:triggerdefs}

  \item \radtau, which requires $\ge$2 \stereo\ tracks and ($\ge$1 \cbmd\ or $\ge$2 \cblo),

  \item \eltrack, which requires $\ge$1 \axial\ track and $\ge$1 \cblo, and

  \item \barrelbhabha, which requires 2 \cbhi\ clusters on opposite
  sides of the calorimeter barrel.

\end{itemize}
To count hadronic \ups\ decays, we select only those events which
satisfied \hadron, \radtau, or \eltrack, the three triggers that are
efficient for hadronic decays.  (This simplifies our efficiency study.)
Note that a minimal condition for these three trigger is
that at least one \axial\ track and one \cblo\ were observed.  This
minimal requirement is exact because \stereo\ tracks, being extensions
of \axial\ tracks, are always less numerous than or equal in number
to \axial\ tracks, and \cbmd\ clusters are also \cblo\ clusters.

Electron and positron beams are circulated in CESR for about an hour
before their currents are exhausted from collisions.  Data collected
during this time is called a run, and is given a unique, ascending
6-digit identifier.  Runs are the basic unit of CLEO data samples; in
lineshape scans, we generally took one run at each \ecm\ point at a
time.

For some studies, we must simulate our entire detector on a computer.
Such Monte Carlo simulations are most important in determining the
efficiency-corrected cross-section for Bhabhas in CLEO, which is
needed to measure the integrated luminosity of our datasets.  While
the total Bhabha cross-section is infinite, the efficiency-corrected
cross-section, defined by observed Bhabhas, is finite and must be
calculated theoretically.  This calculation has two ingredients, the
differential cross-section as a function of $\theta$ and CLEO's
efficiency for Bhabhas as a function of $\theta$.  The first
ingredient is calculated with perturbative Quantum Electrodynamics
(QED), but the second requires specific knowledge of our detector.
While this efficiency may be approximated as a step function, in which
CLEO observes all Bhabhas within a $\theta$ range and misses all
Bhabhas outside of this range, such a simplification would be bought
at a high price in accuracy.  For the 1\% precision demanded by this
analysis, we must consider all effects: detector geometry, electron
propagation and scattering in materials, sub-component response
efficiency, fringe magnetic fields from the CESR magnets, et cetera.
Our Monte Carlo simulation is based on the GEANT framework
\cite{geant}, and is carefully tuned to reproduce the real detector's
output at all levels of analysis.

\chapter{Backgrounds and Event Selection}
\label{chp:backgrounds}

To define a set of hadronic \ups\ decays, we will accept only those
events which satisfy given criteria, or cuts.  We want this set of
events to include as many hadronic \ups\ decays as possible, to
minimize the efficiency correction for lost events.  We therefore only
seek to reduce the backgrounds to a manageable level by cutting out
regions of parameter space where the hadronic \ups\ contribution is
minimal.  We accomplish this with a set of four explicit cuts.

With such an approach, we cannot completely eliminate backgrounds,
especially because continuum \qqbar\ final states are identical to
8.9\% of \ups\ decays.  Instead, we estimate and subtract the
backgrounds that remain after cuts, which we can do very accurately
using control data.  If we can accurately subtract any residual
backgrounds after cuts, why cut at all?  There are two reasons: large
background subtractions introduce large statistical uncertainties, and
the trigger itself selects events in a way which can be hard to
predict, leading to systematic uncertainties.  By imposing more
restrictive event selections with fully-reconstructed data, we can
render the trigger biases insignificant.

\section{Suppressing Backgrounds with Event Selection}

\begin{figure}[p]
  \begin{center}
    \includegraphics[width=\linewidth]{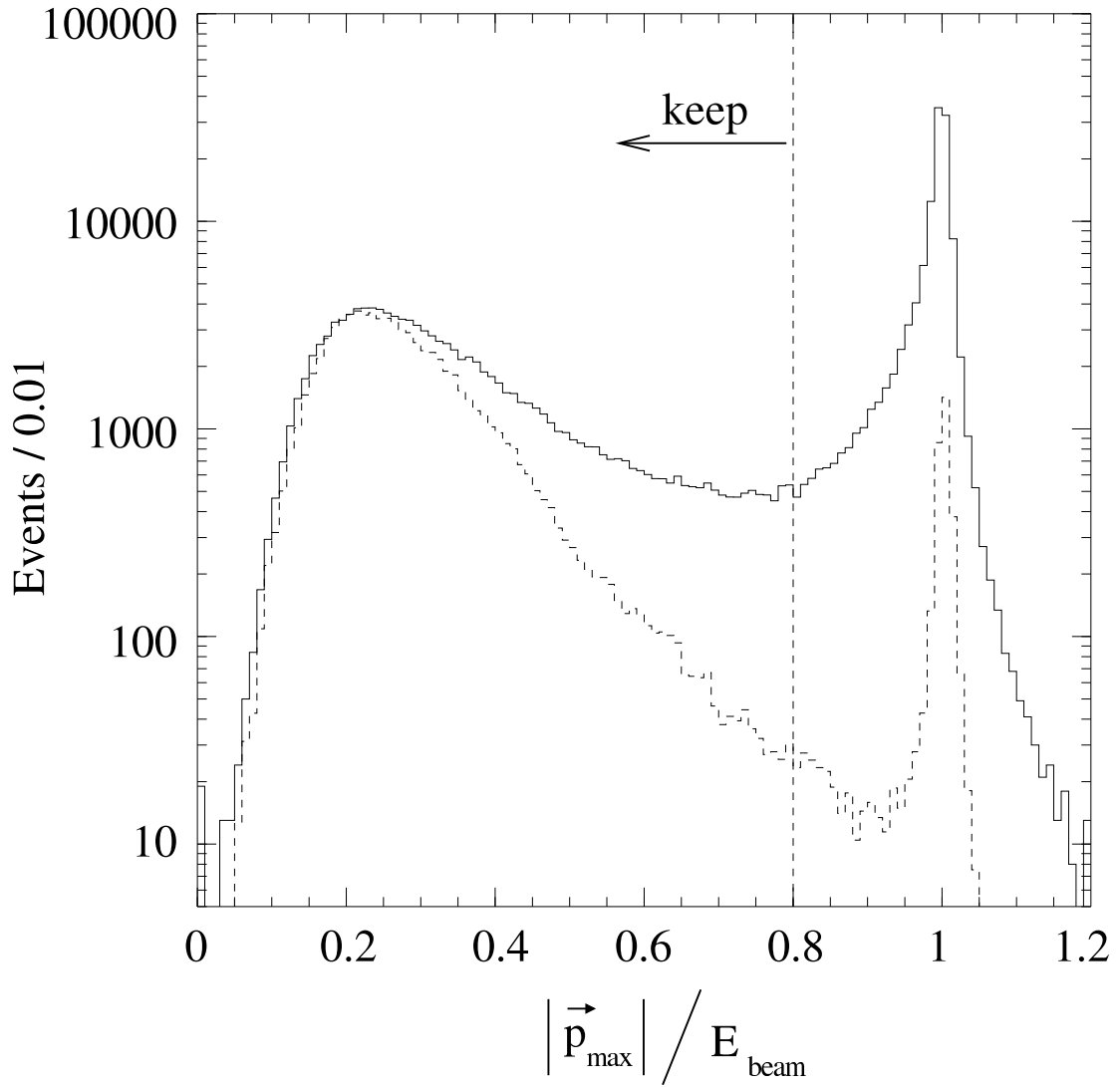}
  \end{center}
  \caption[Largest track momentum distribution]{\label{pmax} Largest
  track momentum (\pmax) in each event for data (solid histogram) and
  Monte Carlo \ups\ decays (dashed), with all other cuts applied.
  Data have large backgrounds from Bhabhas and radiative Bhabhas.
  Hadronic \ups\ decays peak at 20\% of \ebeam, while $\Upsilon \to
  e^+e^-$ and \mumu\ peak at 100\% of \ebeam.}
\end{figure}

Bhabha scattering is our largest potential background before cuts, and
among our largest backgrounds after cuts.  We suppress Bhabhas by
requiring the largest track momentum, \pmax, to be less than 80\% of
\ebeam.  According to Monte Carlo, this rejects 0.15\% of hadronic
\ups\ events, but 99.73\% of $\Upsilon \to e^+e^-$ and \mumu\
(Figure~\ref{pmax}).  We do not make a similar requirement on
calorimeter shower energy, which also peaks at \ebeam\ for each
electron in the Bhabha event, because we find track momentum
measurements to be more stable in time than shower energy
measurements.  Unlike track momentum, which is a geometric measurement
of wires in space, shower energy depends sensitively on the
amplification of the calorimeter read-out.  This amplification is
measured with 0.02--0.06\% precision, but the Bhabha spectrum is so
steep that 5\% of Bhabha showers move across a reasonable
threshold (75\% of \ebeam) with these fluctuations in energy scale.

It is also common to reject Bhabhas by requiring more than two tracks
in the event: according to our simulation, 98.9\% of hadronic events
have more than two tracks and 99.4\% of Bhabhas in the observable
range have exactly two.  We considered this cut at an early stage in
the analysis, but decided against it because we found the number of
tracks distribution difficult to simulate for hadronic events.  At
that time, we intended to determine the cut efficiency with our Monte
Carlo simulation, so this would have contributed significantly to the
systematic uncertainty.  Since then, we have found a way to measure
hadronic efficiency without resorting to simulations, but we did not
re-introduce the cut because we do not need it.  The \pmax\ cut
reduces Bhabha contamination to approximately the same level as
continuum \qqbar, so the Bhabha contribution to the statistical
uncertainty of the background-subtracted count is not dominant.
(Several of our cuts imply that an event must generate at least one
track, but this does not significantly affect the Bhabha background.)

\begin{figure}[p]
  \begin{center}
    \includegraphics[width=\linewidth]{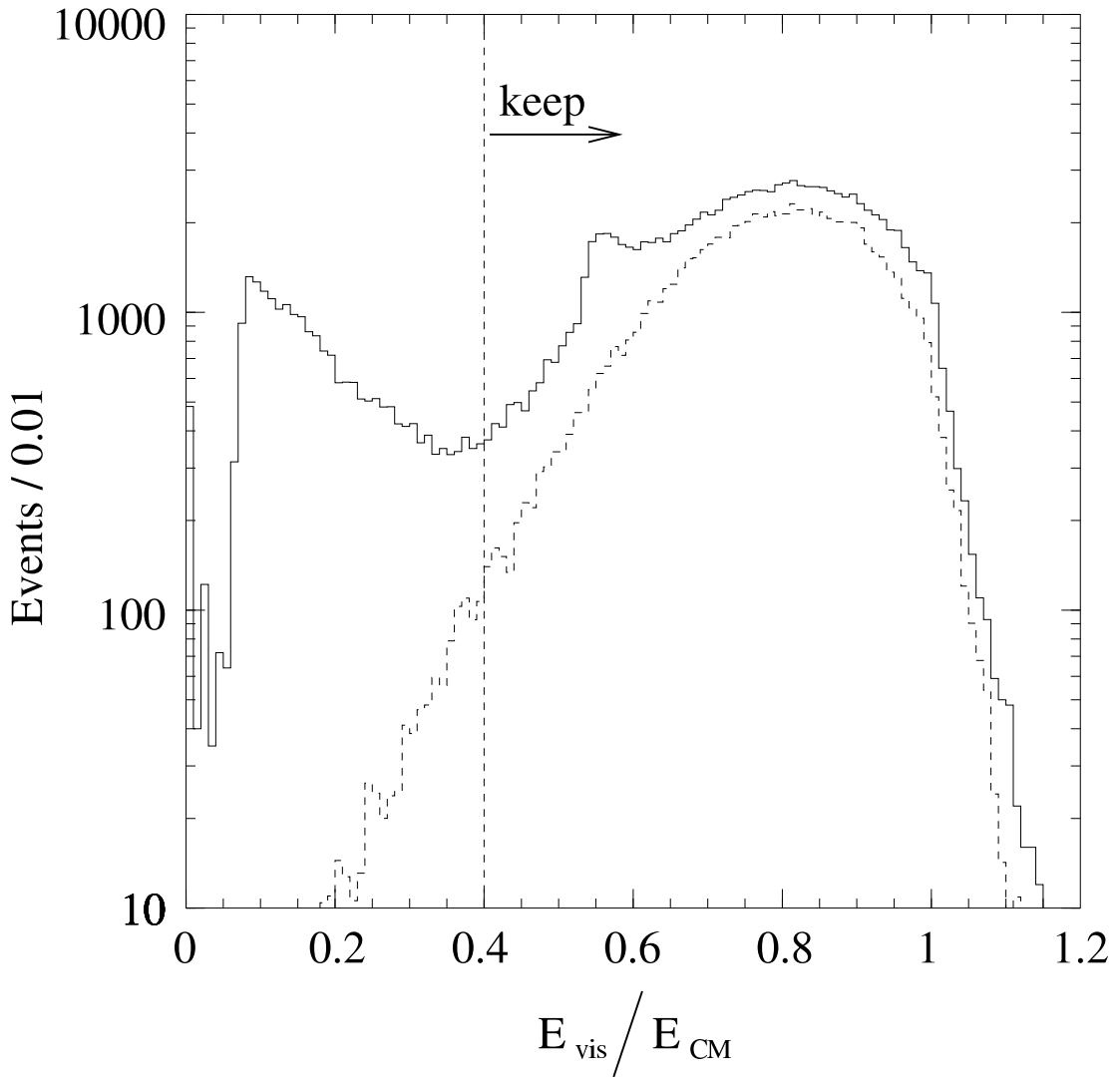}
  \end{center}
  \caption[Visible energy distribution]{\label{visen} Total visible energy (\visen) in each event
  for data (solid histogram) and Monte Carlo \ups\ decays (dashed)
  with all other cuts applied.  At least two-thirds of the peak at
  15\% of \ecm\ in data are two-photon fusion events, and the peak
  above 50\% of \ecm\ are likely to be radiative Bhabhas, missing an
  electron.  See text for a more complete discussion.}
\end{figure}

As previously mentioned, the cross-section of all continuum $e^+e^-
\to f\bar{f}$ processes ($f$ is any fermion) fall off as $1/s$ while
the two-photon fusion cross-section ($e^+e^- \to e^+e^- X$) increases
as $\log s$.  Continuum $f\bar{f}$ may therefore be estimated and
subtracted collectively, while two-photon fusion must be handled
separately, possibly introducing systematic error if it is large.  We
therefore suppress two-photon fusion events by requiring the visible
energy of the event, \visen, to be greater than 40\% of \ecm.  Visible
energy is the sum of all track energies (determined from momentum,
assuming the charged particle to have a mass of 140~MeV) and neutral
shower energies (neutral showers must be at least 7~cm from all
tracks).  If all particles in an event are detected, \visen\ $\approx$
\ecm.  As seen in Figure~\ref{visen}, hadronic \ups\ events peak in
\visen\ at 80\% \ecm, and there is a peak of non-\ups\ events at 15\%
\ecm.  At least two-thirds of the events in this low-\visen\ peak are
two-photon collisions, in which one incident electron has taken most
of the center-of-mass energy, undetected, down the beam-pipe.  We know
this because two-thirds of events with less than 30\% visible energy
contain one low-momentum electron, whose charge and direction are
correlated with the incident beams, and a highly anisotropic
distribution of shower energy, presumably from the boosted hadron
system $X$.  Decays of \tautau\ cover a broad spectrum of \visen, due
to energy lost in one or two neutrinos, extending but not peaking
below our cut threshold.

Rejecting low-\visen\ events also protects our hadron count from
uncertainties associated with trigger thresholds.  In our simulations,
only 0.07\% of hadronic events with \visen\ $>$ 40\% \ecm\ fail to
trigger, so any fluctuations in the electronics will be on this level.
The \visen\ threshold, situated in the flat minimum between the
two-photon peak and the signal peak, is minimally sensitive to
fluctuations in the two-photon background and the signal efficiency.
Since only 0.82\% of simulated hadronic \ups\ decays fail this cut,
any fluctuations in this efficiency will be well under a percent.

In addition to the two-photon peak at an \visen\ of 20\% \ecm, there
is an excess of events with \visen\ just above 50\% \ecm\
(Figure~\ref{visen}).  These events are likely to be radiative Bhabhas
($e^+e^- \to \gamma e^+e^-$) in which one of the two electrons is
lost.  They contain neutral calorimeter energy and an energetic
electron whose charge and direction is correlated with the incident
beams, like the two-photon fusion events.  However, visible energy in
two-photon collisions is expected to be much less than 50\% of \ecm.

The number of background events from a continuum process is
proportional to the integrated luminosity, just like the number of
signal \ups\ decays.  Their contribution to the apparent cross-section
will therefore be purely a function of \ecm.  The same cannot be said
for backgrounds that are not the product of beam-beam collisions.
Beam-gas and beam-wall rates are a function of the individual beam
currents, the gas pressure inside the beam-pipe (for beam-gas) and the
extreme tails of the bunch shape (for beam-wall).  Cosmic rays are
abundant in our detector, and the number of cosmic ray events is only
a function of time.  Integrated luminosity, integrated current, and
time are approximately proportional (within a factor of two), so a
continuum subtraction largely removes these effects, but not entirely.

\begin{figure}[p]
  \begin{center}
    \includegraphics[width=\linewidth]{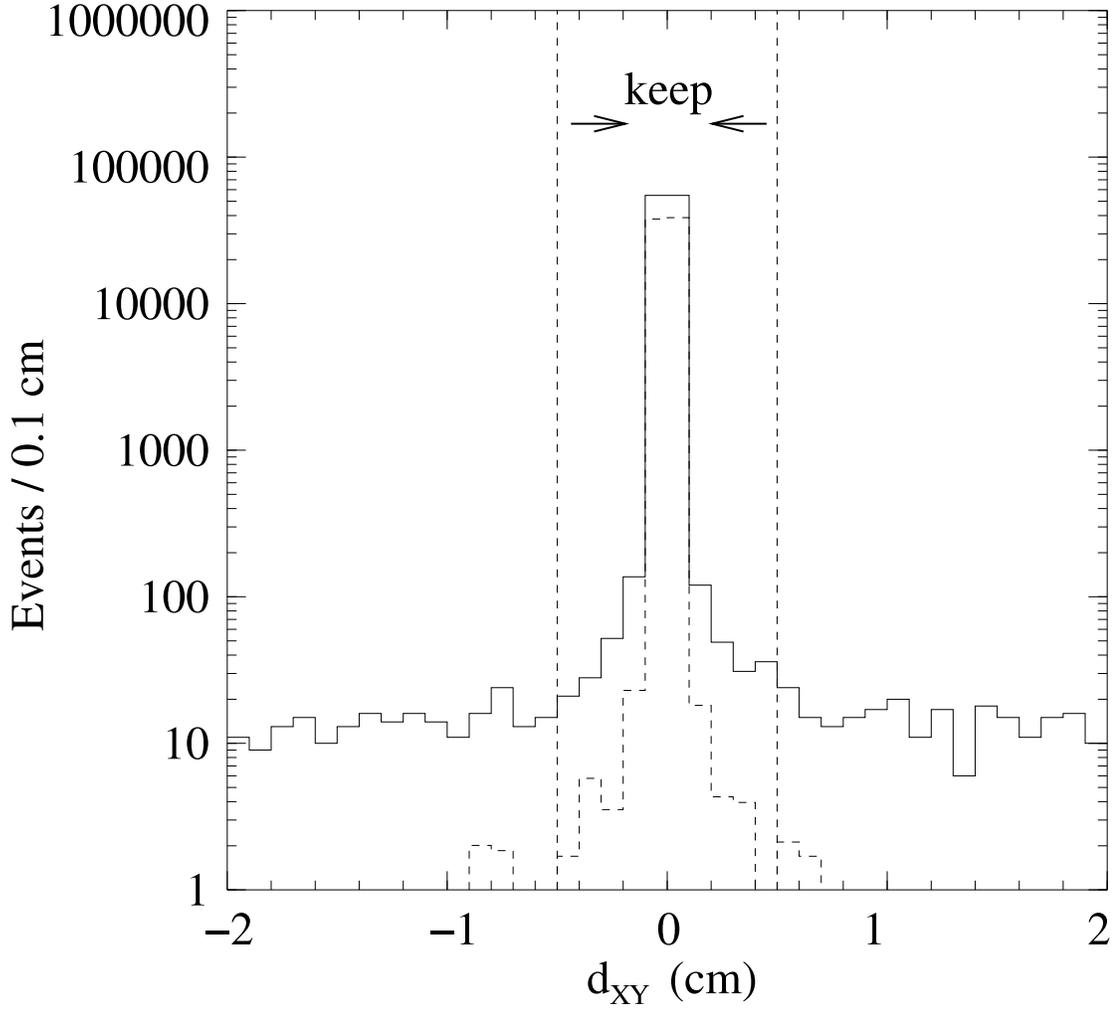}
  \end{center}
  \caption[Distribution of the closest track projection to the
  beam-line]{\label{dxy} Distance from the closest track projection to
  the beam-line for data (solid histogram) and Monte Carlo \ups\
  decays (dashed) with all other cuts applied.  The sign is related to
  the orientation of the track's curvature.  The flat background in
  data is due to cosmic rays.}
\end{figure}

We suppress beam-gas, beam-wall, and cosmic ray events by requiring
the event to originate near the beam-beam crossing point.  To select
events originating near this point in an $x$-$y$ projection, we
require at least one track to extrapolate within 5~mm of the beam-line.
We define \dxy\ as the distance of closest approach of the closest
track to the beam-line, and reject events with \dxy\ $>$ 5~mm.  Tracks
extrapolated from the tracking volume are corrected for momentum loss
in the beam-pipe and silicon detector, and the location of the
beam-beam crossing point is measured independently for each run, using
the first 500 hadronic events.  The \dxy\ distribution
(Figure~\ref{dxy}) is much narrower than our 5~mm threshold: only
0.1\% of beam-beam collision events fail this cut.  This allows for
$\sim$1~mm errors in the beam-beam intersection measurement, which is
far larger than expected.

\begin{figure}[t]
  \begin{center}
    \includegraphics[width=0.6\linewidth]{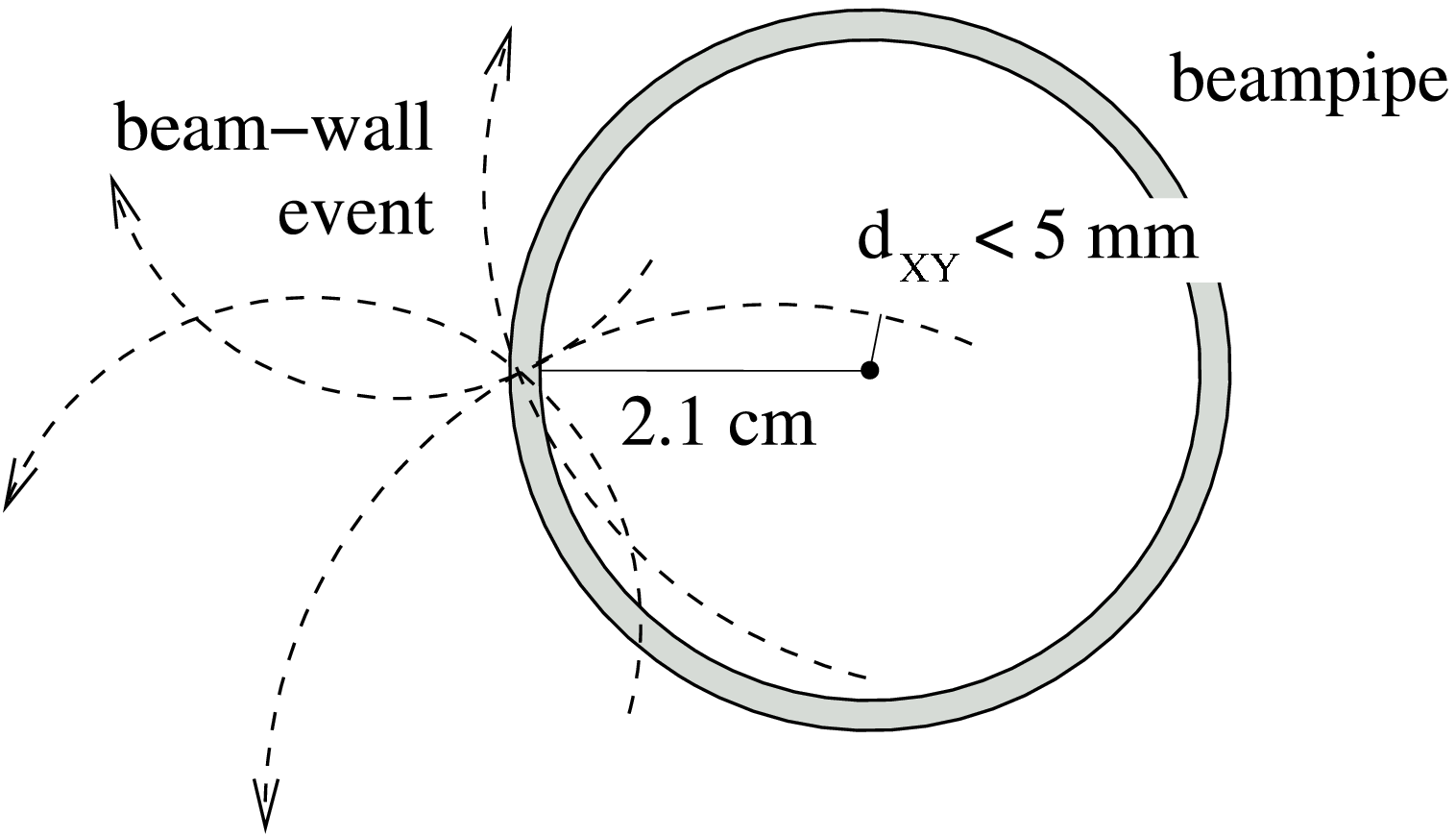}
  \end{center}
  \caption[How beam-wall events project tracks within 5~mm of the
  beam-line]{\label{biastowardzero} Though beam-wall events are
  centered at the beam-pipe, tracks may still project within 5~mm of
  the beam-line, thus passing the \dxy\ cut.}
\end{figure}

Our \dxy\ cut is extremely effective at rejecting cosmic ray events.
Cosmic rays rain uniformly into the detector, generating a uniform
background to \dxy, which extends to 25~cm with our triggers.  Only
cosmic rays that pass within 5~mm of the beam-line survive.  In
principle, beam-wall events should also be eliminated, since they are
generated in the beam-pipe, 2.1~cm from the beam-line.  However,
beam-wall events contain several tracks, any one of which may project
into the accepted \dxy\ region (Figure~\ref{biastowardzero}).  By
placing our requirement on the closest track, we bias this background
to peak within our accepted region, diluting the effectiveness of the
cut.

\begin{figure}[p]
  \begin{center}
    \includegraphics[width=\linewidth]{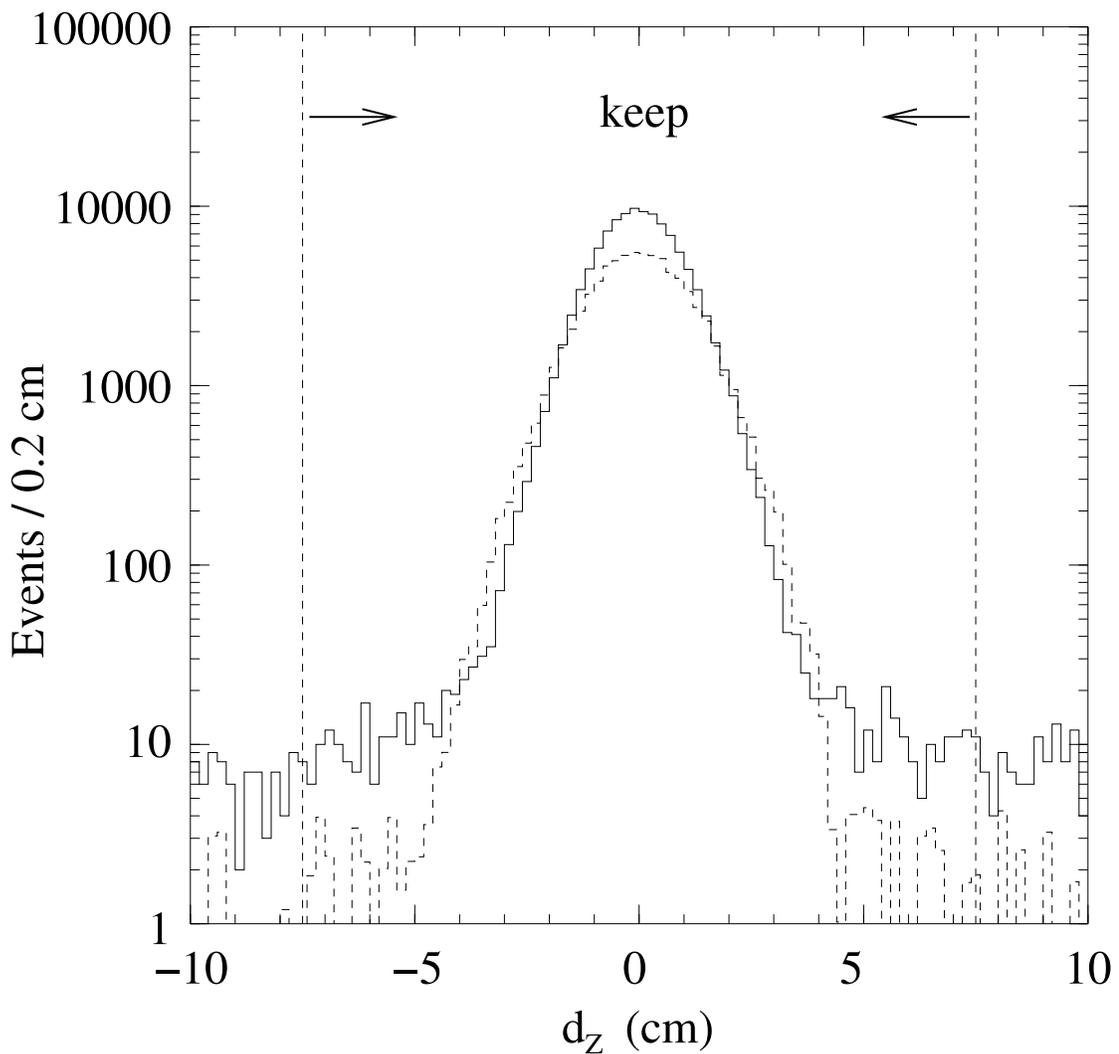}
  \end{center}
  \caption[Distribution of event vertex $z$ positions]{\label{dz}
  Location of the event vertex in $z$, according to our algorithm, for
  data (solid histogram) and Monte Carlo \ups\ decays (dashed) with
  all other cuts applied.  Data and Monte Carlo differ in the
  $z$-length of the beam-beam overlap region.  The flat background in
  data is primarily beam-gas and beam-wall, though Monte Carlo
  indicates that 0.5\% of \ups\ decays are misreconstructed and extend
  beyond the cut threshold.}
\end{figure}

Beam-gas and beam-wall events originate along the beam-line and
beam-pipe, extending beyond beam-beam collisions in $z$.  Placing a
requirement on the closest track to the $z$-collision point would be
ineffective for the same reason as above; beam-gas and beam-wall
events both have many tracks, and the probability that one of these
would project into the signal region (which is several centimeters
wide) is not negligible.  Instead, we reconstruct the $z$ position of
the event vertex using all tracks, and call this quantity \dz.  The
CLEO event vertexing algorithm is not useful because it was designed
for signal reconstruction and fails to fit too many beam-gas and
beam-wall events.  Instead, we developed a simple algorithm of our
own.  Tracking resolution is such that most of the tracks from a
beam-beam collision intersect within 0.1~mm of a common origin in the
$x$-$y$ plane, and the number of intersections near this point grows
rapidly with the number of primary tracks.  Accidental track
intersections far from this point grow more slowly.  We can therefore
determine the event vertex very accurately by averaging the $z$
positions of track-track intersections.  We define the $z$ position of
an $x$-$y$ intersection to be halfway between the $z$ positions of the
two track helices, evaluated at the $x$-$y$ intersection point.  If
the intersection is a true three-dimensional vertex, the tracks' $z$
positions will be nearly equal.  We weight these intersections with
uncertainties propagated from the track uncertainties, the tracks' $z$
separation, and the $x$-$y$ distance to the beam-line added in
quadrature, to prefer true intersections from the primary vertex.  We
plot this \dz\ distribution in Figure~\ref{dz}, and cut very loosely
at 7.5~cm, to allow for errors in the beam-beam intersection
measurement.

\begin{figure}[p]
  \begin{center}
    \includegraphics[width=\linewidth]{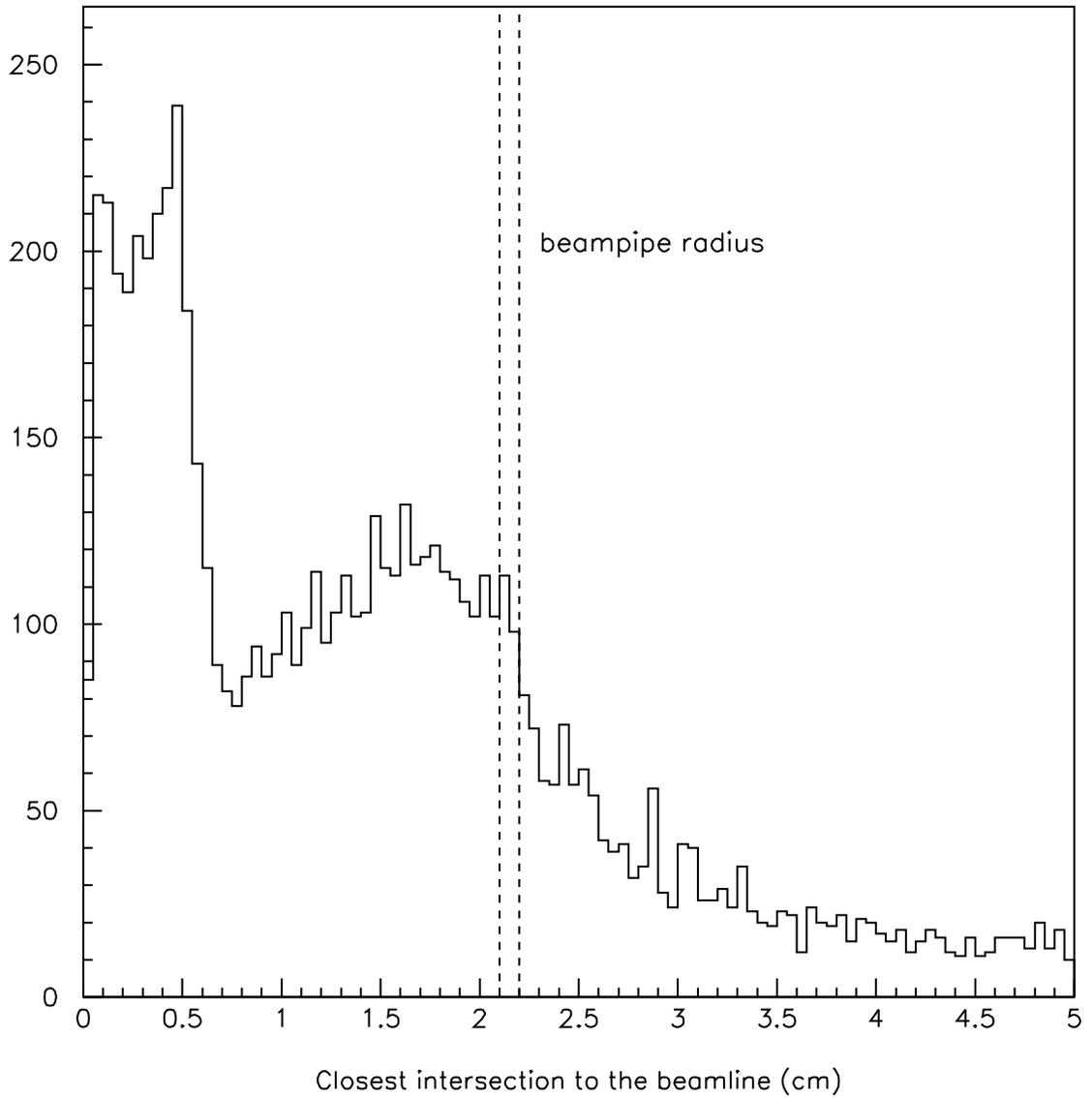}
  \end{center}
  \caption[Relative magnitudes of beam-gas, beam-wall, and cosmic
  rays]{\label{smallbeamwall} Closest track-track intersection to the
  beam-line in special data with one, non-colliding beam in CESR (all
  events are beam-gas, beam-wall, and cosmic rays).  The rough peak
  below 0.6~cm is mostly beam-gas, and the broad peak from 1 to 2~cm
  is due to beam-wall events.}
\end{figure}

It is also possible to use track intersections to distinguish
beam-wall events from beam-gas.  The distance of the closest
track-track intersection to the beam-line will be nearly zero for
beam-gas events, but peak below the beam-pipe radius for
beam-wall events (because selecting the closest intersection to the
beam-line biases the distribution toward zero).  In
Figure~\ref{smallbeamwall}, we plot the distribution of closest
intersections for events with \dxy\ $<$ 5~mm from data with only one
beam in CESR.  We see that the \dxy\ cut reduces beam-wall to the
extent that it is approximately as common as beam-gas.  A more
sophisticated average of intersections could help to discriminate
between beam-gas and beam-wall, but as we will see in
Subsection~\ref{sec:bgbwcr}, the two processes combined are a small
contamination, about 0.2\% of the continuum for most runs.  We
therefore will not attempt to correct for beam-wall and beam-gas
separately.

\section{Data Quality Requirements}
\label{sec:quality}

Not all data were collected under ideal conditions, so we applied some
general criteria for rejecting bad runs.  As the data were collected,
two CLEO operators inspected the data for hardware failure.  In the
most serious cases, these data were eliminated from all CLEO analyses,
but if the effect was limited, it was listed in a ``bad runs'' file
({\tt /home/dlk/Luminosity/badruns3S}).  We rejected any runs that
were flagged with drift chamber, silicon vertex detector, or CsI
calorimeter problems.

\begin{figure}[p]
  \begin{center}
    \includegraphics[width=\linewidth]{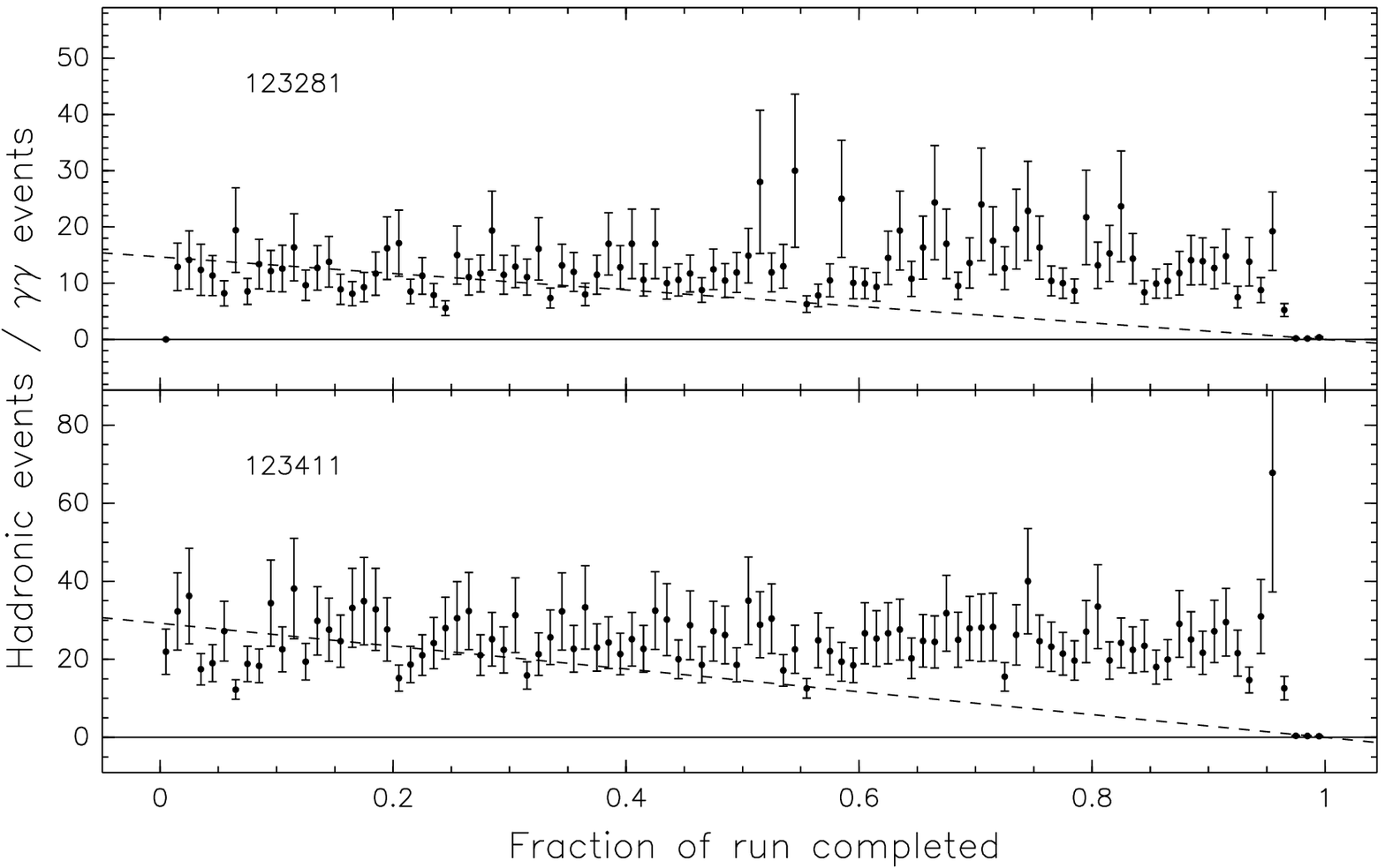}
  \end{center}
  \caption[Two runs in which the hadronic cross-section dropped at the
  end of the run]{\label{crashruns} Ratio of hadronic events to \gamgam\ events
  as a function of time through two runs.  The hadronic cross-section
  drops to zero in the last 3\% of these runs (the dashed lines are
  linear fits to the data.)}
\end{figure}

We want a robust measurement of cross-section, and cross-section is
constant with time, even as the beam currents are depleted during a
run.  We therefore checked for variations in cross-section during each
run by comparing hadronic events and \gamgam\ events in hundredths of
each run.  This ratio fluctuates statistically, but we found two
examples in which the drift chamber lost sensitivity to tracks before
the calorimeter lost sensitivity to showers in the last few minutes of
the run (Figure~\ref{crashruns}).  Most likely, the drift chamber lost
high voltage just before the end of the run.

\begin{figure}[p]
  \begin{center}
    \includegraphics[width=0.95\linewidth]{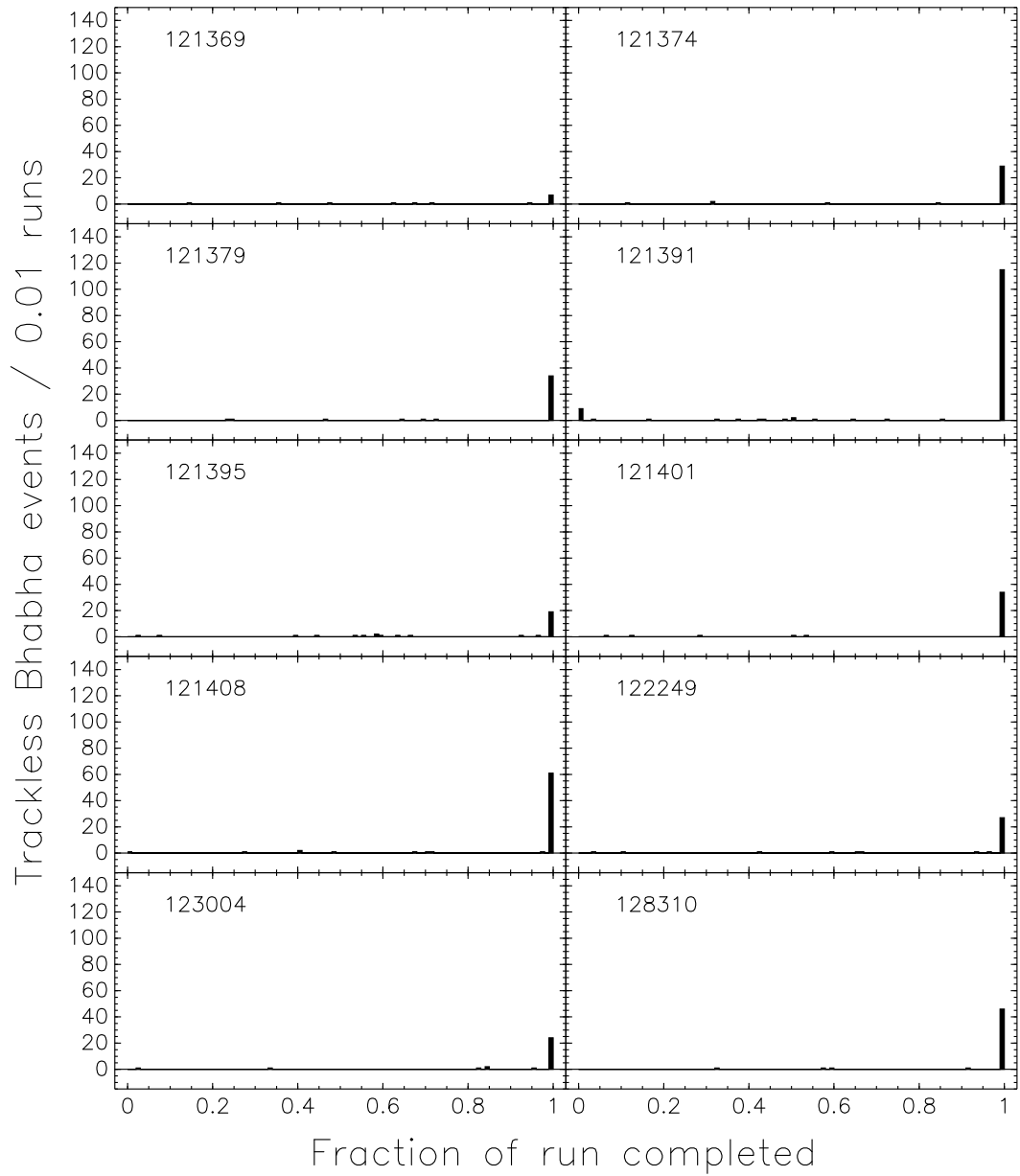}
  \end{center}
  \caption[Ten runs in which the trackless Bhabha fraction increased
  at the end of the run]{\label{crashruns2} Bhabha events with no
  observed tracks as a function of time through ten runs.  In all of
  these cases, there is an excess in the last 1\% of the run.}
\end{figure}

To catch more instances of this kind of failure, we also compared the
rate of trackless Bhabhas to total Bhabhas.  We recognize the \ee\
final state by the two beam-energy showers it produces in the
calorimeter, curved 0.1 radians away from perfect collinearity by the
magnetic field.  Twenty-five runs had high trackless Bhabha rates
(above 0.3\%), and all of the trackless Bhabha excesses were in the
same hundredth of a run (usually the last).  Ten of these (presented
in Figure~\ref{crashruns2}) were crucial to the resonance scans and
therefore not rejected.  Instead, we determined the cross-section from
the first 99\% of these runs.  The twenty-seven runs with drift
chamber failures are listed in Table~\ref{tab:runfailures}.

\begin{table}
  \caption[Runs rejected for hardware/calibration
  reasons]{\label{tab:runfailures} Runs rejected for
  hardware/calibration reasons.}
  \begin{center}
    \begin{tabular}{p{0.5\linewidth} p{0.4\linewidth}}
      \hline\hline
      Drift chamber failed at the end of the run & 121476, 121748, 121822, 121847, 122685, 123281, 123411, 123436, 123847, 123873, 124816, 124860, 124862, 125367, 126273, 126329, 127280 \\
      \barrelbhabha\ trigger inefficiency & 121928, 121929, 121953, 127951, 127955, 130278, 121710, 121930, 121944, 121954, 123884 \\
      Overestimated track momenta & 124452, 124454, 124456, 124458, 124462, 124464, 124465, 124466, 124467, 124469, 124472, 124473, 124474, 124475, 124477, 124478, 124479, 124480 \\
      Overestimated barrel shower energies & 122331, 122335, 122336, 122339, 122341, 122342, 122344, 122345, 122349, 122350, 122352 \\
      Large cosmic ray/beam-gas backgrounds & 122353, 126341, 129522 \\
      Large, unidentified backgrounds & 121595, 122093, 122330, 126510 \\
      Too little data for tests & 123013, 123014 \\\hline\hline
    \end{tabular}
  \end{center}
\end{table}

The \gamgam\ final state, which we use for some diagnostic checks, is
accepted only by the specialized \barrelbhabha\ trigger.  We studied
the efficiency of this trigger with Bhabha events and discovered
eleven runs with very low efficiency, which we rejected, though these
failures would only have affected our cross-checks.  They are also
listed in Table~\ref{tab:runfailures}.

\begin{figure}[p]
  \begin{center}
    \includegraphics[width=0.85\linewidth]{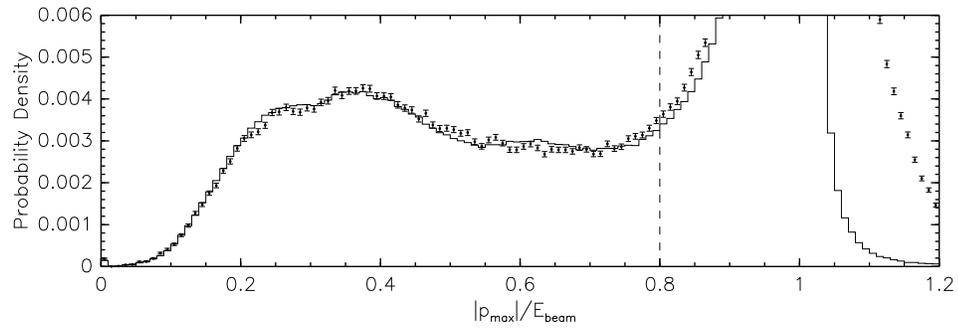}
  \end{center}
  \caption[Distribution of largest track momenta on March 7,
  2002]{\label{highptracks} The largest track momentum in
  off-resonance \us\ (solid histogram) and March 7, 2002 runs (points
  with errorbars).}
\end{figure}

\begin{figure}[p]
  \begin{center}
    \includegraphics[width=0.85\linewidth]{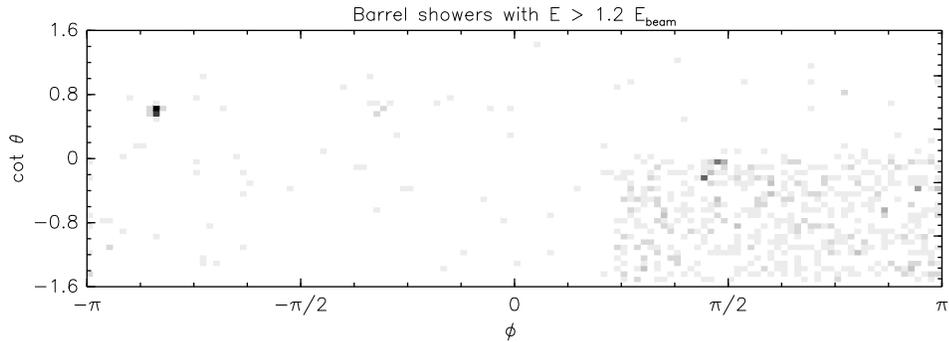}
  \end{center}
  \caption[Geometry of unphysically energetic showers on December 16,
  2001]{\label{higheshowers} The locations of unphysical shower
  energies on December 16, 2001.}
\end{figure}

\begin{figure}[p]
  \begin{center}
    \includegraphics[width=0.85\linewidth]{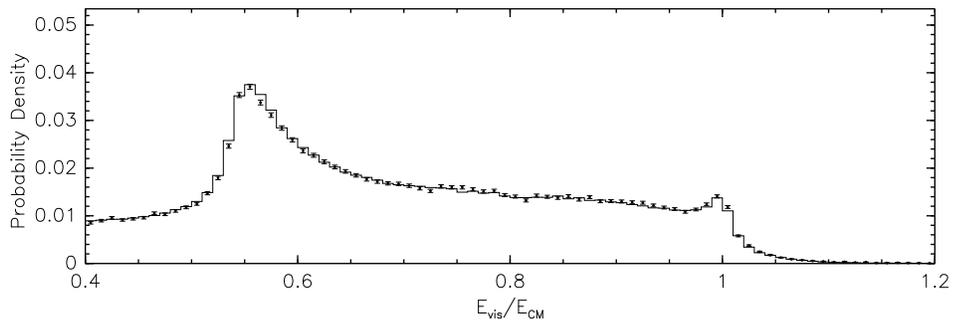}
  \end{center}
  \caption[Visible energy distribution on December 25--29,
  2001]{\label{higheendcapvisen} The visible energy spectrum in
  off-resonance \usss\ (solid histogram) and December 25--29, 2001
  runs (points with errorbars).}
\end{figure}

We also tested the quality of the drift chamber and calorimeter output
by counting unphysically high-energy tracks and showers.  In good
data, less than 1\% of Bhabhas will generate a track or a shower with
momentum or energy above 120\% \ebeam.  In a contiguous block of data
on March 7, 2002, the fraction of high-momentum tracks abruptly
increased to 3\%.  We see that the Bhabha peak for these runs has a
high-energy tail (Figure \ref{highptracks}), which suggests that the
momentum in a fraction of tracks is overestimated.  If this hypothesis
applies to tracks with lower momenta, events may fail the \pmax\ cut
due to anomalous momentum measurements, changing the cut efficiency.
We exclude these runs.  On a separate occasion, December 16, 2001, the
rate of high-energy showers abruptly rose to 3\%.  In this case, we
observed that most of the unphysical showers occupy a regular block in
the calorimeter barrel, indicating a read-out issue (Figure
\ref{higheshowers}).  Only our \visen\ cut depends on shower energies,
and in particular, only showers that cannot be associated with any
track, so the influence of calorimeter malfunctions on our hadronic
efficiency is limited.  However, we will use showers in the
calorimeter barrel to identify Bhabhas and \gamgam\ events, so we
exclude these runs as well.  Another calorimeter malfunction, this
time in the endcap, occured on December 25--29, 2001.  The \visen\
spectrum for off-resonance runs in this time period is not distorted
by excess background from a high-side tail on the two-photon fusion
peak (see Figure~\ref{higheendcapvisen}), so we do not exclude these
runs.  All rejected runs are listed in Table~\ref{tab:runfailures}.

We rejected a handful of runs due to high background rates.  From
Figure~\ref{backgroundsvsrun}, we set a 5\% upper limit on acceptable
cosmic ray yields relative to the continuum yield, and an upper limit
of 2\% on beam-gas.  Three runs failed these criteria.  We also
noticed that the fractions of hadronic, Bhabha, \gamgam, and \mumu\
events dropped abruptly in the middles of four runs, indicating a
sudden turn-on of some large background.  We rejected these, too.
Finally, two runs had so little data (16,695 events total) that it was
difficult to perform any of the above tests.  We rejected them for
convenience.

This analysis combines small ``scan'' datasets, taken on the \ups\
resonances but not at its maximum, with off-resonance and ``peak''
data taken at the maximum cross-sections.  The scan data were acquired
specifically for this analysis and therefore were not rejected
lightly.  (Only one run in Table~\ref{tab:runfailures} is a scan run:
124452.)  The peak data are less valuable, and even after the
selections described above, far more is available than is necessary.
A measurement of the area of an \ups\ lineshape (i.e.\ \gee) can be
conceptually decomposed into width measurements and height
measurements, in which the fractional uncertainty in the area is the
sum of the fractional uncertainty in the width and in the height, in
quadrature.  Scan data constrain both the width and the height, while
peak data constrain only the height.  Adding peak data to a fit will
always reduce the statistical uncertainty, though this reaches an
asymptotic limit as the uncertainty comes to be dominated by the width
measurements.  However, as the beam energy calibration drifts with
time, cross-sections slightly off the peak of the resonance may be
represented as being exactly on-resonance, thereby biasing the height
measurement.  We accepted no more peak data than what is necessary to
bring the statistical uncertainties within 5\% of their limiting
values.  Since we are concerned with potential drifts with time, we
re-expressed this limit as a time limit: we only include peak data in
a lineshape fit if this data were taken less than 48 hours after the
beginning of a scan.  We imposed no limit on off-resonance data.

\begin{figure}[t]
  \begin{center}
    \includegraphics[width=0.8\linewidth]{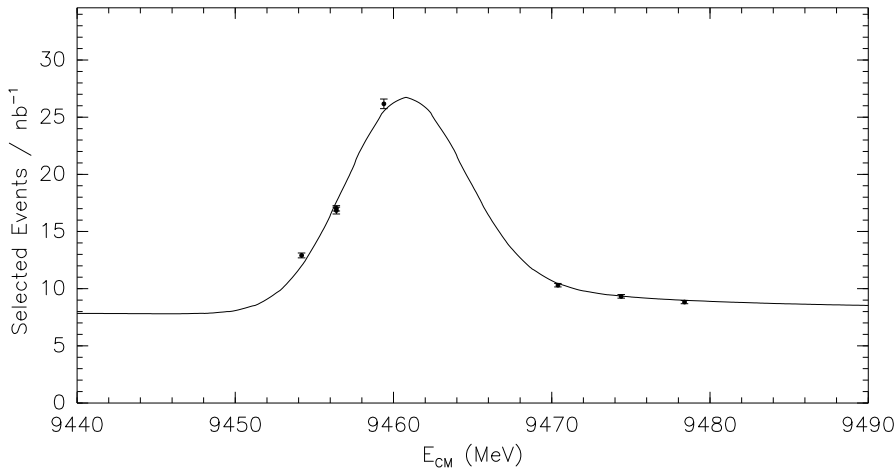}
  \end{center}
  \caption[The rejected April 3, 2002 scan]{\label{apr03scan} The
  April~3, 2002 lineshape scan, overlaid by a fit to all other \us\
  scans.  No data significantly constrain the high-energy side of the
  peak.}
\end{figure}

We rejected a \us\ scan, acquired on April 3, 2002.  This scan is
missing key cross-section measurements on the high-energy side of the
peak (Figure~\ref{apr03scan}), which makes it difficult to assess
uncertainties in the beam energy and the beam energy spread.  This
scan does include cross-section measurements well above the \ups\
mass, and may have been the victim of miscommunicated beam energy
requests.  (Requests are made relative to the \ups\ mass, and
single-beam energies used by CESR differ from our center-of-mass
energies by a factor of two.)  Its exclusion from the \us\ fit affects
the fit result by 0.12\% with no appreciable difference in
uncertainty.

\section{Subtracting Residual Backgrounds}

\begin{figure}[p]
  \begin{center}
    \includegraphics[width=0.9\linewidth]{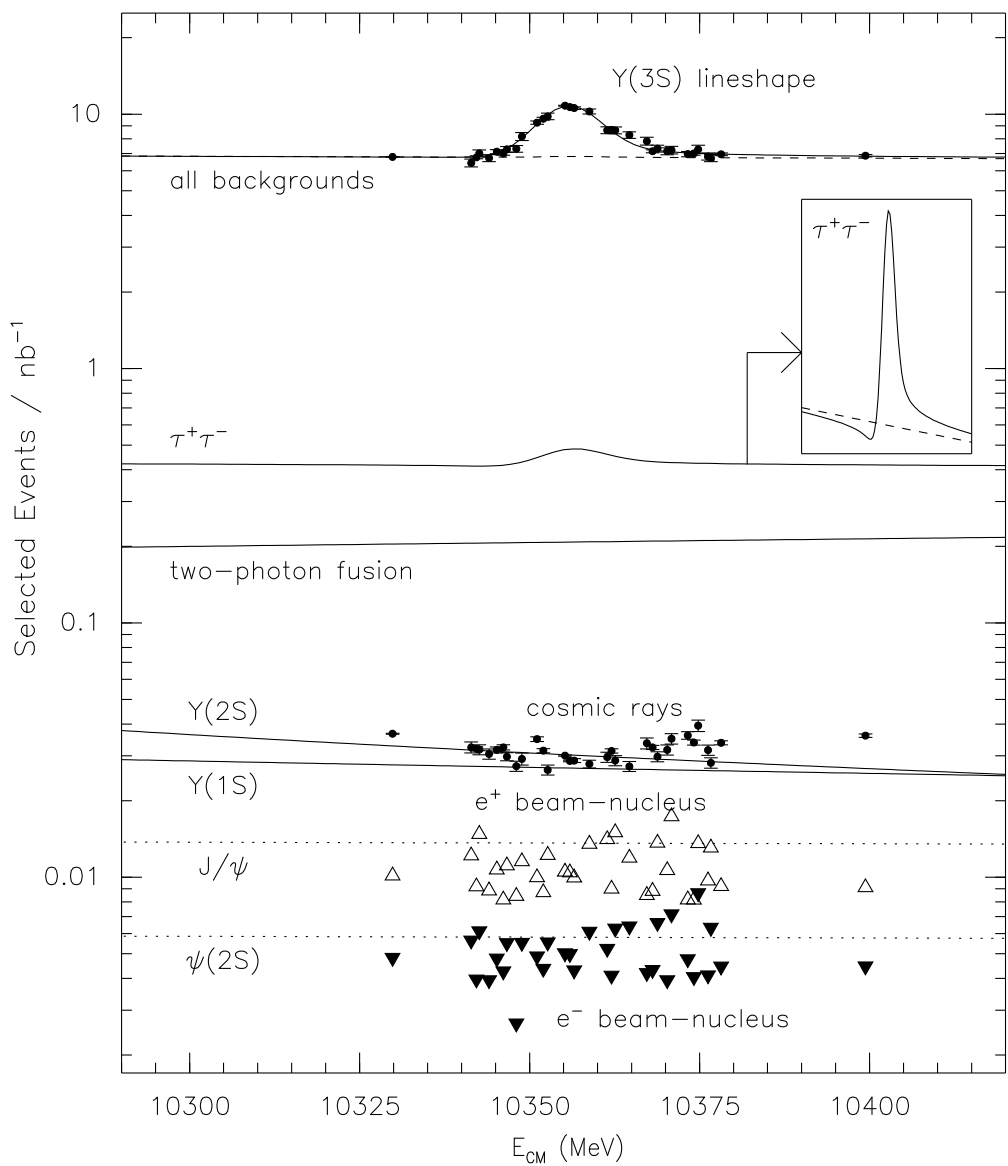}
  \end{center}
  \caption[Illustration of backgrounds in log scale]{\label{awesome}
  The \usss\ lineshape in log scale, to illustrate backgrounds.  The
  top dashed curve represents the sum of all backgrounds, which is
  dominated by $1/s$ continuum processes.  The solid curves and data
  points below this are non-$1/s$ corrections included in ``all
  backgrounds.''  Dashed curves represent ISR tails from charmonium
  resonances which are included in the two-photon fusion curve.  The
  overlap of ISR tail curves and non-beam-beam counts is accidental.}
\end{figure}

Backgrounds remaining after our cuts are summarized in Figure~\ref{awesome}.  We will discuss each of these, and their subtractions,
in the subsections that follow.

\subsection{Backgrounds that Vary Slowly with Beam Energy}
\label{sec:varyslowly}

After our cuts, radiative Bhabhas and continuum \qqbar\ dominate the
background, adding a flat, 8~nb plateau below our three \ups\ peaks
(18~nb, 7~nb, and 4~nb, respectively) in apparent cross-section versus
\ecm.  All continuum processes except for two-photon fusion
evolve as $1/s$, so we include such a function in our lineshape fits.
The magnitude of this term is determined independently for the \us,
\uss, and \usss\ by the large off-resonance samples taken only 20~MeV
below each \ups\ mass.  The $1/s$ curve is the dashed line near the top
of Figure~\ref{awesome}.

\begin{figure}[p]
  \begin{center}
    \includegraphics[width=\linewidth]{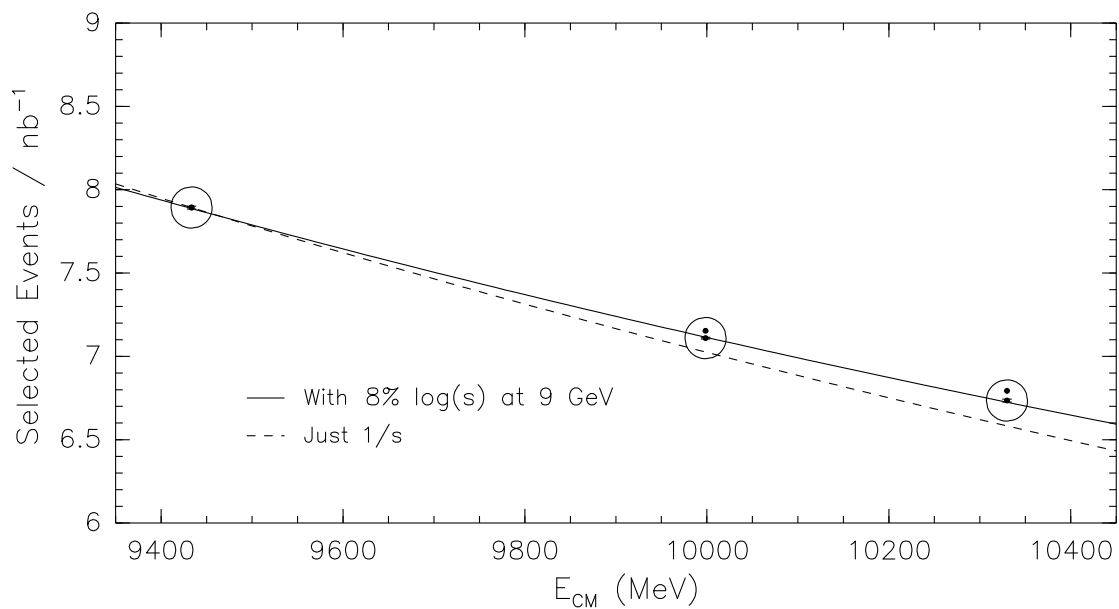}
  \end{center}
  \caption[Off-resonance effective cross-section as a function of
  \ecm]{\label{logsfit} Off-resonance cross-section measurements
  versus \ecm, with and without ISR tail corrections.  (Corrected data
  are at the centers of the circles.)  The solid curve is the best fit
  to $A/s+B\log s$, and the dashed curve is $1/s$ only, constrained to
  pass through the first data point.}
\end{figure}

The first correction to the background curve is to add lower-energy
\ups\ resonances, which have a $1/(\sqrt{s} - M_\Upsilon)$
distribution.  The magnitude of an ISR tail is set by the magnitude of
the \ups\ resonance.  We therefore fit \us, \uss, and \usss\ in
ascending order to obtain tail corrections from the previous fits.
The \us\ and \uss\ ISR tails under the \usss\ peak are labeled in
Figure~\ref{awesome}, and Figure~\ref{logsfit} shows \uss\ and \usss\
off-resonance cross-sections with and without this tail correction.

\begin{figure}[p]
  \begin{center}
    \includegraphics[width=\linewidth]{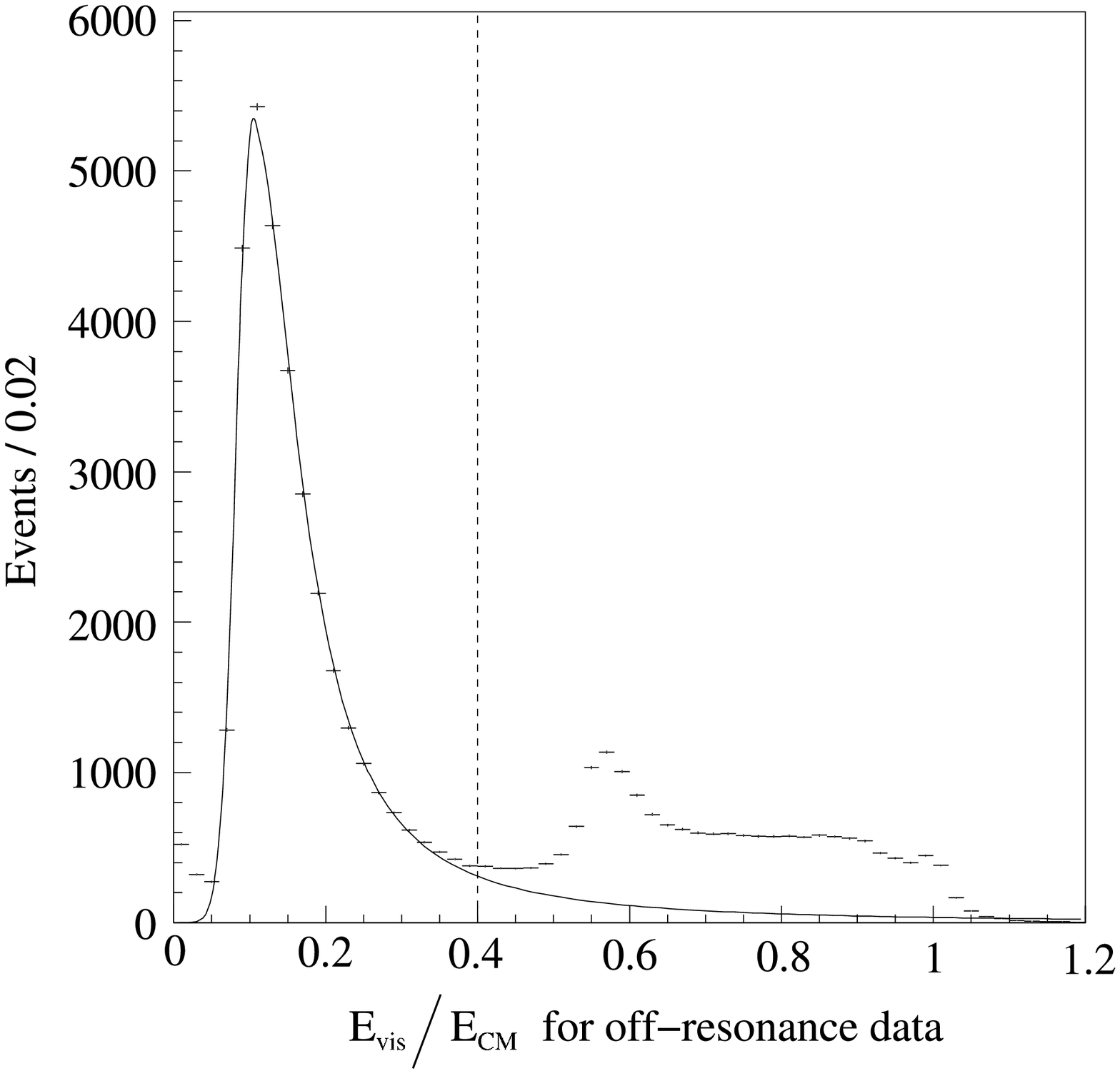}
  \end{center}
  \caption[Extrapolation of two-photon peak into the accepted visible
  energy region]{\label{extrapolatevisen} The visible energy (\visen)
  of off-resonance data, overlaid with a fit to the low-\visen\ peak.
  The fit is Gaussian on the low-energy side and Lorentzian on the
  high-energy side, and is used to roughly estimate the two-photon
  fusion events which survive after the 40\% of \ecm\ cut (dashed
  vertical line).}

\end{figure}

\label{pag:logs}
To parameterize the $\log s$ correction for residual two-photon
fusion, we fit the three off-resonance cross-sections to $A/s + B\log
s$ and present this fit in Figure~\ref{extrapolatevisen}.  We find
(8.0 \PM\ 0.5)\% of the apparent cross-section at 9~GeV to be due to
the $\log s$ component.  To see if this is plausible, we roughly
estimate the two-photon background surviving our cuts by extrapolating
the two-photon peak above our cut threshold in \visen\ (see
Figure~\ref{extrapolatevisen}), yielding a two-photon fraction of 6\%.
This is consistent with our $A/s+B\log s$ fit.  Other effects may
contribute to part of the $\log s$ term, such as \ecm\ dependence in
our cut efficiency for continuum events, a slow variation in the
hadronic continuum cross-section, and ISR tails from charmonium
resonances ($J/\psi$ and $\psi'$, see Figure~\ref{awesome} for scale).
All of these effects vary slowly with \ecm, so our parameterization
for large differences in \ecm\ (900~MeV from \us\ to \usss) applies to
small differences in \ecm\ as we project the \us, \uss, and \usss\
off-resonance cross-sections below each peak.  The difference in
cross-section between a pure $1/s$ curve and the fully parameterized
curve is only 0.04\% at the peak.

\subsection{Continuum-Resonance Interference}

As discussed in Section \ref{sec:earlyinterference}, resonant
$\Upsilon \to q\bar{q}$ interferes with continuum \qqbar.  We must
therefore also add a $\tilde{\sigma}_\subs{int}(E_\subs{CM})$ term to
our fit function (see Equation \ref{eqn:yint}).

In this analysis, we assume that $e^+e^- \to q\bar{q} \to$ hadrons
interferes with $e^+e^- \to \Upsilon \to q\bar{q} \to$ hadrons but not
$e^+e^- \to \Upsilon \to ggg \to$ hadrons, though the latter may share
some final states which are indistinguishable from \qqbar\ decays.
(Interference from \gggamma\ is negligible because its branching
fraction is only 3\% of $ggg$ and most \gggamma\ events have a
distinctive, high-energy photon.)  For interference between \qqbar\
and $ggg$ decays, quantum states must remain coherent through the
hadronization process.  This effect has been observed in $J/\psi$ and
$\psi' \to \pi^+\pi^-$ and $K^+K^-$, but it is unclear if the effect
is significant when summed over all final state amplitudes, since they
may cancel.  The phase difference between \qqbar\ and $ggg$ for the
inclusive process is also unknown, and some phase differences cannot
be constrained by our lineshape fits.  We will therefore only assume
parton-level interference, and discuss full hadronic interference as a
fitting issue in Chapter \ref{chp:fitting}.

\subsection{Backgrounds from \boldmath \ups}

Since we are selecting hadronic \ups\ events, $\Upsilon \to e^+e^-$,
\mumu, and \tautau\ are backgrounds which peak under the hadronic
\ups\ signal.  We have no control sample for leptonic \ups\ modes, so
we estimate these with a Monte Carlo simulation: negligible \ee\ and
\mumu\ survive the \pmax\ cut (0.22\% and 0.25\%), even with
final-state radiation ($\Upsilon \to \gamma e^+e^-$ and $\gamma
\mu^+\mu^-$) modeled by PHOTOS.  Our cuts and trigger are 57\%
efficient for \tautau, however.  A tau lepton may decay into several
hadrons, making it difficult to distinguish from hadronic \ups\
decays.  Tau-pairs are rejected primarily by the \visen\ cut, as their
visible energy spectrum is very broad due to neutrinos in the final
state.

We will need to subtract \tautau\ events from the hadronic \ups\
count.  The \ecm\ dependence of $\Upsilon \to \tau^+\tau^-$ is the
same as $\Upsilon \to$ hadronic, though the magnitudes of the resonant
and interference terms both differ.  The resonant \tautau\
contribution is a factor of ${\mathcal B}_{\tau\tau}/{\mathcal
B}_\subs{had}$ times smaller than the hadonic resonance, and the
\tautau\ interference term has a $\alpha_\subs{int}$ of 0.20, 0.37,
and 0.27 for the \us, \uss, and \usss, respectively.  Continuum
\tautau, like continuum \qqbar, is included in the $1/s$ term.  When
we estimate systematic uncertainties in the lineshape parameterization
in Section~\ref{pag:dontneedepsilon} (page
\pageref{pag:dontneedepsilon}), \label{pag:dontneedepsilonb} we will
note that the uncertainty in \btt\ overwhelms the uncertainty in
\tautau\ efficiency and \tautau\ $\alpha_\subs{int}$, so only the
branching fraction uncertainty must be propagated.

\subsection{Beam-Gas, Beam-Wall, and Cosmic Rays}
\label{sec:bgbwcr}

The non-beam-beam backgrounds are not a strict function of integrated
luminosity, so we will need to explicitly subtract them from the
hadronic \ups\ count for each run.  To do this, we identify cosmic ray
events, beam-gas, and beam-wall events in every run with special cuts.
We then use control samples containing only cosmic rays or cosmic
rays, beam-gas, and beam-wall to determine how to relate the number of
non-beam-beam backgrounds that we counted to the number that survive
our hadronic cuts.  We then subtract this excess.

\begin{figure}[p]
  \begin{center}
    \includegraphics[width=\linewidth]{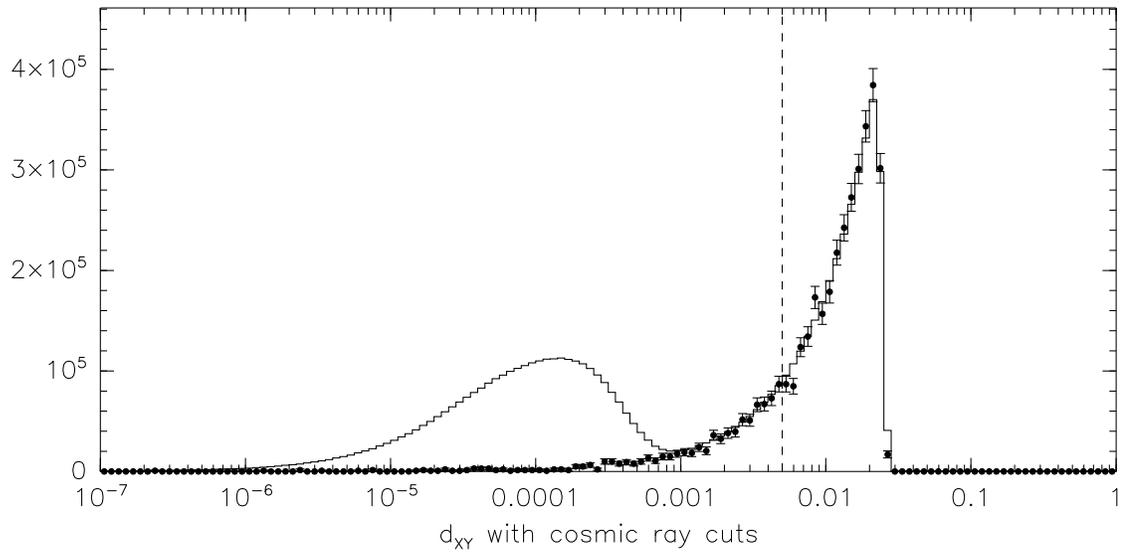}
  \end{center}
  \caption[Projecting cosmic ray magnitude into accepted data using
  the no-beam sample]{\label{dxydzcontaminationa} Distance from the
  beam-line ($|d_{XY}|$ in $\log x$-scale, where 1 = one meter) with
  all other cosmic ray cuts applied.  Data with beam-beam collisions
  are the solid histogram with a peak at 0.0001 (0.1~mm) from
  collisions and 0.01 (10~cm) from cosmic rays.  Data from the no-beam
  sample are the points with error bars, normalized to equal numbers
  of cosmic rays.  The dashed vertical line is the cut boundary at
  5~mm.  Triggers fail to accept cosmic rays beyond 25~cm.}
\end{figure}

To identify cosmic rays, we require the following.
\begin{itemize}

  \item No track may project within 5~mm of the beamspot ($|d_\subs{XY}|$ $>$
    5~mm).

  \item The event must contain at least two tracks, since our track
    reconstruction algorithm identifies the descending-radius part of
    the cosmic ray as one track and the ascending-radius part as
    another.

  \item The normalized dot product of the two largest track momenta
    ($\vec{p}_1 \cdot \vec{p}_2 / |\vec{p}_1| |\vec{p}_2|$) must be
    less than -0.999 or greater than 0.999, since the angles of these
    two tracks differ only due to tracking resolution (though the
    orientation may be confused by hits with unexpected drift times).

  \item The total calorimeter energy must be less than 2~GeV,
    consistent with two minimally-ionizing muon showers, and

  \item \visen\ $>$ 4\% of \ecm\ for less sensitivity to trigger
    thresholds.

\end{itemize}
These cuts are, by design, much more efficient for cosmic rays than
our hadronic cuts, but the number of identified cosmic rays and the
number of cosmic rays contaminating our hadron count is proportional.
To determine this constant of proportionality, we apply both sets of
event selection criteria to a data sample acquired with no beams in
CESR.  (The no-beam runs are listed in Table~\ref{tab:controls}.)
Figure~\ref{dxydzcontaminationa} superimposes cosmic ray candidates
from this no-beam sample on cosmic ray candidates from a large
beam-beam sample, indicating a clear separation between cosmic rays
and beam-beam collisions in \dxy.
We assume that all events in the no-beam dataset which pass our
hadronic cuts are cosmic rays, so the desired constant is just a ratio
of the cosmic ray count to the hadronic event count in this sample.
The effective cross-section of cosmic rays are plotted with
uncertainties in Figure~\ref{awesome}.

\begin{table}
  \caption[Run numbers for beam-gas, beam-wall, and cosmic ray control
    datasets]{\label{tab:controls} Run numbers for beam-gas,
    beam-wall, and cosmic ray control datasets.}
  \begin{center}
    \begin{tabular}{p{0.45\linewidth} p{0.45\linewidth}}
      \hline\hline
      no-beam & 128706 128736 128741 128748 \\
      electron single-beam & 126828 126920 126922 \\
      positron single-beam & 126785 \\ \hline\hline
    \end{tabular}
  \end{center}
\end{table}

Beam-gas and beam-wall events are hard to distinguish from one
another, but they are both small backgrounds which depend on the
electron and positron beam currents.  This dependence is not
identical, since beam-gas rates are proportional to the gas pressure
inside the beam-pipe while beam-wall is not.  However, the
contamination from beam-gas and beam-wall combined is typically 0.2\%
of the continuum.  Furthermore, our beam-gas and beam-wall cuts have a
small background from beam-beam data, meaning that our estimate is too
large.  Instead of subtracting all of this estimate, we inflate our
uncertainty.

To identify beam-gas and beam-wall events (which we will call
beam-nucleus), we require
\begin{itemize}

  \item $|d_\subs{XY}|$ $<$ 5~mm, $|d_\subs{Z}|$ $>$ 7.5~cm,

  \item $|\vec{p}_1 \cdot \vec{p}_2| / |\vec{p}_1| |\vec{p}_2|$ $<$
    0.9 to further reject cosmic rays,

  \item at least two tracks, and \visen\ $>$ 4\% of \ecm.

\end{itemize}
To distinguish between electron-induced beam-nucleus and
positron-induced beam-nucleus events, we also cut on the net
$z$-momentum of all tracks ($p_z^\subs{tr}$).  For a positron-induced
event, we require $p_z^\subs{tr}$ $>$ 10\% of \ebeam\ because incident
positron momentum is in the positive $z$ direction (see Figure~\ref{coordinatesystem}).  Electron-induced events must have
$p_z^\subs{tr}$ $<$ $-$10\% of \ebeam.

\begin{figure}[p]
  \begin{center}
    \includegraphics[width=\linewidth]{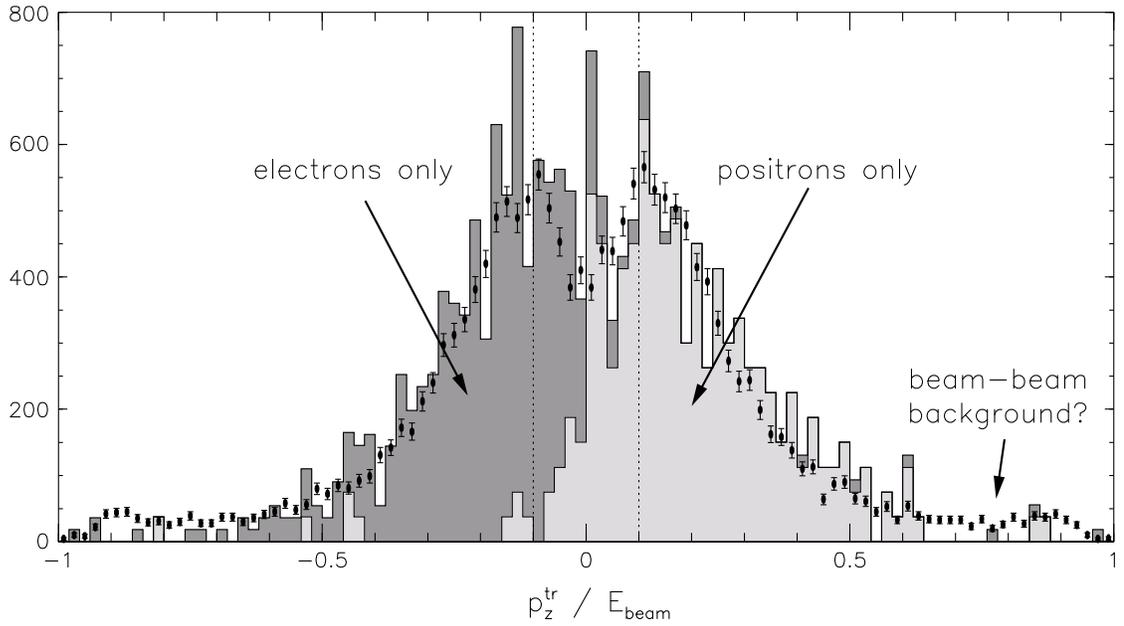}
  \end{center}
  \caption[Distinguishing electron- and positron-induced beam-nucleus
  events with $z$ momentum]{\label{beamgaspz} Net $z$-momentum of all
  tracks ($p_z^\subs{tr}$) with all other beam-nucleus cuts applied.
  Data with only positrons in CESR are lightly-shaded, electrons-only
  are darkly-shaded and stacked on the positrons-only histogram, and
  data from collisions are represented by points with error bars.  The
  boosts imparted by the incident beams are evident, and dotted
  vertical lines at $\pm$10\% of \ebeam\ indicate cuts for electron-
  and positron-induced beam-nucleus events.  The beam-beam data do not
  exactly reproduce the combined distribution, though the
  electrons-only and positrons-only histograms have been normalized to
  the same totals above and below $\pm$10\% of \ebeam.}
\end{figure}

\begin{figure}[p]
  \begin{center}
    \includegraphics[width=\linewidth]{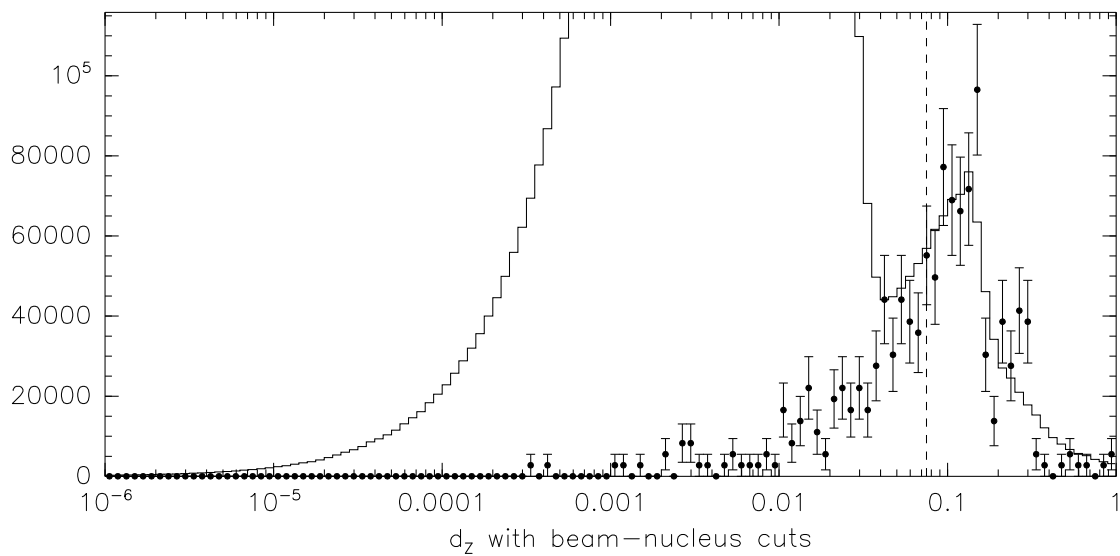}
  \end{center}
  \caption[Projecting beam-nucleus magnitude into accepted data using
  the single-beam samples]{\label{dxydzcontaminationb} Distance of the $z$ event
  vertex from the center of the beam-beam distribution ($|d_Z|$ in
  $\log x$-scale, where 1 = one meter).  Data with beam-beam
  collisions are the solid histogram with a peak above the plot window
  at 0.01 (1~cm) from beam-beam collisions and a peak at 0.1 (10~cm)
  from beam-nucleus collisions.  Data from the single-beam samples are
  the points with erorr bars, normalized to equal numbers of
  beam-nucleus events.  Cosmic rays have been subtracted from both
  samples.  The dashed vertical line is the cut boundary at 7.5~cm.}
\end{figure}

\begin{figure}[p]
  \begin{center}
    \includegraphics[width=0.85\linewidth]{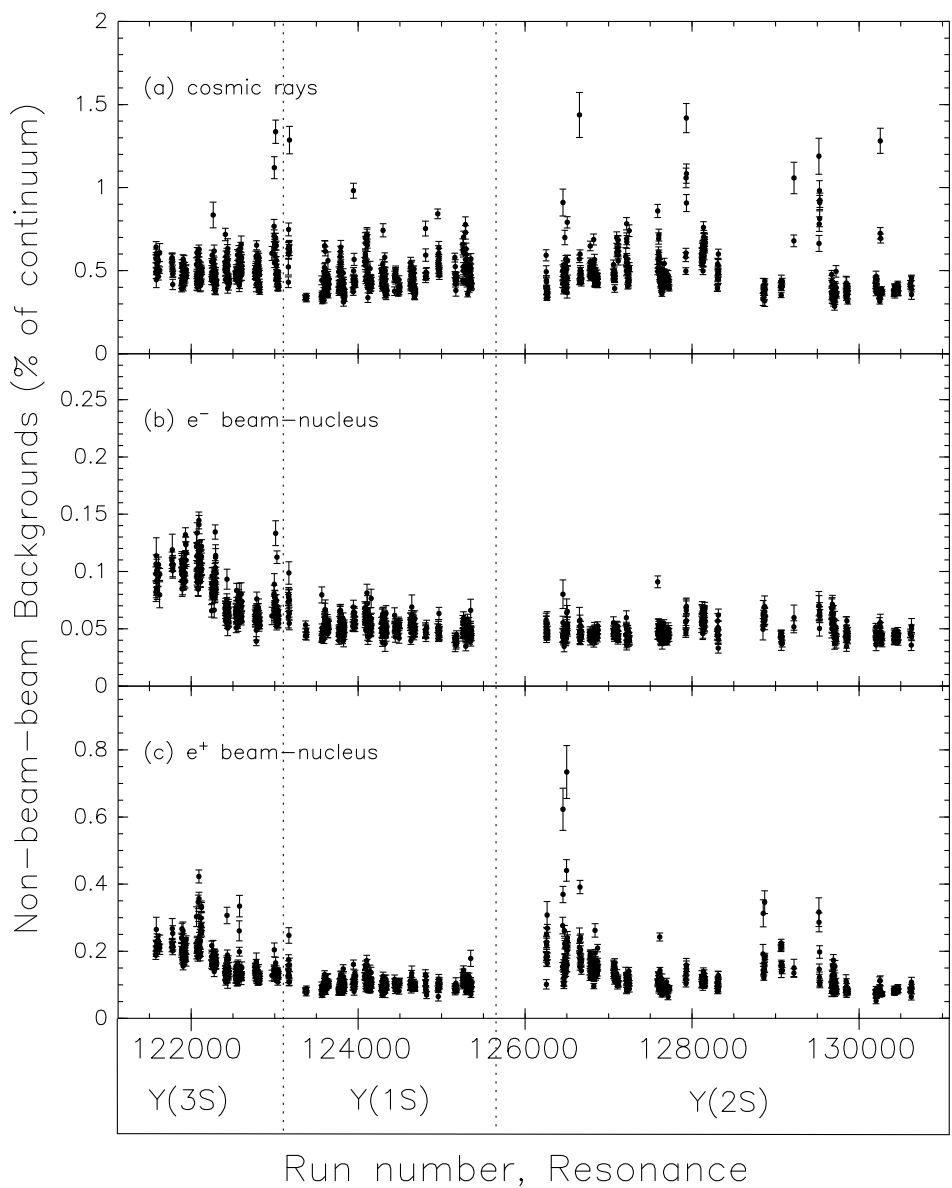}
  \end{center}
  \caption[Fraction of non-beam-beam backgrounds in every
  run]{\label{backgroundsvsrun} Total counts of non-beam-beam
  backgrounds as a fraction of the continuum level and a function of
  run number.  Dashed vertical lines separate \usss, \us, and \uss\
  data-taking periods.  Note that all three plots have different
  vertical scales: cosmic rays (a) are the most abundant, and the
  proportion of positron-induced beam-nucleus events (c) are typically
  twice that of electron-induced events (b).}
\end{figure}

To relate the number of identified beam-nucleus events to the number
that contaminate our hadronic event count, we employ data samples
acquired with only one beam in CESR (also listed in
Table~\ref{tab:controls}).  To use these samples, we must first
subtract the cosmic rays using the technique described above.
Figure~\ref{beamgaspz} demonstrates the separation of electron- and
positron-induced beam-nucleus by their net $z$-momenta.  This Figure
also indicates that a small fraction, perhaps 10\%, of our
beam-nucleus candidates are contaminated by beam-beam events, probably
two-photon fusion with a misreconstructed \dz.  The potential for
contamination is also evident in Figure~\ref{dxydzcontaminationb}.
Therefore, a beam-nucleus correction in analogy with the cosmic ray
correction would be an over-subtraction of about 10\%.  The
beam-nucleus estimates are typically only 0.1\% of the continuum
(Figure~\ref{backgroundsvsrun}), so we subtract 50\% \PM\ 50\% of the
electron- and positron-induced beam-nucleus estimates.  The effective
cross-section of beam-nucleus estimates are plotted near the bottom of
Figure~\ref{awesome}.  The cosmic ray and beam-nucleus estimates for
every run we used to determine \gee\ are plotted in
Figure~\ref{backgroundsvsrun}.

\chapter{Hadronic Efficiency}
\label{chp:efficiency}

\section{Motivation for the Data-Based Approach}

Inefficiency is in some sense the opposite of the problem of
backgrounds: after removing the events which should not be in our
hadronic \ups\ sample, we need to add in the events that are missing.
In this Chapter, we will determine the probability that a hadronic
\ups\ decay is included in our count, for each of the three
resonances.  These efficiencies are high, about 97\% for each
resonance.

Often, efficiencies are determined from Monte Carlo simulations.  One
simulates all known decay modes, and constructs an aggregate efficiency
\begin{equation}
  \epsilon = \sum_i \, \epsilon_i \, {\mathcal B}_i \mbox{,}
  \label{eqn:agregate}
\end{equation}
where $\epsilon_i$ is the efficiency of each mode.  We don't directly
use this method for two reasons.
\begin{enumerate}

  \item Hadronic decays are the result of the hadronization of bare
    quarks and gluons.  This is a non-perturbative process which is
    only empirically approximated by LUND/JetSet in the Monte Carlo.
    If we assume a non-perturbative QCD model to determine a
    non-perturbative QCD parameter, we would introduce a circular
    dependence that would have to be quantified.

  \item Our definition of hadronic \ups\ decays includes potentially
    unknown modes whose efficiencies may be very different from the
    hadronic modes we simulate.  For instance, it is possible that
    \ups\ decays into invisible $\mbox{\sc wimp}$s with zero
    efficiency or that unknown QCD resonances may enhance decays to
    $K_L$ or neutrons, which fail our \visen\ cut with greater
    probability.

\end{enumerate}

Instead, we take advantage of our 1.3 fb\inv\ sample of \uss\ decays
to study \twotoone\ transitions.  The \us\ mesons in these decays are
produced nearly at rest, and decay as they would from direct $e^+e^-
\to \Upsilon(1S)$.  However, the $2S \to 1S$ cascade events
additionally include two charged pions which may satisfy a trigger and
cause the event to be recorded, regardless of how the \us\ decays.  As
an extreme example, we can use this technique to collect events
featuring invisible $\Upsilon(1S) \to \nu\bar{\nu}$ or $\mbox{\sc wimp
} \overline{\mbox{\sc wimp}}$ decays, which would be impossible with a
direct \us\ sample.

We exploit this broad access to \us\ decays to measure the \us\
efficiency.  From \twotoone\ cascades, we select a subset in which the
\pipi\ by itself guarantees that the trigger will accept the event, so
that we know that the trigger did not rely on decay products of the
\us.  Assuming that there is no correlation between the kinematics of
the \pipi\ and the branching fractions of the \us, this subset of
\pipi\ is accompanied by a generic set of \us\ decays: \us\ decay
modes are represented in the data sample with the same proportions as
in nature.  We may then apply our cuts to the \us\ decay products in
our sample to determine the fraction which succeed.  This is the
efficiency.

\section{Hadronic Efficiency of the \boldmath \us}
\label{sec:usefficiency}

\begin{figure}[t]
  \begin{center}
    \includegraphics[width=\linewidth]{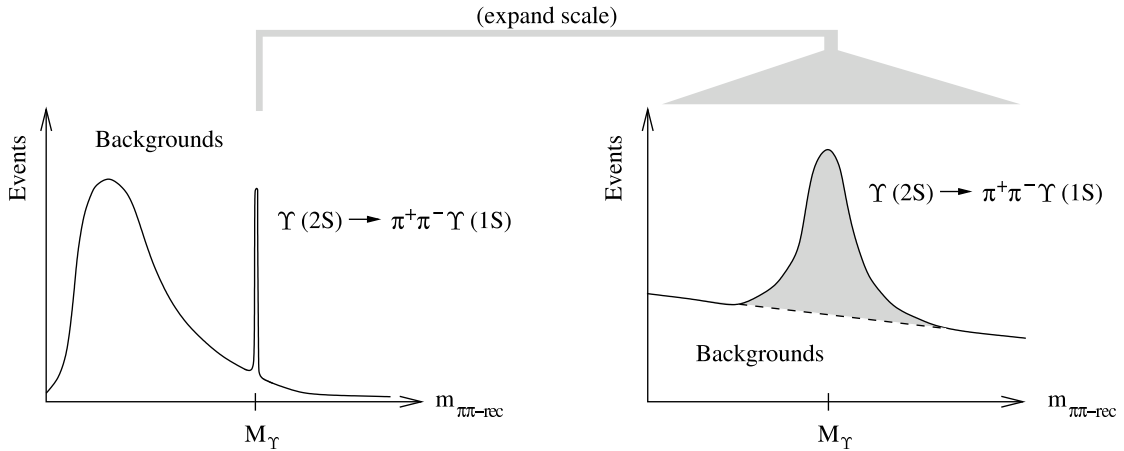}
  \end{center}
  \caption[Identifying \twotoone\ events by \pipi\ recoil
  mass]{\label{cascadescartoon} Distinguishing the recoil mass of
  \pipi\ in kinematically-constrained \twotoone\ events from other,
  accidental track-track combinations (backgrounds).  The background
  distribution is not simple, but it has no structure in the narrow
  region of our interest.}
\end{figure}

We have two goals for our event selection in this study, to identify
\twotoone\ candidates and to choose \pipi\ candidates which are
sufficient to satisfy the trigger.  The \pipi\ in \twotoone\ are
kinematically constrained by the mass difference between the \uss\ and
the \us, so the mass of the system recoiling against the two pions,
\begin{equation}
  {m_\subs{$\pi\pi$-rec}}^2 = \left(M_{\Upsilon(2S)} - \sqrt{|\vec{p}_1|^2 + {m_\pi}^2}
  - \sqrt{|\vec{p}_2|^2 + {m_\pi}^2}\right)^2 - |\vec{p}_1 + \vec{p}_2|^2 \mbox{,}
  \label{eqn:mrec}
\end{equation}
peaks at the \us\ mass.  This allows for excellent background
rejection, because the peak from \twotoone\ has a 3~MeV resolution
while the background spectrum is much broader (see
Figure~\ref{cascadescartoon}).  We require \pipi\ candidates to have a
recoil mass ($m_\subs{$\pi\pi$-rec}$) between 9.441 and 9.480~GeV.  We
further suppress background by requiring the track helices that we
identify as \pipi\ to intersect in the $x$-$y$ plane within 5~mm of
the nominal beam-beam collision point.  The track helices at this
$x$-$y$ point must also be less than 2.5~cm from each other in $z$,
and their average $z$ must be within 5~cm of the beam-beam collision
point.  The momenta used in Equation~\ref{eqn:mrec} are evaluated at
the intersection point.

To satisfy the trigger, we select \pipi\ track candidates with more
than 150~MeV of momentum perpendicular to the $z$ axis ($p_\perp$), so
that their trajectories reach beyond the sixteenth layer in the drift
chamber and satisfy the geometric requirements for \axial\ tracks.  A
study of CLEO's trigger response to hadronic tracks revealed that the
probability for a track with $p_\perp$ $=$ 150~MeV to be detected as
an \axial\ track is 99.96\%, and this probability grows with $p_\perp$
\cite{inga}.  We can therefore be at least 99.92\% certain that events
accepted by the \twotrack\ trigger (which requires two \axial\ tracks)
with this cut on the \pipi\ did not rely on \us\ decay products to be
accepted.

To simplify the process of excluding the \pipi\ candidates when we
apply our hadronic cuts, we additionally require each pion track to
have more than 60~MeV of $z$-momentum magnitude.  As discussed in
Section \ref{pag:spiraling} on page \pageref{pag:spiraling}, such
trajectories exit the detector before completing one half-orbit in the
magnetic field.  This protects our sample from events with pions which
spiral in the tracking volume, potentially generating many tracks,
only one of which is identified as a pion from \twotoone.  With this
cut, we can assume that each pion is responsible for only one track
and the calorimeter showers associated with that track.

There is occasionally more than one pair of tracks which satisfy these
\pipi\ criteria in a single event.  In this case, we choose a \pipi\
candidate randomly.  If we were to choose the \pipi\ candidate that
best reconstructs the \us\ mass, would bias the non-cascade
background to peak under the \twotoone\ signal, complicating the
background subtraction.  Since the \pipi\ candidate is chosen
randomly, the background distribution has no structure on the scale of
tens of MeV, and we can approximate it with a low-order polynomial.

Selecting \twotoone\ candidates in this manner, from events accepted
by the \twotrack\ trigger, and applying our hadronic cuts to the \us\
events in the peak, we find a hadronic \us\ efficiency that is
consistent with 100\% with 3\% uncertainty, which is not satisfactory
for our analysis.  Hadronic efficiency is a factor in \geehadtot, so
the fractional uncertainty in hadronic efficiency adds to the
\geehadtot\ fractional uncertainty in quadrature.

The main culprit is the prescaled \twotrack\ trigger, which randomly
rejects 94.7\% of the events that satisfy its two \axial\ track
criteria.  We can circumvent this loss of data and improve the
precision of our result by splitting the \us\ hadronic efficiency into
two factors.  Define an \us\ decay as ``visible'' if the trigger
records one \axial\ track from the \us\ decay products and one \cblo\
cluster from either the \us\ decay or the \pipi.  Define \evis\ to be
the probability that an \us\ decay is visible, and \ecuts\ to be the
probability that a visible \us\ decay is selected by our triggers and
cuts.  The hadronic \us\ efficiency is the product of \evis\ and
\ecuts.  We will determine \evis, a number which is very close to
100\%, using the \twotrack\ trigger, and \ecuts\ using the \hadron\
trigger, which selects visible \us\ decays accompanied by \pipi.
Since $(1-\epsilon_\subs{vis})$ is such a small inefficiency, a large
fractional uncertainty in $(1-\epsilon_\subs{vis})$ propagates to a
small fractional uncertainty in \evis.

Our definition of ``visible'' is tailor-made for the \hadron\ trigger
in \twotoone\ events.  The \hadron\ trigger requires three \axial\
tracks and one \cblo\ cluster.  We have selected \pipi\ kinematics
such that each pion must generate one \axial\ track, so the \us\ decay
is responsible for the third track and possibly the \cblo.  Whatever
the probability is that the \cblo\ comes from the \pipi, rather than
the \us, it is the same while measuring \evis\ as it is while
measuring \ecuts.  This gives us the freedom to determine the cut
efficiency with a large set of events from the \hadron\ trigger,
knowing that we can correct for the bias it introduces with \evis.

In our article in Physical Review Letters \cite{me}, we discuss this
technique more succinctly using two factors, $\epsilon_\subs{htrig}$
and \ecuts.  In that article, $\epsilon_\subs{htrig}$~=~\evis,
and \ecuts\ has the same meaning in both.

\subsection{Determination of \boldmath \evis}

\begin{figure}[p]
  \begin{center}
    \includegraphics[width=\linewidth]{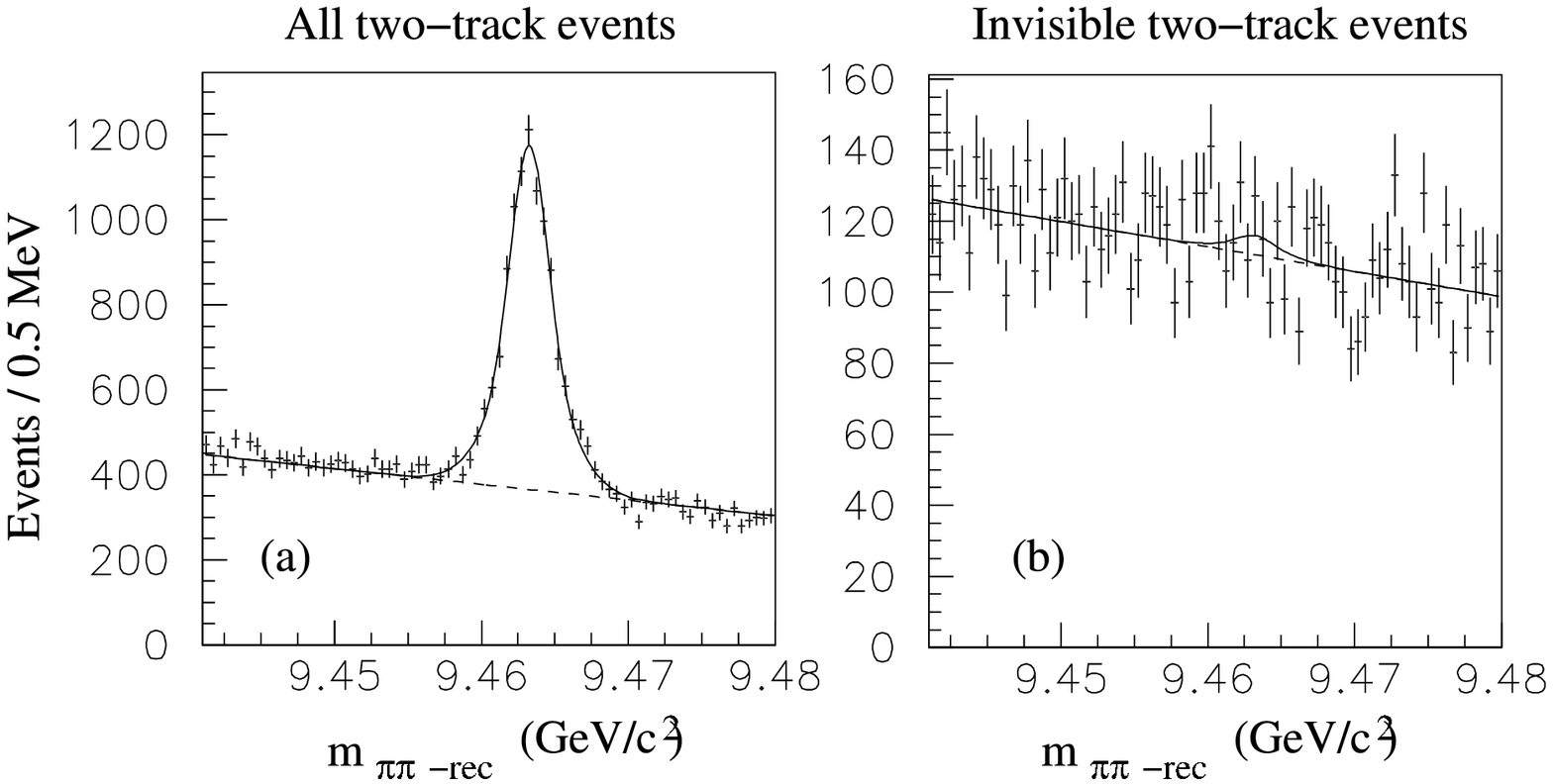}
  \end{center}
  \caption[\pipi\ recoil mass for events from the \twotrack\
  trigger]{\label{pipitwotrack} Recoil mass of \pipi\ candidates for
  (a) all events accepted by the \twotrack\ trigger, and (b) only
  those in which the rest of the event is ``invisible.''  True
  \twotoone\ events are contained in the peak near 9.46~GeV, which is
  highly suppressed in (b), indicating that very few \us\ decays are
  invisible.  The solid curve is a fit to the data, and the dashed
  curve is the linear background contribution.}
\end{figure}

To determine \evis, we apply the above procedure, replacing our full
set of cuts for the ``visible'' condition we have just defined.  The
recoil mass distribution of \pipi\ candidates accepted by the
\twotrack\ trigger is shown in Figure~\ref{pipitwotrack}(a).  The real
\twotoone\ events peak near the \us\ mass of 9.46~GeV, and backgrounds
are distributed linearly underneath.  To measure the fraction of \us\
decays that are ``invisible,'' we select from this sample events that
fail the \hadron\ trigger.  These are plotted in Figure~\ref{pipitwotrack}(b), revealing no significant peak.  This indicates
that nearly all of the \us\ decays are visible.

To quantify this statement, we fit the data in Figures
\ref{pipitwotrack}(a) and \ref{pipitwotrack}(b) to the same function and
extract the ratio of \twotoone\ yields.  We fit the full \twotrack\
dataset (Figure~\ref{pipitwotrack}(a)) to a double Gaussian with a 1:4
ratio of areas and a linear background.  When we fit the
invisible-\us\ data (Figure~\ref{pipitwotrack}(b)), we fix the mean and
widths, imposing the assumption that momentum resolution of the \pipi\
is independent of the visibility of the \us.  This very reasonable
assumption purchases much of the statistical precision in the
measurement.  The ratio of fit yields, which is the probability that
an \us\ decay will be invisible, is (0.67 \PM\ 0.62)\%.

\label{pag:invisible} The invisible $\Upsilon \to \nu\bar{\nu}$ and $\Upsilon \to \mbox{\sc
wimp} \overline{\mbox{\sc wimp}}$ we discussed earlier are included in
$(1-\epsilon_\subs{vis})$, though the 0.05\% $\Upsilon \to
\nu\bar{\nu}$ branching fraction is much smaller than the statistical
uncertainty.  This measurement does, however, place a
model-independent upper bound on unknown, neutral decays of \us, which
may be interesting for constraining models that feature new neutral
particles such as {\sc wimp}s or large branching fractions for
all-neutral hadronic events.  This is a constraint on neutral decays
because our definition of ``visible'' requires at least one track with
$|\cos\theta| < 0.93$ and $p_\perp > 150$~MeV.  The \us\ branching
fraction to events with no such tracks is less than 1.01\% at 90\%
confidence level.

This ratio of fit yields represents the fraction of \us\ decays that
are invisible, including leptonic \us\ decays.  Leptonic decays,
particularly \ee\ and \mumu, are more likely to be invisible than
hadronic decays, since their final state consists of only two
particles which are geometrically back-to-back: if one lepton
disappears down the beam-pipe, the other probably will do so on
the other side.  We determine the probability for leptons to be
invisible from leptonic Monte Carlo simulations (which include the
$(1+\cos\theta)$ angular distribution for leptonic decays through a
virtual photon), and this probability is (10.91 \PM\ 0.01)\%.  We use
Equation \ref{eqn:agregate} to determine the hadronic visibility
efficiency, \evis, from the total visibility efficiency
($\epsilon_\subs{vis}^\subs{tot} = 1 - 0.0067$) and the leptonic
visibility efficiency ($\epsilon_\subs{vis}^\subs{lep} = 1 - 0.1091$).
\begin{equation}
  \epsilon_\subs{vis} = \frac{\epsilon_\subs{vis}^\subs{tot} - \epsilon_\subs{vis}^\subs{lep} \times (3{\mathcal B}_{\mu\mu})}{(1 - 3{\mathcal B}_{\mu\mu})}
  = (100.02 \pm 0.62)\% \mbox{.}
  \label{eqn:leptoniccorrection}
\end{equation}
Slightly more than half of this probability distribution is above
100\%, which is impossible for a real efficiency, so we truncate the
part above 100\% and normalize the remaining distribution to obtain an
asymmetric uncertainty: \evis\ = (99.59~$^{+0.29}_{-0.45}$)\%.

\subsection{Determination of \boldmath \ecuts}

\begin{figure}[p]
  \begin{center}
    \includegraphics[width=\linewidth]{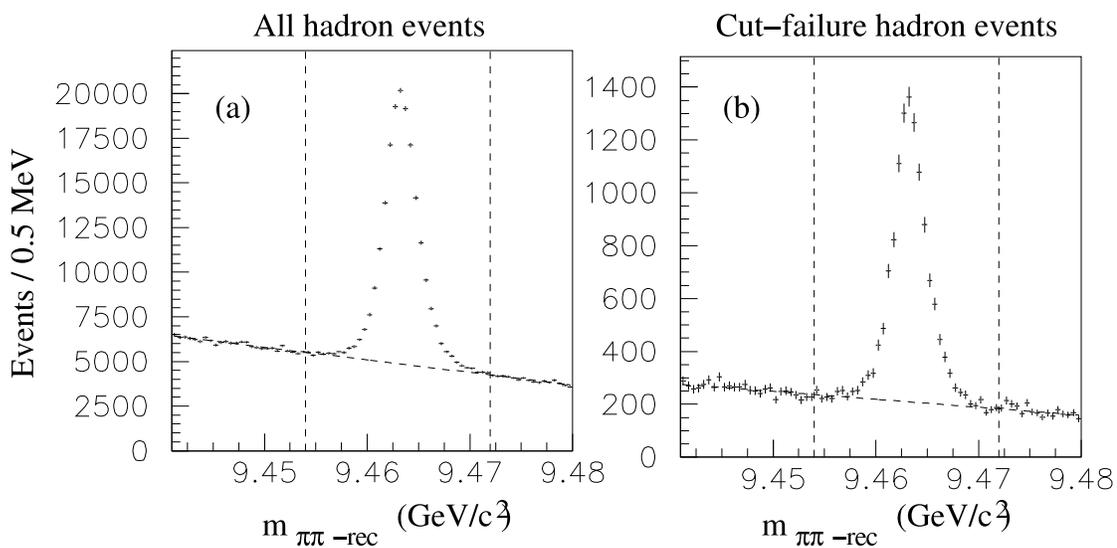}
  \end{center}
  \caption[\pipi\ recoil mass for events from the \hadron\
  trigger]{\label{pipihadron} Recoil mass of \pipi\ candidates for (a)
  all events accepted by the \hadron\ trigger, and (b) only those in
  which the rest of the event fails our cuts (``cut-failure'' events).
  The dotted vertical lines identify the signal region, and the dashed
  curve is a fit to the background, used to subtract background events
  from the signal region.}
\end{figure}

We select \pipi\ candidates in the same way to determine \ecuts,
except that we choose events accepted by the \hadron\ trigger rather
than the \twotrack\ trigger.  Figure~\ref{pipihadron}(a) presents the
recoil mass of these \pipi\ candidates, recoiling against visible \us\
decays.  We apply our hadronic cuts (but not the trigger requirements)
to this sample, excluding tracks and showers associated with the
\pipi, and plot the recoil mass for these in Figure~\ref{pipihadron}(b).  These ``cut-failure'' events have a prominent
peak at the \us\ mass due to all the visible \us\ events which failed
our cuts.  (Most of them are leptonic decays.)

To obtain the ratio of cut-failure \us\ events to all visible \us\
events, we avoid the fit procedure because of the potential for
systematic uncertainties in the fit parameterization.  In the \evis\
study, the statistical uncertainty in the number of invisible events
was almost as large as the number of invisible events itself, and this
overwhelmed any bias introduced by the fit function shape.  Here, the
number of cut-failures is significantly greater than zero and ought to
be measured more precisely.  Instead of fitting, we count events with
a recoil mass between 9.454 and 9.472~GeV and subtract the 
backgrounds, which have been determined by a linear fit to the sidebands
(between 9.441 and 9.480 GeV, excluding the signal region).  To
estimate the systematic uncertainty in yield due to assuming a linear
background, we repeat the procedure with the largest quadratic term
allowed by the data.

The ratio of cut-failure \us\ events to all visible \us\ events is
(92.58 \PM\ 0.13 \PM\ 0.02)\%, where the first uncertainty is
statistical and the second is due to the parameterization of the
background distribution.  The leptonic modes account for most of this
7\% inefficiency: correcting for leptonic modes as above
(Equation~\ref{eqn:leptoniccorrection}) yields \ecuts\ = (98.32 \PM\
0.21)\%.

The apparent \us\ efficiency, measured in \twotoone\ cascades, may
differ from the efficiency of direct $e^+e^- \to \Upsilon(1S)$ events
because of the slight relativistic boost of the \us\ ($\gamma =
1.005$) or the potential for \pipi\ showers to overlap \us\ showers in
cascade decays.  To address these possibilities, we generate hadronic
Monte Carlo for the direct and the cascade cases, applying the same
procedure to determine the hadronic efficiency from the cascade
simulation.  We observe no significant difference: the ratio of direct
efficiency to cascade efficiency is 1.0014 \PM\ 0.0022.  We apply this
as a correction primarily to propagate the uncertainty in this study.

\begin{figure}[p]
  \begin{center}
    \includegraphics[width=0.9\linewidth]{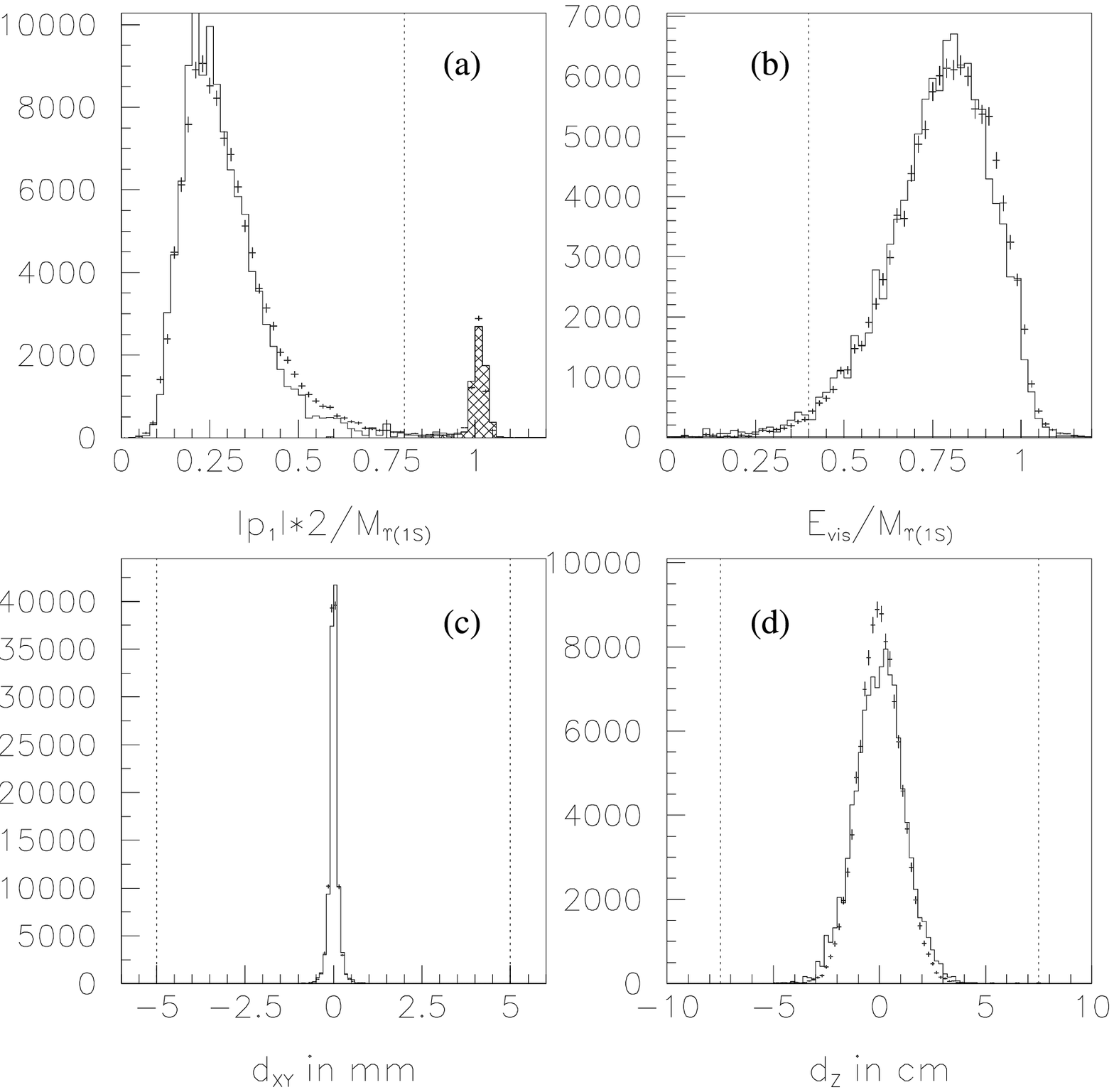}
  \end{center}
  \caption[Test of the Monte Carlo simulation in \us\ recoils from
  \twotoone\ decays]{\label{cascadeagreement} Our four cut variables,
  as seen in background-subtracted \twotoone\ events.  Points with
  errorbars are data, the solid histograms are \twotoone\ Monte Carlo
  simulations with the same procedure applied, and the dotted vertical
  lines are the cut thresholds.  (a) Largest track momentum (\pmax),
  divided by $M_{\Upsilon(1S)}/2$, the equivalent of \ebeam\ if this
  were a direct decay.  The peak at 1 is due to \ee\ and \mumu\
  decays, cross-hatched in the Monte Carlo.  (b) Visible energy
  (\visen), divided by $M_{\Upsilon(1S)}$, the equivalent of \ecm.
  (c) The distance of the closest track to the beam-line (\dxy).  (d)
  The $z$-vertex of the event (\dz).}
\end{figure}

Incidentally, the cascade Monte Carlo prediction of \ecuts, (98.54
\PM\ 0.22)\%, agrees with our data-based measurement of (98.32 \PM\
0.21)\%.  We would have been correct, if not justified, if we had
derived our cut efficiency directly from the Monte Carlo.  We also
extract \pmax, \visen, \dxy, and \dz\ from our cascade data and
cascade Monte Carlo, and find that they agree fairly well (Figure~\ref{cascadeagreement}).

\begin{figure}[p]
  \begin{center}
    \includegraphics[width=\linewidth]{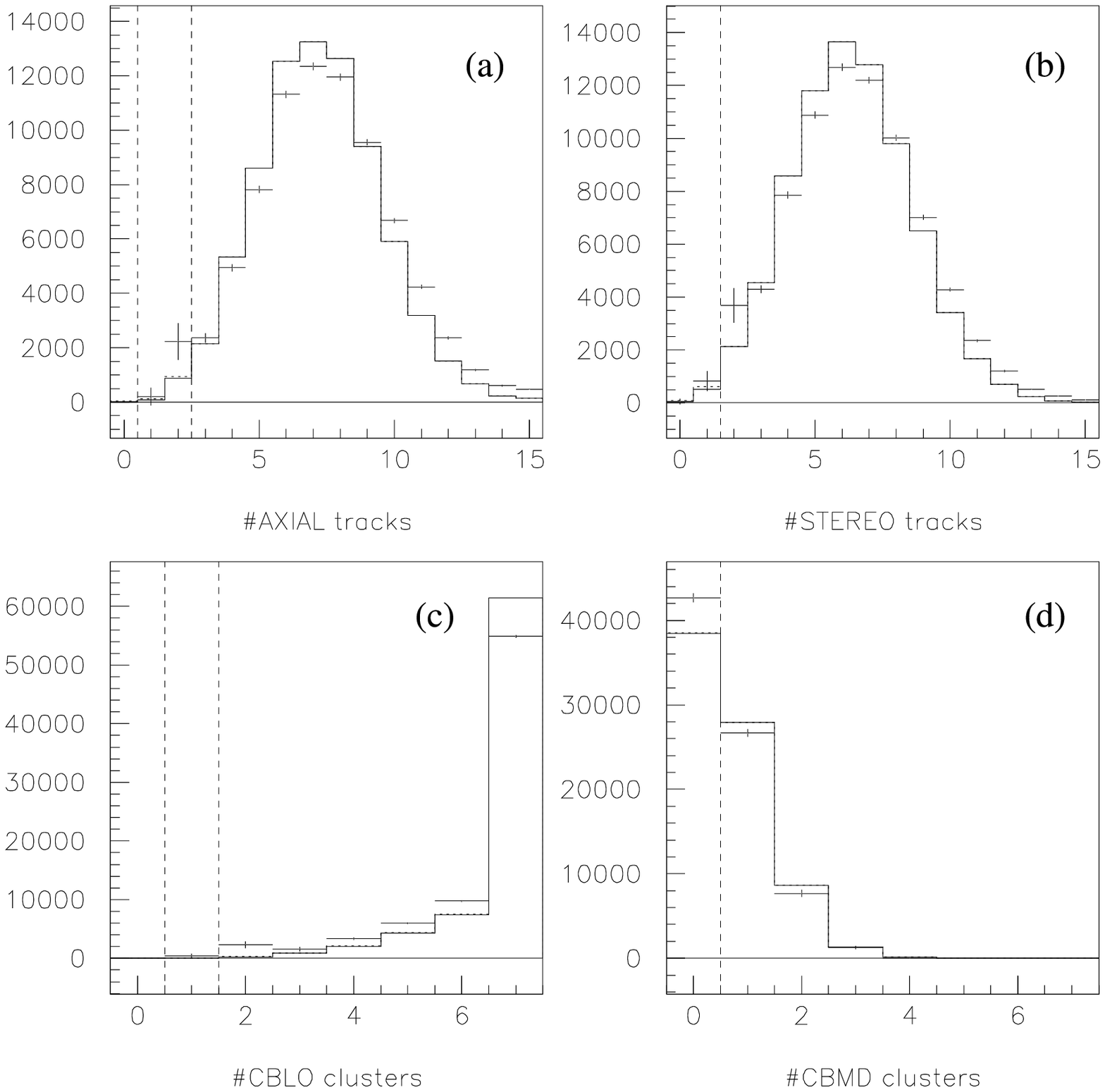}
  \end{center}
  \caption[Test of the Monte Carlo trigger simulation in $e^+e^- \to
  \Upsilon(1S)$]{\label{triggeragreement} The four cut variables used
  in \hadron, \radtau, and \eltrack\ trigger decisions (defined on
  page \pageref{pag:triggerdefs}).  Points with errobars are
  continuum-subtracted data, histograms are Monte Carlo simulations,
  and the dashed vertical lines are cut thresholds.  Trigger
  selections have been applied to data and Monte Carlo, though Monte
  Carlo without trigger selections are overlaid as dotted histograms
  (barely visible).}
\end{figure}

One correction is still missing from the \ecuts\ we have derived:
\ecuts\ must include the trigger efficiency, but we excluded trigger
requirements from our measurement.  While we can remove the \pipi\
tracks and showers from our fully reconstructed data, it would be
impractical to apply an analogous procedure on our trigger data for
technical reasons.  We therefore use Monte Carlo to determine the
efficiency of the trigger once our cuts have been applied, a value of
99.87\%.  Figure~\ref{triggeragreement} overlays data and Monte Carlo
\axial, \stereo, \cblo, and \cbmd\ distributions in direct \us\
decays, showing fairly good agreement.  The predicted inefficiency of
0.13\% can therefore be trusted within 100\% of itself, so we
conservatively assign this as the systematic uncertainty.

After all of these corrections, the efficiency of visible hadronic
\us\ decays is \ecuts\ = $(0.9832)(0.9987)(1.0014)$ = (98.33 \PM\
0.33)\%.  The efficiency of all hadronic \us\ decays is
\evis~$\times$~\ecuts\ = (97.93 $^{+0.44}_{-0.56}$)\%.

\section{Hadronic Efficiency of the \boldmath \uss\ and \usss}
\label{sec:ussusssefficiency}

The $\Upsilon(3S) \to \pi^+\pi^- \Upsilon(2S)$ and $\Upsilon(4S) \to
\pi^+\pi^- \Upsilon(3S)$ rates are too low to accumulate large samples
to study \uss\ and \usss\ efficiencies in the same way that we did
\us.  Instead, we derive correction factors from the Monte Carlo that
allow the \us\ efficiency to be applied to the \uss\ and \usss.

The decays of the \uss\ and \usss\ differ from those of the \us\ in
two ways.  The \uss\ and \usss\ decay products are slightly more
energetic, as they originate in a state of higher \ecm, and the \uss\
and \usss\ can decay via transitions to lower $b\bar{b}$ resonances.
The first correction is very small because our \pmax\ and \visen\ cut
thesholds are constant fractions of \ecm, and the difference in \ecm\
from \us\ to \usss\ is only 10\%.  In the Monte Carlo simulation, the
\uss\ and \usss\ efficiency for $ggg$, \gggamma, and \qqbar\ is only
0.2\% lower than the \us\ efficiency.

Most \uss\ and \usss\ transition decays have the same efficiency as
$ggg$, \gggamma, and \qqbar, so they have no impact on total hadronic
efficiency.  The exceptions are transitions that result in a lower
\ups\ resonance decaying into \ee\ or \mumu.  According to the Monte
Carlo simulation, these ``cascade-to-leptons'' decays have (0.69 \PM\
0.22)\% efficiency for \uss\ and (0.38 \PM\ 0.19)\% efficiency for the
\usss--- almost zero.  Therefore the efficiency correction will be
approximately $(1 - {\mathcal B}_\subs{cas})$, where \bcas\ is the
branching fraction for these modes.

\begin{figure}[p]
  \begin{center}
    \includegraphics[width=\linewidth]{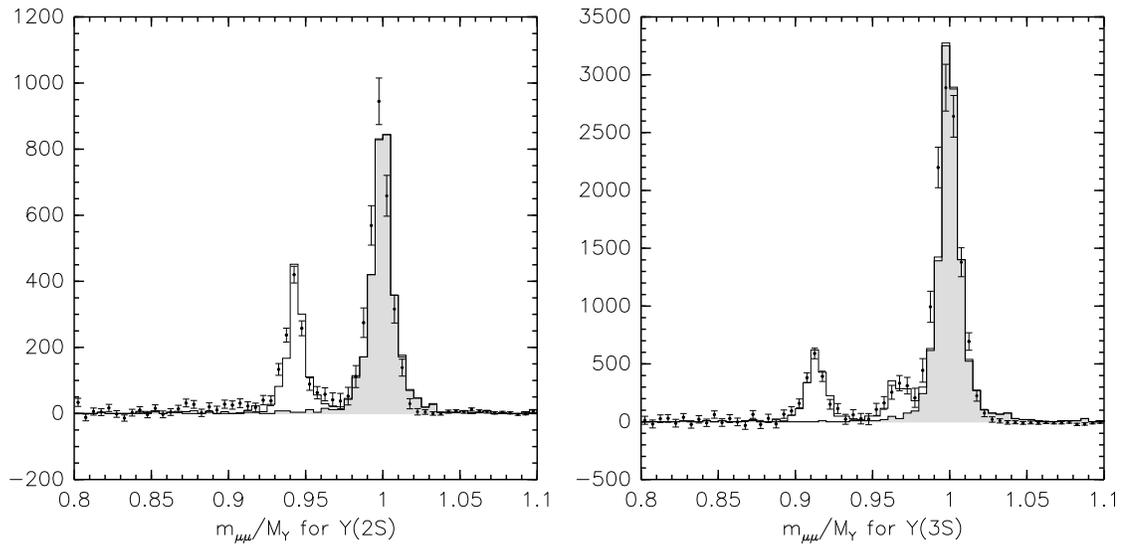}
  \end{center}
  \caption[Determining ${\mathcal B}_\subs{cas}$ from \mumu\ invariant
  mass in \uss\ and \usss]{\label{invariantmumass} Invariant mass
  spectra of \mumu\ in events with two energetic muons.  Points with
  errorbars are continuum-subtracted data, and histograms are Monte
  Carlo simulations, with prompt $\Upsilon \to \mu^+\mu^-$ shaded.
  The horizontal shift in energy scale is consistent with the
  uncertainty in our magnetic field estimate.}
\end{figure}

We determine this branching fraction from the data by counting
cascade-to-leptons events relative to $\Upsilon \to \mu\mu$ in the
full \uss\ and \usss\ datasets.  We select all events that have two
high-momentum tracks ($|\vec{p}| >$ 70\% \ebeam) without associated
high-energy showers ($E_\subs{max} <$ 70\% \ebeam), that is,
consistent with two high-energy muons accompanied by anything.  We
plot the invariant mass of these two muons in Figure~\ref{invariantmumass}, after subtracting continuum processes using the
off-resonance data.  Muon pairs from direct \ups\ decays are easily
distinguished from cascade-to-leptons events.  The Monte Carlo
reproduces the invariant mass spectrum with only tiny errors in the
calibration of the magnetic field (a horizontal shift in the plot).

We measure \bcas\ relative to \bmm\ by fitting Monte Carlo
cascade-to-muon pairs and Monte Carlo direct muon pairs to the data in
Figure~\ref{invariantmumass}.  This fit has two free parameters, the
magnitude of \bcas\ and the magnitude of \bmm.  We assign conservative
10\% uncertainties to this procedure, which overwhelm \bmm\
uncertainties and yield only 0.1\% uncertainties in the final
efficiency determination.  The resulting \bcas\ for \uss\ is (1.58
\PM\ 0.16)\% and for \usss\ is (1.34 \PM\ 0.13)\%, accounting for the
factor of two from cascade-to-electron pairs.

Applying these corrections to the \uss\ and \usss\ efficiencies, we
obtain (96.18 $^{+0.44}_{-0.56}$ \PM\ 0.15)\% and (96.41
$^{+0.44}_{-0.56}$ \PM\ 0.13)\% for the \uss\ and \usss, respectively.
The first uncertainty is derived from the \twotoone\ study and is
common to all three resonances.  The second uncertainty arises from
our \bcas\ measurements and is independent for the \uss\ and \usss.
The first uncertainty therefore cancels in ratios of \gee, while the
second does not.

\chapter{Integrated Luminosity}
\label{chp:luminosity}

Now that we can reliably count the number of hadronic \ups\ decays at
every \ecm\ we sampled, we need to determine the hadronic
cross-section this implies by measuring the integrated luminosity of
the same data.  We get the integrated luminosity from a count of
Bhabha events, since these are plentiful and their rate can be
accurately calculated from perturbative QED.  Integrated luminosity is
the ratio of Bhabha events counted to the efficiency-weighted Bhabha
cross-section, and the hadronic \ups\ cross-section is the ratio of
hadronic \ups\ events to the integrated luminosity.

Any theoretically calculable process can be used to determine the
integrated luminosity; Bhabhas were chosen primarily for their
abundance, since this minimizes statistical uncertainty.  A Bhabha
count is complicated by the fact that $\Upsilon \to e^+e^-$ is
indistinguishable from Bhabhas on an event-by-event basis, and this
background peaks under the resonance.  Alternatively, one could
determine the integrated luminosity from $e^+e^- \to \gamma\gamma$
because $\Upsilon \not\to \gamma\gamma$ (Electromagnetic decays do not
violate parity).  Unfortunately, this comes at the cost of a factor of
8.6 in statistics.  The $\Upsilon \to e^+e^-$ contribution
can easily be controlled, and the additional statistical power is
valuable when determining ratios of \gee$(nS)$/\gee$(mS)$, so we
determine integrated luminosities from Bhabhas and use \gamgam\ events
as a cross-check.

\section{Bhabha Count and \boldmath \gamgam\ count}

\begin{figure}[p]
  \begin{center}
    \includegraphics[width=0.4\linewidth]{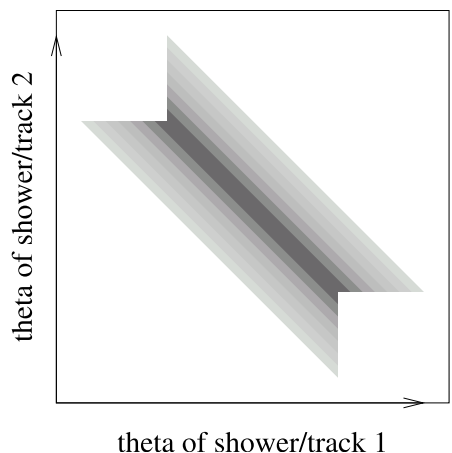}
  \end{center}
  \caption[Distribution provided by an asymmetric
  cut]{\label{asymmetriccartoon} The distribution of two
  anti-correlated variables with an asymmetric cut (the masked square
  regions).  Horizontal and vertical smearing around the central
  diagonal is independent; we cut each track or shower in a way that
  doesn't imply a cut on the other.}
\end{figure}

To select Bhabha events, we require
\begin{itemize}

  \item two or more tracks with momenta between 50\% and 110\% of
    \ebeam, and an energy sum (including showers from bremsstrahlung
    radiation as the electrons propagate through the detector) of more
    than 90\% \ecm.

  \item The larger (smaller) track $|\cos\theta|$ must be less than
    0.766 (0.707), and

  \item each track must be associated with a calorimeter shower, with the
    larger (smaller) shower energy divided by track momentum
    ($E_\subs{shower}/|\vec{p}_\subs{track}|$) being greater than 80\%
    (50\%).

\end{itemize}
With these cuts, backgrounds other than $\Upsilon \to e^+e^-$ are
negligible.  Different thresholds are set for the larger and smaller
angles and calorimeter energies to reduce sensitivity to the threshold
values, and possible efficiency variation with time.  See Figure~\ref{asymmetriccartoon} for an illustration of this kind of cut.

\begin{figure}[p]
  \begin{center}
    \includegraphics[width=\linewidth]{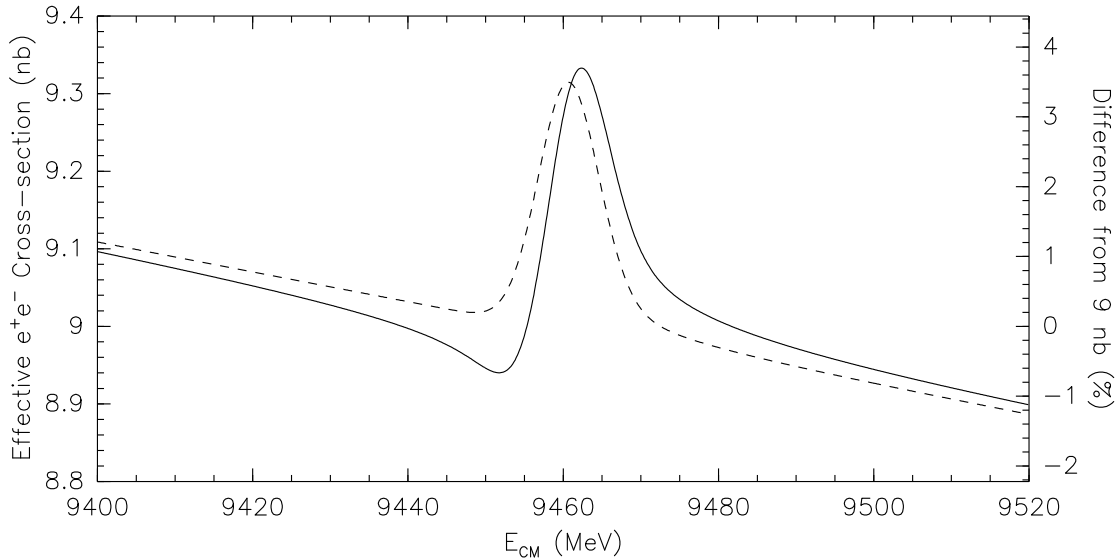}
  \end{center}
  \caption[Influence of $\Upsilon \to e^+e^-$ on effective \ee\
  cross-section]{\label{allee} The influence of $\Upsilon \to e^+e^-$
  on effective \ee\ cross-section, with interference (solid) and
  without interference (dashed).  Note that the vertical axis is
  zero-suppressed.}
\end{figure}

Contamination from $\Upsilon \to e^+e^-$ at the resonance peaks is
3.8\%, 1.4\%, and 1.0\%, respectively.  This background is readily
calculated for any \ecm\ by multiplying the \ups\ lineshape by \bee\
(we assume \bee\ = \bmm\ for greater precision) and the cut efficiency
for $\Upsilon \to e^+e^-$ (which has a different angular distribution
than Bhabhas).  Since the \ups\ lineshapes are derived from
cross-section measurements, this is a circular dependence, so we
applied an iterative procedure, starting with \gamgam\ luminosity.
The \ups\ and continuum \ee\ interfere, so we also calculate an
interference term (Equation \ref{eqn:yint}) with $\alpha_\subs{int}$ =
0.60, 0.87, and 0.69 for \us, \uss, and \usss, respectively.  The
effective cross-section for \ee\ as a function of \ecm, including the
continuum and \ups\ contributions, is presented in Figure~\ref{allee},
with and without the interference term.  The presence of the
interference term has negligible impact on the lineshape fit results.

To select \gamgam\ events for our cross-checks, we require
\begin{itemize}

  \item two showers (subscripted 1 and 2) with energies higher than
    70\% of \ebeam,

  \item $|\cot\theta_1 + \cot\theta_2| < 0.1$ (showers back-to-back in
    $\theta$, the polar angle), and

  \item $|\sin(\phi_1 - \phi_2)| < 0.04$ (showers back-to-back in
    $\phi$, the azimuthal angle).

  \item The larger (smaller) $|\cot\theta|$ must be less than 1.28 (1.18)
    (this is within the calorimeter barrel),

  \item the larger (smaller) $|\cot\theta|$ must be greater than 0.15
  (0.05) (avoiding the central region, where trigger efficiency is
  low), and

  \item there must be no tracks in the event.

\end{itemize}
When we select \gamgam\ events, we are dependent on only one trigger,
\barrelbhabha, since this is the only trigger that doesn't
require any tracks.  We studied the efficiency of this trigger with
Bhabhas, and found calorimeter tiles whose efficiencies dropped for a
significant fraction of the data-taking period.  Rather than applying
a run-dependent efficiency, we masked out these regions with our cuts.
The largest shower on the western side of the detector must not be
found in any of these regions:
\begin{itemize}

  \item $-\frac{14}{64}\pi < \phi < \frac{9}{64}\pi$ and
    $|\cot\theta_1 + \cot\theta_2| < 1.08$,

  \item $-\frac{53}{64}\pi < \phi < -\frac{14}{64}\pi$ and
    $|\cot\theta_1 + \cot\theta_2| > 1.90$, and

  \item $-0.4 < \phi < -0.3$.

\end{itemize}
Given these angular cuts, the \barrelbhabha\ trigger is 99.67\%
efficient with only statistical deviations.  (The efficiency of every
run is above 99.2\%.)

\section{Overall Luminosity Scale}

The efficiency-weighted Bhabha cross-section is the second ingredient
in the luminosity measurement.  This sets the luminosity scale for all
Bhabha counts.  The scale factor, a number of nb\inv\ per observed
Bhabha, is the inverse of the efficiency-weighted Bhabha cross-section
(nb).

The efficiency-weighted cross-section is the cross-section of observed
Bhabhas.  Expressed as an integral over a single variable for clarity,
\begin{equation}
  \sigma_\subs{eff} = \int_0^\pi \frac{d\sigma}{d\theta} \,
  \epsilon(\theta) \, d\theta
  \label{eqn:effectivecrosssection}
\end{equation}
where $\epsilon(\theta)$ is our detector's Bhabha cut efficiency at a
given polar angle $\theta$.  Rather than exhaustively simulating the
detector's response to Bhabhas in $\theta$ bins, we generate Bhabhas
with an angular cut-off beyond the detector's geometric acceptance
(where $\epsilon(\theta) = 0$), calculate the cross-section
this represents ($\sigma_0$), and multiply it by the efficiency of
these simulated events ($\epsilon_0$), determined by passing them
through the detector simulation.  Both $\sigma_0$ and $\epsilon_0$
depend on our choice of cut-off, but the product doesn't.  This
product is the desired efficiency-weighted cross-section because
\begin{equation}
  \sigma_0 = \int_{\theta_\subs{min}}^{\theta_\subs{max}}
  \frac{d\sigma}{d\theta'} \, d\theta'
  \mbox{\hspace{0.5 cm} and \hspace{0.5 cm}}
  \epsilon_0 = \int_{\theta_\subs{min}}^{\theta_\subs{max}}
  \frac{dP}{d\theta} \, \epsilon(\theta) \, d\theta \mbox{.}
  \label{eqn:effectivecrosssectiona}
\end{equation}
where $dP/d\theta$ is the probability distribution of Bhabhas with
polar angle $\theta$ in our simulation, which is normalized:
\begin{equation}
  \frac{dP}{d\theta} = \left(\frac{d\sigma}{d\theta}\right) \left(
  \int_{\theta_\subs{min}}^{\theta_\subs{max}}
  \frac{d\sigma}{d\theta'} \, d\theta' \right)^{-1}_{\mbox{\normalsize .}}
  \label{eqn:effectivecrosssectionb}
\end{equation}
Equation~\ref{eqn:effectivecrosssection} may be derived from
Equations~\ref{eqn:effectivecrosssectiona} and \ref{eqn:effectivecrosssectionb}.

We simulate Bhabhas with the Babayaga event generator, which
calculates $dP/d\theta$ and $\sigma_0$ to fourth order in the fine
structure constant \cite{babayaga}, with an angular cut-off of
$|\cos\theta_\subs{max}|$ = 0.819.  Passing these simulated events
through our full detector simulation, we calculate an
efficiency-weighted cross-section of 8.993 \PM\ 0.035 nb\inv\ at an
\ecm\ of 9.43~GeV, 7.945 \PM\ 0.031 nb\inv\ at 10.00~GeV, and 7.361
\PM\ 0.032~nb\inv\ at 10.33~GeV, which are the off-resonance \us,
\uss, and \usss\ energies, respectively.

\begin{figure}[p]
  \begin{center}
    \includegraphics[width=0.7\linewidth]{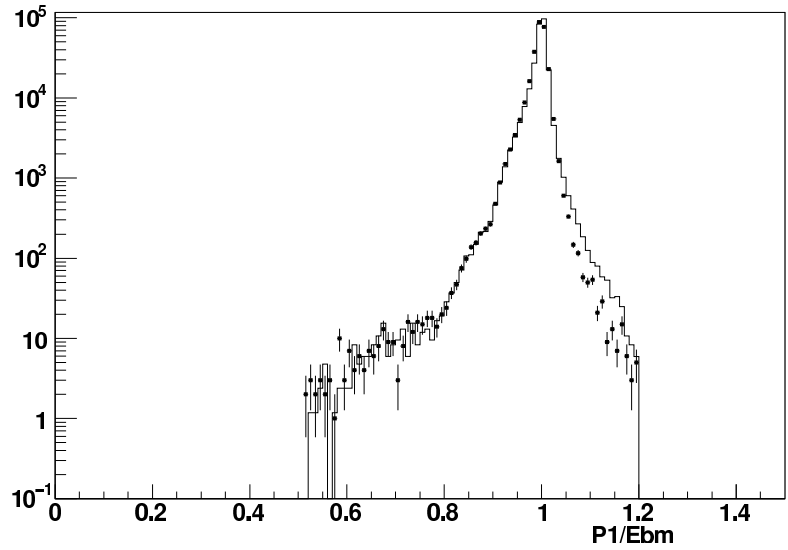}
  \end{center}
  \caption[Largest track momentum distribution in Bhabha events]{\label{eeagreementa} Largest track momentum divided by
  \ebeam\ in the 10.33~GeV data (points) and Monte Carlo (histogram)
  with other cuts applied.  Data between 0.5 and 1.1 are accepted.}
\end{figure}

\begin{figure}[p]
  \begin{center}
    \includegraphics[width=0.7\linewidth]{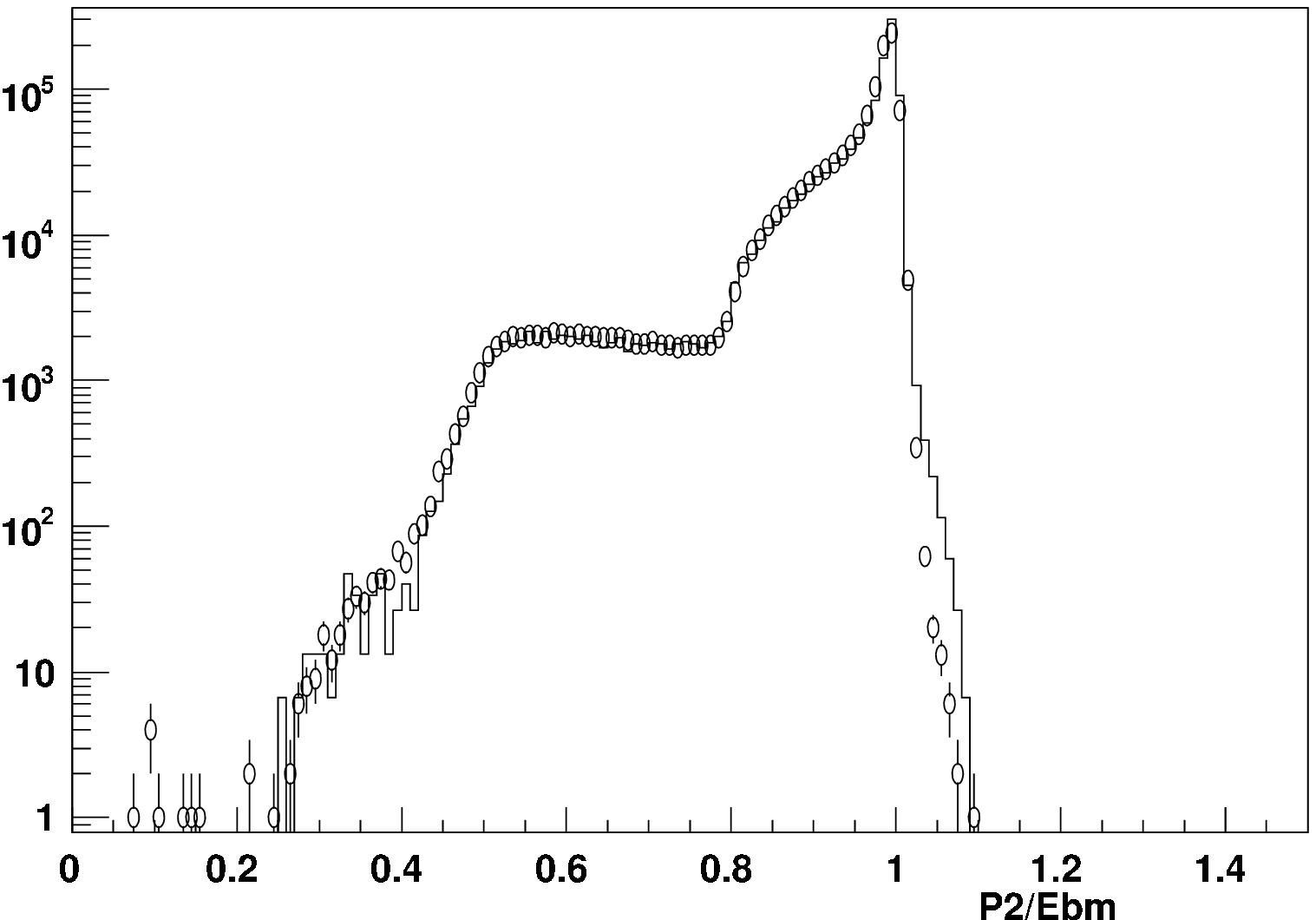}
  \end{center}
  \caption[Second-largest track momentum distribution in Bhabha events]{\label{eeagreementb} Second-largest track momentum divided
  by \ebeam\ in the 10.33~GeV data (points) and Monte Carlo
  (histogram) with other cuts applied.  Data between 0.5 and 1.1 are
  accepted.}
\end{figure}

\begin{figure}[p]
  \begin{center}
    \includegraphics[width=0.7\linewidth]{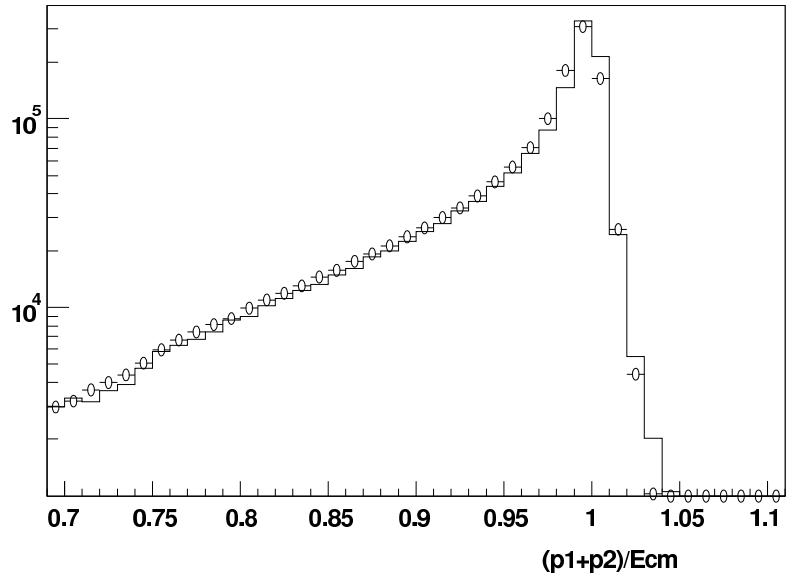}
  \end{center}
  \caption[Sum of the two largest-momentum tracks in Bhabha events]{\label{eeagreementc} Sum of two largest-momentum track
  energies and associated bremsstrahlung showers divided by \ecm\ in
  the 10.33~GeV data (points) and Monte Carlo (histogram) with other
  cuts applied.  Data above 0.9 are accepted.}
\end{figure}

\begin{figure}[p]
  \begin{center}
    \includegraphics[width=0.7\linewidth]{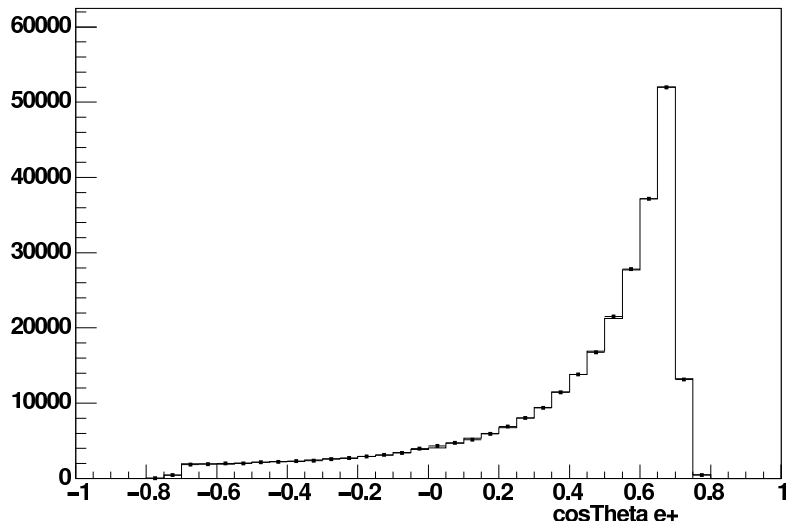}
  \end{center}
  \caption[Positron $\cos\theta$ distribution in Bhabha events]{\label{eeagreementd} Positron $\cos\theta$ (polar angle)
  distribution in the 10.33~GeV data (points) and Monte Carlo
  (histogram) with other cuts applied.  The larger (smaller)
  $|\cos\theta|$ of the two electrons must be below 0.766 (0.707) for
  acceptance.}
\end{figure}

\begin{figure}[p]
  \begin{center}
    \includegraphics[width=0.7\linewidth]{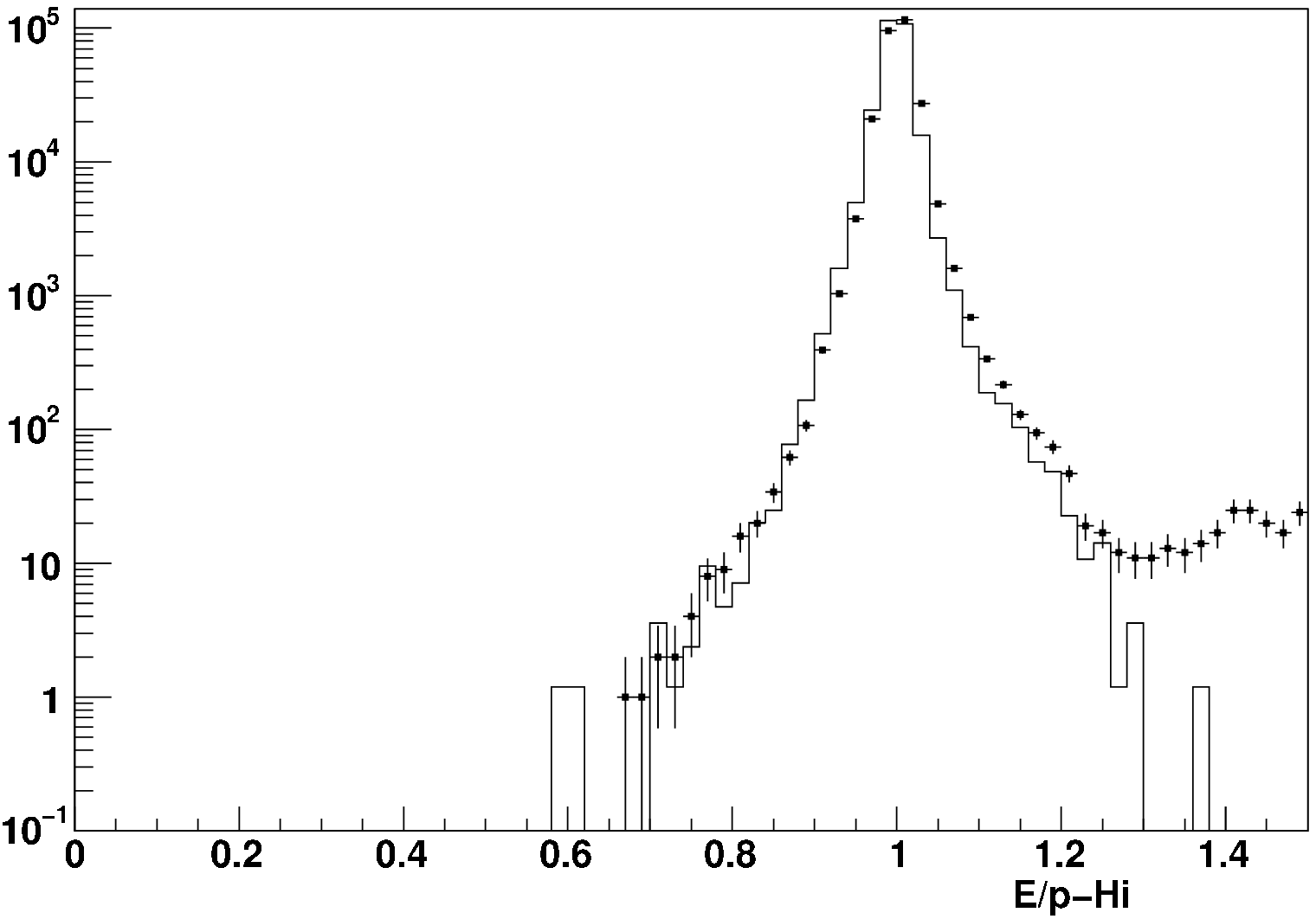}
  \end{center}
  \caption[Largest $E_\subs{shower}/|\vec{p}_\subs{track}|$ in Bhabha events]{\label{eeagreemente} Largest
  $E_\subs{shower}/|\vec{p}_\subs{track}|$ divided by \ebeam\ in the
  10.33~GeV data (points) and Monte Carlo (histogram) with other cuts
  applied.  Data above 0.8 are accepted.}
\end{figure}

\begin{figure}[p]
  \begin{center}
    \includegraphics[width=0.7\linewidth]{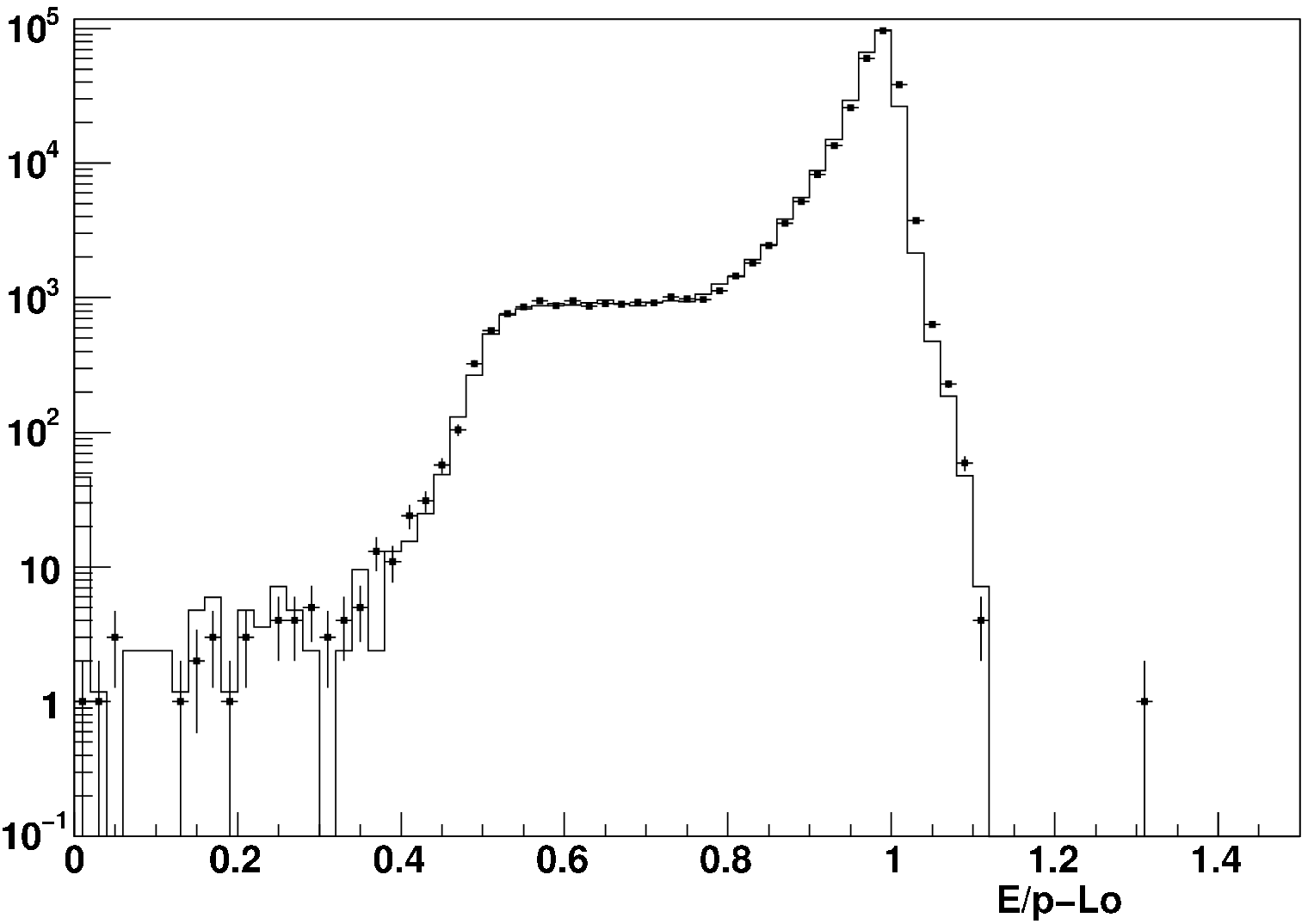}
  \end{center}
  \caption[Second-largest $E_\subs{shower}/|\vec{p}_\subs{track}|$ in Bhabha events]{\label{eeagreementf} Second-largest
  $E_\subs{shower}/|\vec{p}_\subs{track}|$ divided by \ebeam\ in the
  10.33~GeV data (points) and Monte Carlo (histogram) with other cuts
  applied.  Data above 0.5 are accepted.}
\end{figure}

\begin{figure}[p]
  \begin{center}
    \includegraphics[width=0.7\linewidth]{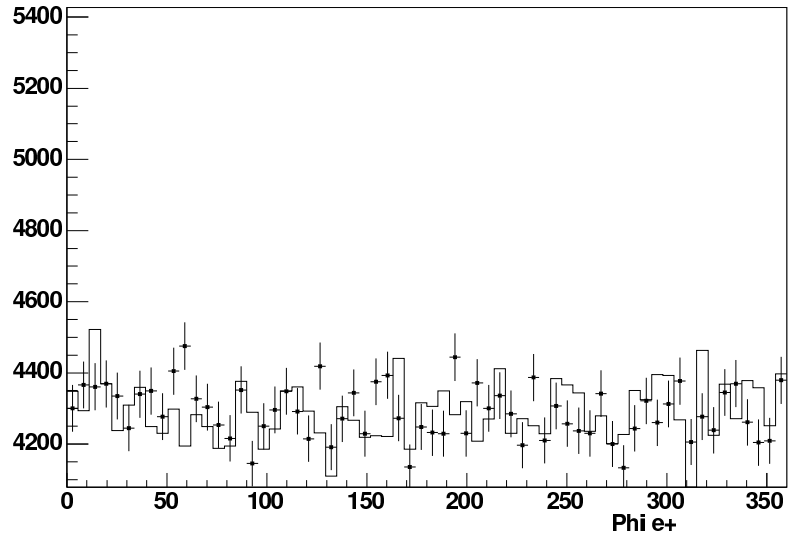}
  \end{center}
  \caption[Positron $\phi$ distribution in Bhabha events]{\label{eeagreementg} Positron $\phi$ (azimuthal angle)
  distribution in the 10.33~GeV data (points) and Monte Carlo
  (histogram) with other cuts applied.  This variable is not used for
  cuts.}
\end{figure}

\begin{figure}[p]
  \begin{center}
    \includegraphics[width=0.7\linewidth]{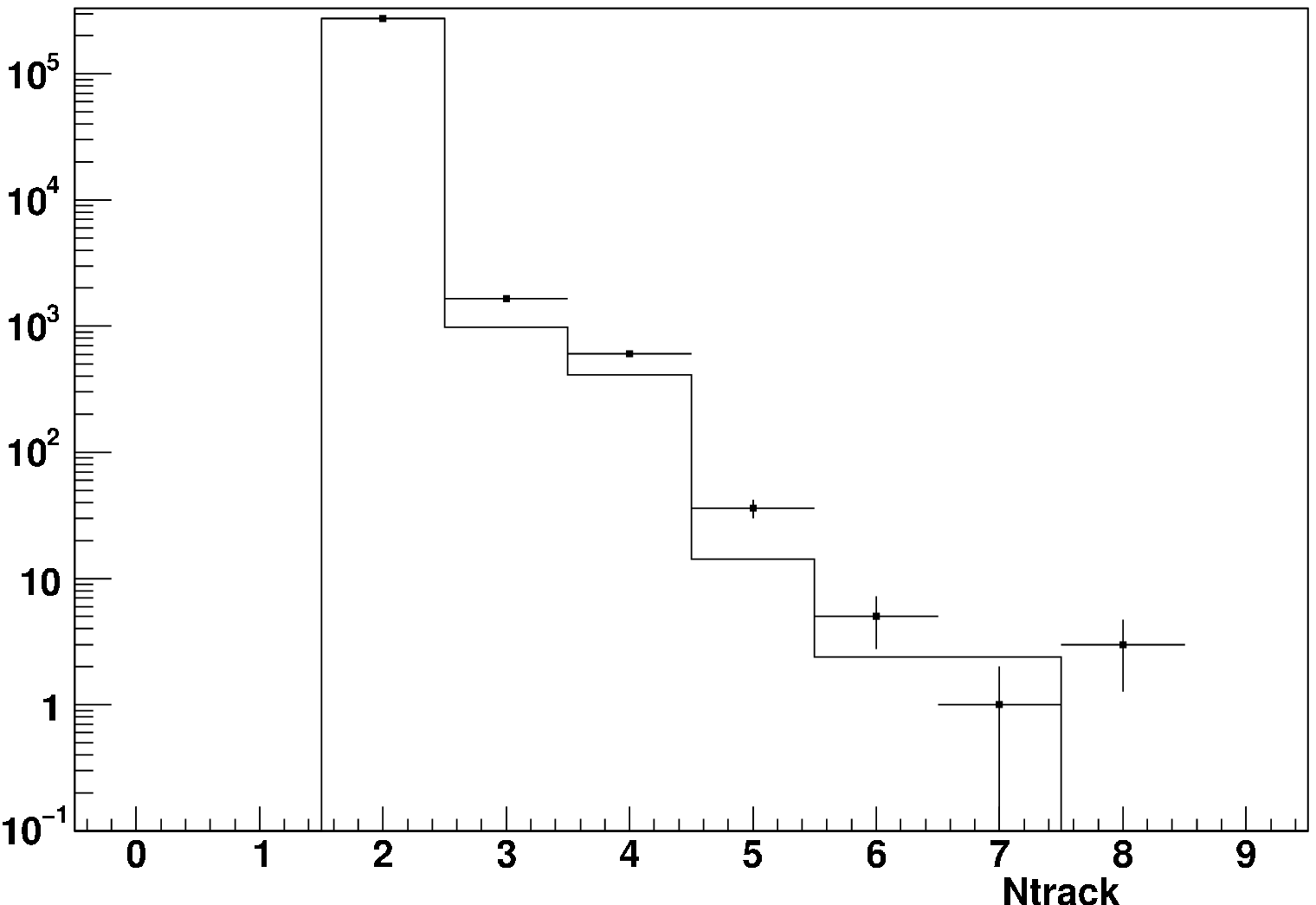}
  \end{center}
  \caption[Number of charged tracks in Bhabha events]{\label{eeagreementh} Number of charged tracks in the
  10.33~GeV data (points) and Monte Carlo (histogram) with other cuts
  applied.  This variable is not used for cuts.}
\end{figure}

The Monte Carlo reproduces data distributions well at all three
energies, which is demonstrated for the 10.33~GeV data in eight
relevant distributions in
Figures~\mbox{\ref{eeagreementa}--\ref{eeagreementh}}.

The efficiency-weighted Bhabha cross-sections at 9.43, 10.00, and
10.33~GeV differ by more than $1/s$ because the Monte Carlo finds
these Bhabha cuts, particularly the requirement that the energy sum of
the two tracks be greater than 90\% of \ecm, to be energy-dependent.
We checked this claim by comparing Bhabha counts with \gamgam\ counts,
and by loosening the energy sum cut.  Both methods reveal the same 2\%
per GeV energy dependence from \us\ to \usss.

We additionally determine the efficiency-weighted cross-sections of
$e^+e^- \to \mu^+\mu^-$ and $e^+e^- \to \gamma\gamma$ to reduce
systematic uncertainties by comparing the luminosity predicted for the
same dataset by different processes.  We follow the method of
\cite{oldlumi} to assign 1.6\% systematic uncertainties for Bhabha
efficiency-weighted cross-section, 1.6\% for \mumu, and 1.8\% for
\gamgam.  The sources of these uncertainties are tabulated in
Table~\ref{tab:lumisyst}: \ee\ and \mumu\ uncertainties are dominated
by the degree of resonance interference and the track-finding
efficiency, while \gamgam\ uncertainties are dominated by the
photon-finding efficiency and angular resolution.

\begin{table}
  \caption[Fractional systematic uncertainties in efficiency-weighted
  cross-sections of \ee, \mumu, and \gamgam]{\label{tab:lumisyst}
  Fractional systematic uncertainties in our determinations of the
  efficiency-weighted cross-sections of \ee, \mumu, and \gamgam.  All
  values are percentages.}
  \begin{center}
    \begin{tabular}{p{0.5\linewidth} | c c c}
      \hline\hline
      & \ee & \mumu & \gamgam \\\hline
      Finite Monte Carlo sample & 0.4 & 0.5 & 0.6 \\
      Radiative corrections & 0.5 & 0.5 & 0.5 \\
      Resonance interference & 1.0 & 1.0 & \\
      Trigger efficiency & 0.1 & 0.1 & 0.7 \\
      Track-finding efficiency & 1.0 & 1.0 & \\
      Photon-finding efficiency & & & 1.0 \\
      Dependence on cuts & 0.5 & 0.3 & 1.0 \\
      Cosmic ray backgrounds & & 0.2 & \\
      ISR tail backgrounds & & 0.1 & \\\hline
      Total & 1.6 & 1.6 & 1.8 \\\hline\hline
    \end{tabular}
  \end{center}
\end{table}

\begin{figure}[p]
  \begin{center}
    \includegraphics[width=\linewidth]{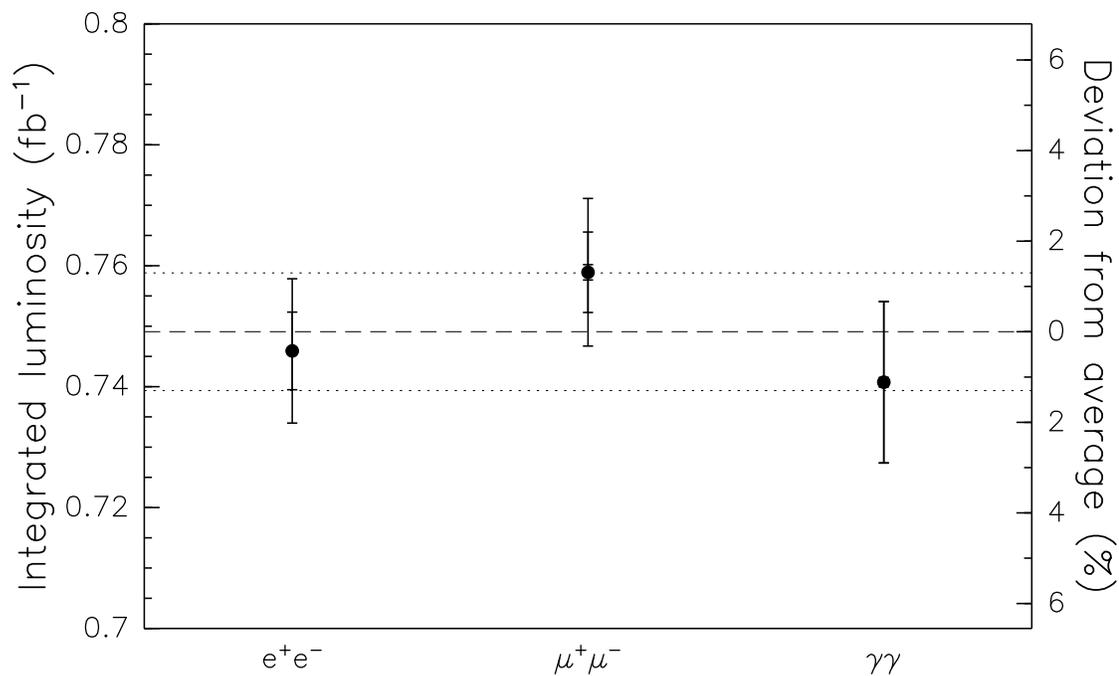}
  \end{center}
  \caption[Integrated luminosity in \ee, \mumu, and \gamgam\
  measurements]{\label{comparelumis} The integrated luminosity of all
  off-resonance data combined, as determined from \ee, \mumu, and
  \gamgam\ counts.  Outermost errorbars include all systematic
  uncertainties, and the second errorbars on \ee\ and \mumu\ have
  common tracking and resonance interference systematics removed.
  Innermost errorbars (visible only for \mumu) are statistical-only.
  The weighted average and RMS of the three measurements are
  represented by dashed and dotted horizontal lines.}
\end{figure}

We can compare these as luminosity measurements by dividing the \ee,
\mumu, and \gamgam\ counts from the same sample--- all available
off-resonance data--- by their efficiency-weighted cross-sections.  We obtain
the integrated luminosities plotted in Figure~\ref{comparelumis},
which are in good agreement, considering their systematic
uncertainties.  The \ee\ and \mumu\ modes share the same track-finding
efficiency, so we draw error bars with this systematic removed in the
Figure.

The weighted average luminosity from \ee, \mumu, and \gamgam\ is only
0.2--0.6\% higher than the \ee\ luminosity alone (depending on
dataset), so we modify our luminosity determination to return the
average luminosity from the three processes.  The integrated
luminosity is
\begin{eqnarray}
  \mbox{\hspace{-1 cm}} && 0.1114 \mbox{ } (E_\subs{CM}/\mbox{9.43~GeV})^2 \mbox{ nb\inv\ per observed Bhabha event,} \\
  \mbox{\hspace{-1 cm}} && 0.1266 \mbox{ } (E_\subs{CM}/\mbox{10.00~GeV})^2 \mbox{ nb\inv\ per observed Bhabha event, and} \\
  \mbox{\hspace{-1 cm}} && 0.1361 \mbox{ } (E_\subs{CM}/\mbox{10.33~GeV})^2 \mbox{ nb\inv\ per observed Bhabha event}
\end{eqnarray}
in the three datasets.

Without knowing that \mumu\ and \gamgam\ measurements reproduce the
Bhabha result, we would have a 1.6\% uncertainty common to all three
luminosity scale factors.  However, we can incorporate this
information by assigning the \ee, \mumu, and \gamgam\ RMS differences
from the average as the common uncertainty.  Thus, the luminosity
scale factors have an uncertainty of 1.3\%.  \label{sec:luminosity}
The luminosity scale factor is a factor in \geehadtot, so the
fractional uncertainty in this scale factor adds to the \geehadtot\
fractional uncertainty in quadrature.

The systematic uncertainties in the three luminosity scale factors are
not independently 1.3\%: they will partly cancel in ratios.  We can
see this by considering the minimum information necessary to determine
$\Gamma_{ee}(ns)/\Gamma_{ee}(mS)$.  One must know the ratio of Bhabha
cut efficiencies near the $\Upsilon(nS)$ and the $\Upsilon(mS)$ and
the scaling of Bhabha cut efficiency with \ecm, but the conversion
from the number of Bhabha events to inverse nanobarns will cancel.
Thus, the dominant uncertainty in ratios of luminosity scale factors
is the ratio of Bhabha cut efficiencies, determined with 0.5\%
uncertainty by Monte Carlo.

To blind our analysis, we determined the overall luminoisty scale
factors last, after all background (Chapter \ref{chp:backgrounds}),
efficiency (Chapter \ref{chp:efficiency}), and beam energy studies
(Chapter \ref{chp:beamenergy}) were completed.  Until that point, we
only knew the ratios of luminosities of our data samples, and
therefore only $\Gamma_{ee}(nS)/\Gamma_{ee}(mS)$.  We incorporated
these scale factors (using \gamgam\ event counts, rather than Bhabha
event counts) just before presenting preliminary \gee\ results at the
European Physical Society meeting in July, 2005.

\section{Consistency of Bhabhas with \boldmath \gamgam}

Above, we took advantage of the \gamgam\ luminosity measurement's very
different systematic uncertainties to check and correct the Bhabha
scale factors off-resonance, but we have not yet used the fact that
\gamgam\ counts are unaffected by \ups\ decays.  In this Section, we
will compare Bhabha and \gamgam\ rates as a function of \ecm\ through
the \ups\ resonances, to test the $\Upsilon \to e^+e^-$ correction.
We tune the \gamgam\ luminosity measurement to yield the same results
as Bhabhas off-resonance and measure the \gamgam\ and Bhabha
luminosities in \ups\ data.  Both methods should yield the same
integrated luminosity, within statistical uncertainties.

\begin{figure}[p]
  \begin{center}
    \includegraphics[width=\linewidth]{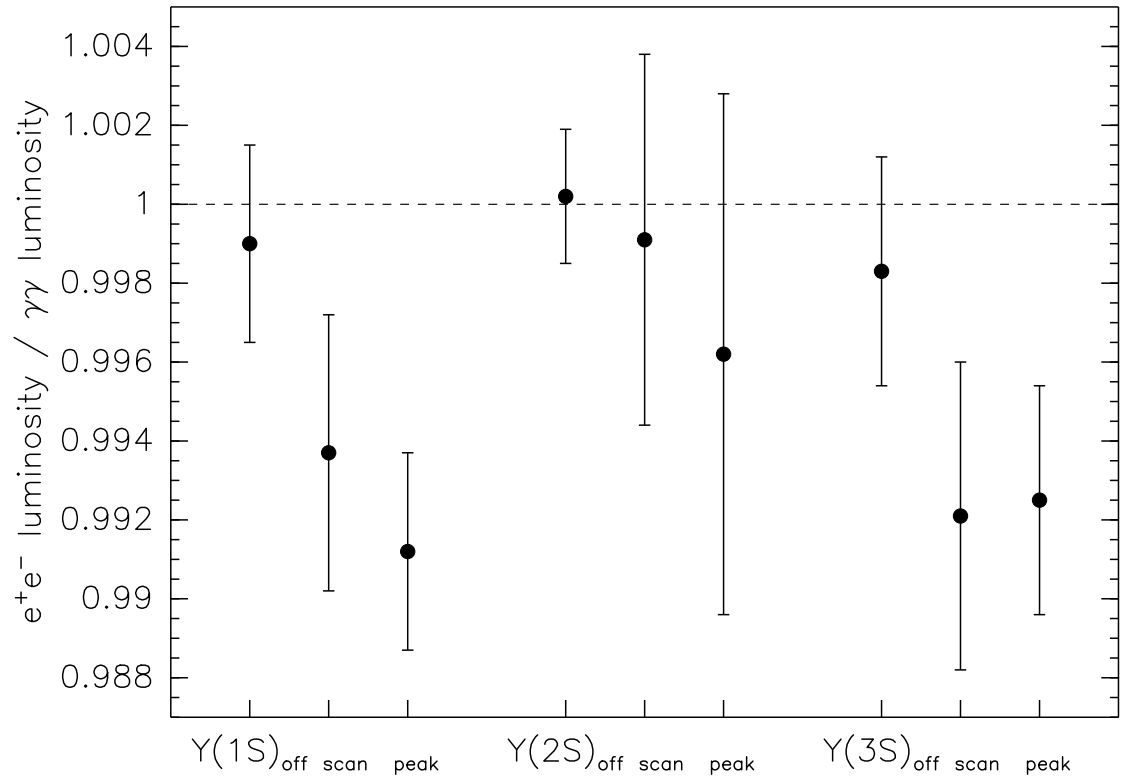}
  \end{center}
  \caption[Ratio of Bhabha luminosity to \gamgam\ luminosity through
  each resonance]{\label{lumigamgam} Ratio of Bhabha luminosity to
  \gamgam\ luminosity, with the weighted average continuum ratio set
  to unity.}
\end{figure}

\label{sec:luminosityconsistency}
In Figure~\ref{lumigamgam}, we plot ratios of luminosity measurements
for off-resonance, scan, and peak data on the \us, \uss, and \usss,
and observe a discrepancy on the \us\ and \usss.  The \gamgam\
luminosity scale factor is tuned to reproduce Bhabha luminosity
off-resonance, but not separately for the three off-resonance
datasets, so the discrepancy we observe occurs only at the \ups\
resonances.  It is not, therefore, linear with respect to \ecm, as
would be expected if the discrepancy were due to energy-dependent cut
efficiencies.  (The discrepancy is also 20 times larger than the
energy dependence in the Bhabha cuts, and the \gamgam\ cuts have no
energy dependence on this scale.)

The direction of this effect, and the fact that it only influences
\ups\ data, could be explained if we were over-subtracting $\Upsilon
\to e^+e^-$ from our Bhabha count.  However, the relative magnitudes
of the \us\ and the \usss\ discrepancies cannot be accounted for.  An
over-subtraction error would be proportional to peak hadronic
cross-section times \bee, which is 0.45~nb for the \us\ and 0.096~nb
for the \usss, but the discrepancies appear to be equal.  Even
considering the uncertainties in these measurements, the
over-subtraction hypothesis is ruled out by 8.7 standard deviations.

We do not know whether the discrepancy is due to an error in the
Bhabha measurement or in the \gamgam\ measurement, so we apply half of
this discrepancy as a correction and add half the discrepancy and its
uncertainty in quadrature to the total uncertainty in luminosity, to
cover the ambiguity.  This is only a 0.4\% uncertainty in the
luminosity of each resonance, and therefore a small contributor to the
total \gee\ uncertainty.

\chapter{Beam Energy Measurement}
\label{chp:beamenergy}

Having presented all the details necessary to properly measure the
hadronic \ups\ cross-section, the vertical axis in our lineshape plot
(Figure~\ref{cartoon}), we now consider our measurement of \ecm, the
horizontal axis.  In Chapter \ref{chp:hardware}, we discussed the
mechanism which determines the beam energy of each run.  While this
reckoning differs from the true beam energy by 18~MeV near the \ups\
resonances, we will show in this Chapter that measurements of beam
energy differences are robust enough for the precision demands of this
analysis.

\begin{figure}[p]
  \begin{center}
    \includegraphics[width=\linewidth]{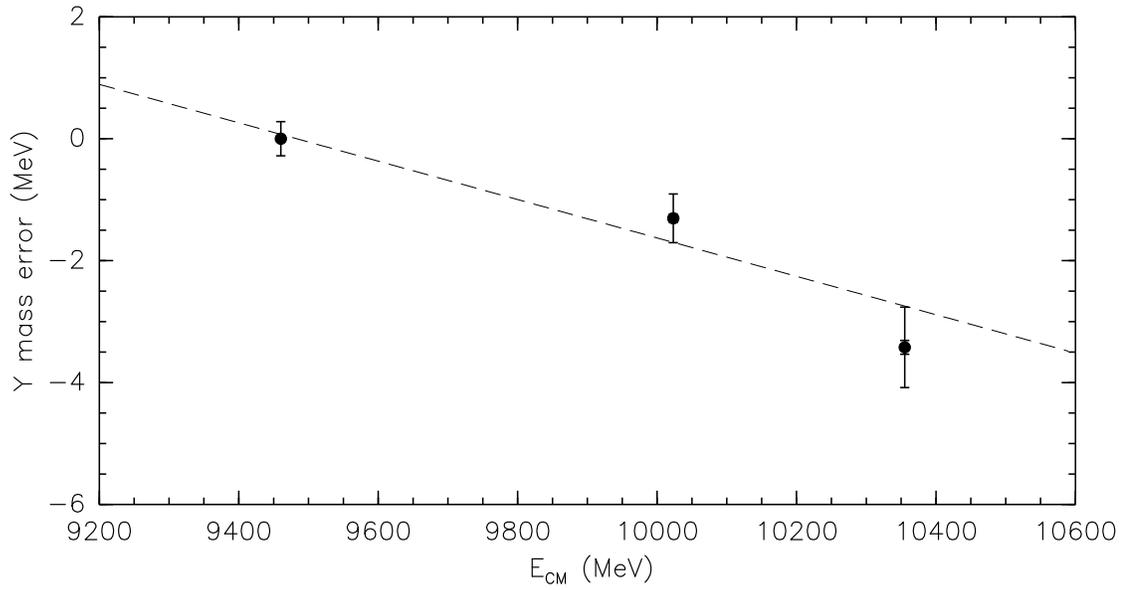}
  \end{center}
  \caption[Beam energy calibration as a function of
  \ecm]{\label{energycalibration} Difference between the measured mass
  and the true mass of the \us, \uss, and \usss, where the \us\
  measurement has been shifted to the true mass.  Outer errorbars
  represent the RMS of all scan measurements at each resonance, and
  inner errorbars are statistical-only.  The dashed curve is a linear
  fit.}
\end{figure}

The masses of the \ups\ resonances have been measured with
0.3--0.5~MeV precision at Novosibirsk \cite{novomass}, so we can use
the \ups\ lineshapes as calibrating markers in beam energy.  If we
correct our \ecm\ measurements by a constant shift at the \us, we
almost reproduce the Novosibirsk \uss\ and \usss\ masses: our average
\uss\ mass is 0.65 \PM\ 0.18~MeV too low, and our average \usss\ mass
is 1.60 \PM\ 0.30~MeV too low (Figure~\ref{energycalibration}).  If
this error is purely a function of beam energy and is not related to
alterations in CESR's configuration between the \usss, \us, and \uss\
data-taking periods, then our beam energy measurements differ from the
true beam energies by a linear transformation, which we determine with
a fit to the three apparent \ups\ masses in
Figure~\ref{energycalibration}.  If this tranformation applies to
small changes in beam energy (scan point differences, about 10~MeV) as
well as large (300~MeV between resonances), the resonance scans are
too narrow, and \geehadtot\ is too low, by only 0.2\%.  Because of the
uncertainties noted above, we do not apply this as a correction.

\begin{figure}[p]
  \begin{center}
    \includegraphics[width=\linewidth]{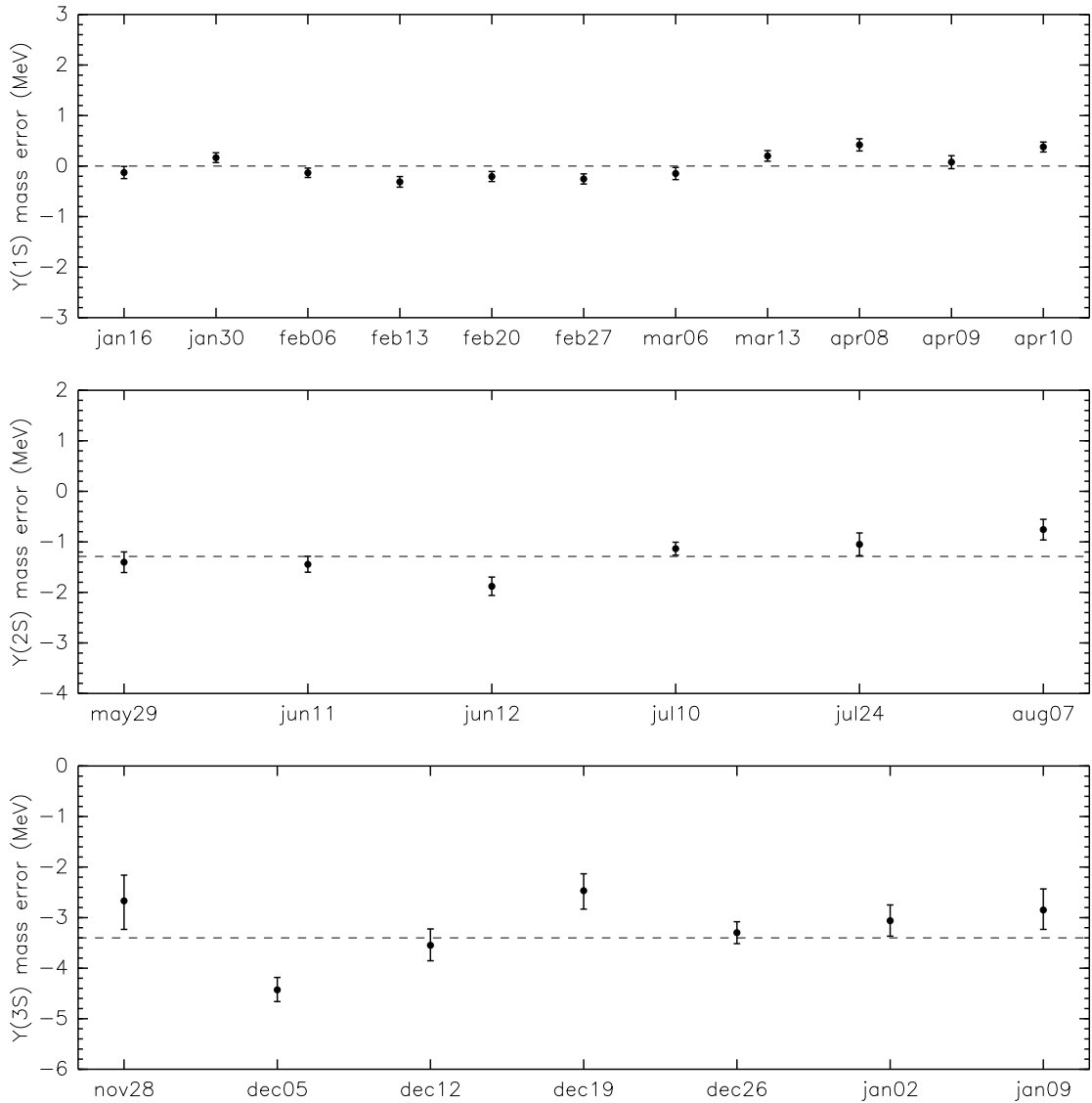}
  \end{center}
  \caption[Beam energy calibration as a function of
  date]{\label{beamenergydrift} Difference between the measured mass
  and the true mass of the \us, \uss, and the \usss\ (top to bottom),
  for each individual scan.  Dashed horizontal lines represent the
  weighted averages.}
\end{figure}

Potentially more worrisome are variations in the beam energy with
time.  The measurement of beam energy differences is sensitive to the
placement of the NMR probes in the test magnets, because at the 0.05\%
level (corresponding to 1~MeV in \ecm), the magnetic field in the test
magnets is a function of position.  Every week, these probes are
exposed to the possibility of movement as a consequence of CESR
machine studies, since these studies may involve reconfiguring the
test magnet.  To protect our lineshape scans from discrete shifts in
beam energy calibration, we divided our data-taking into short,
independent 10-hour scans, plus 38~hours of subsequent peak running,
separated by about a week.  The test magnets were not disturbed during
these dedicated scans or during the peak data-taking associated with
each scan.  Changes in the apparent \ups\ mass from one week to the
next alert us to shifts in the beam energy measurement between scans,
which we observe as a slow drift on the order of 0.5~MeV per month
(Figure~\ref{beamenergydrift}).  To be insensitive to these drifts, we
include the apparent \ups\ mass of each 10+38-hour scan as an
independent parameter in the lineshape fit. \label{pag:massfloats}
Values of \ecm\ in the lineshape fit are only relative to the apparent
\ups\ mass from the week the data were taken.

We are only sensitive to random changes in the beam energy calibration
(true \ebeam\ $-$ measured \ebeam), or jitter, on a 10-hour timescale.
We have already ensured our insensitivity to week-by-week
fluctuations.  If the beam energy fluctuates on very short timescales,
much less than an hour (the length of a run), then the jitter only
contributes to beam energy spread.  (In this sense, the beam energy
fluctuates by about 4~MeV with every collision!)  We reduce our
sensitivity to monotonic 10-hour calibration drifts by alternating
scan points above and below the \ups\ peak, so that a drift would not
systematically widen or narrow the lineshape.

\begin{figure}[p]
  \begin{center}
    \includegraphics[width=\linewidth]{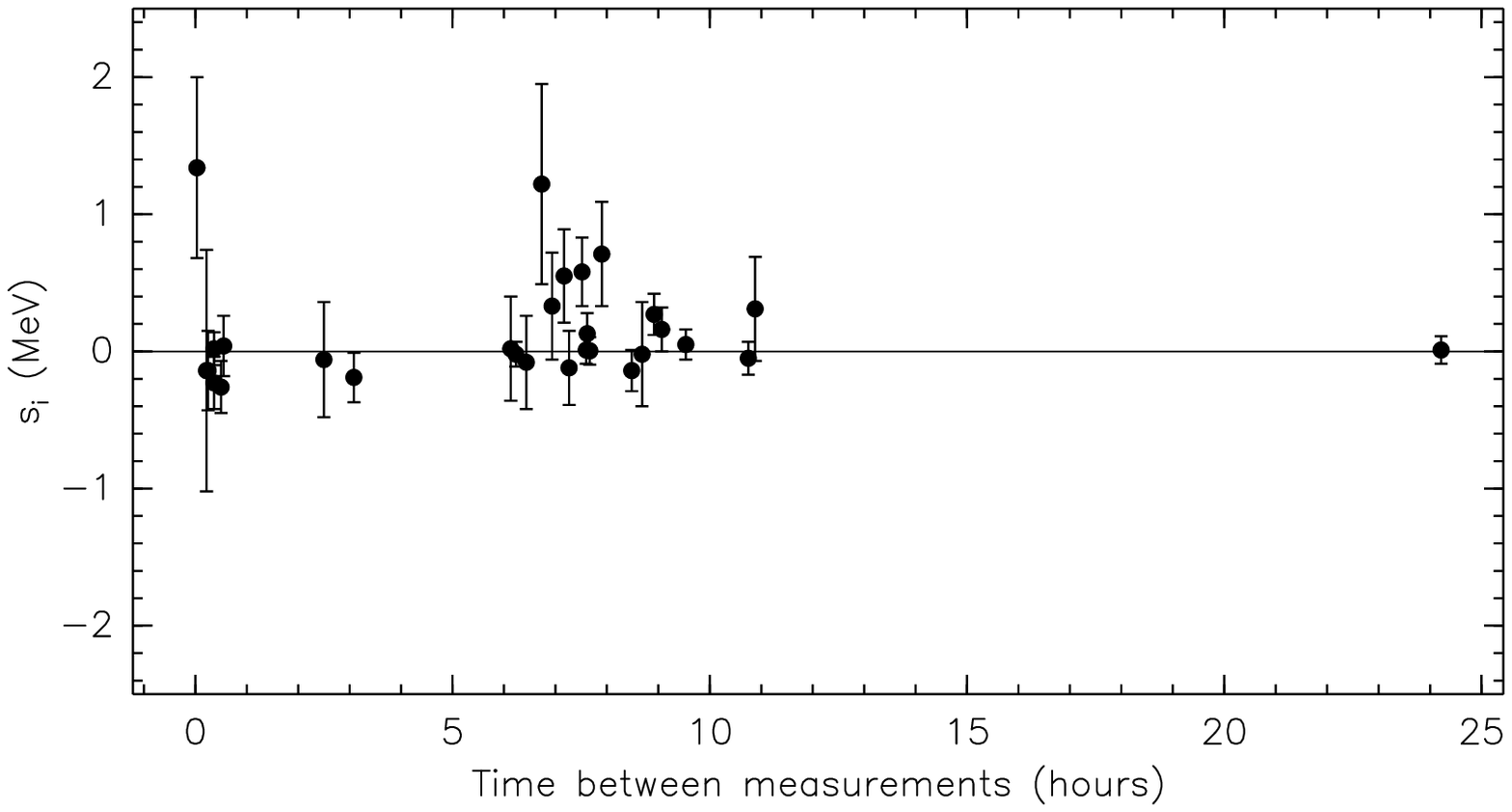}
  \end{center}
  \caption[Calibration shifts from repeated cross-section
  measurements]{\label{miscalhours} Beam energy shifts ($s_i$)
  determined by pairs of repeated cross-section measurements, plotted
  with respect to the time between the first and the second
  measurement.  The weighted mean of $\{s_i\}$ is 0.02 \PM\ 0.03~MeV,
  so the apparent vertical asymmetry of this distribution is an
  illusion.}
\end{figure}

\begin{figure}[p]
  \begin{center}
    \includegraphics[width=\linewidth]{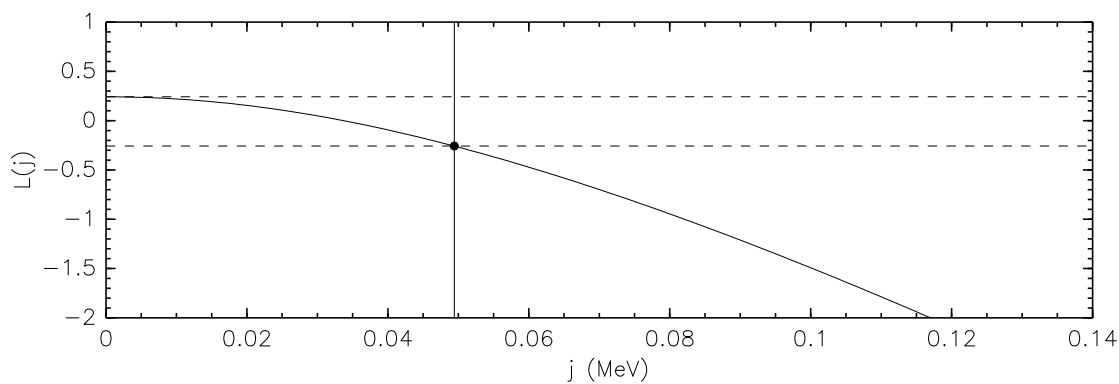}
  \end{center}
  \caption[Log-likelihood of beam energy jitter]{\label{energylj}
  Log-likelihood ($L(j)$) as a function of jitter ($j$) attains a
  maximum near $j=0$ and is reduced by 1/2 at $j=0.05$~MeV.}
\end{figure}

To check for changes in the beam energy calibration on the
10-hour timescale, we measure the cross-section at a high-derivative
point on the lineshape twice, usually at the beginning and end of a
scan.  Since the derivative is high, significantly different
cross-section results at the same reported beam energy would be an
indication that the actual beam energy has shifted between the two
measurements.  Any cross-section difference ($\sigma_1-\sigma_2$) can
be converted into a true beam energy difference ($E_1-E_2$) at a given
lineshape derivative ($d\sigma/dE$) by
\begin{equation}
  \left. E_1 - E_2 = \frac{\sigma_1 - \sigma_2}{d\sigma/dE} \right._{\mbox{.}}
\end{equation}
To find possible shifts in calibration, we subtract the true beam
energy difference from the reported difference, and propagate
uncertainties from the cross-section measurement.  We call this a
shift ($s_i \pm \delta_{s_i}$).  The shifts (plotted in
Figure~\ref{miscalhours}) are not quite consistent with zero, as the
total $\chi^2 = \sum_i (s_i/\delta_{s_i})^2$ is 31.2 for 29 degrees of
freedom (a 0.64\% confidence level).  This broadening of the $\{s_i\}$
distribution can be accomodated by a uniformly-random jitter $j$.  To
quantify $j$, we construct a log-likelihood function of the $\{s_i\}$
data, assuming them to be Gaussian-distributed with statistical and
jitter components to the width.
\begin{equation}
  L(j) = \sum_{i=1}^{29} \ln \left(
  \frac{1}{\sqrt{2\pi({\delta_{s_i}}^2 + j^2)}} \exp\left(\frac{-{s_i}^2}{2
  ({\delta_{s_i}}^2 + j^2)} \right) \right)
\end{equation}
Only jitters larger than 0.05~MeV are inconsistent with the data at
the 68\% confidence level (the value of $j$ necessary to lower the
log-likelihood by 1/2; see Figure~\ref{energylj}).  Alternatively, we
could have constructed an $S$-factor in analogy with the Particle Data
Group's method of calculating uncertainty from a set of
mutually-inconsistent experiments:
\begin{equation}
  \left. S(j) = \sum_{i=1}^{29} \frac{{s_i}^2}{{\delta_{s_i}}^2 + j^2} \right._{\mbox{.}}
\end{equation}
In this formulation, the maximally-allowed jitter is the value which
reduces $S(j)$ to unity, which is $j=0.07$~MeV.  Both methods yield
roughly the same value, so we can be confident that the beam energy
calibration jitter is approximately 0.07~MeV, which is 7 parts per
million.

To learn what fluctuations of this size imply for uncertainty in
\geehadtot, we simulated the lineshape fits with random perturbations
in the measured beam energy.  The nominal \ecm\ and luminosity of each
simulated cross-section measurement were copied from the real data
sample, but the cross-sections themselves were derived from an ideal
curve with statistical errors.  Without perturbing the simulated beam
energy measurements, the fits return our input \geehadtot\ with
perfectly-distributed fit $\chi^2$ values.  When \ecm\ is randomly
perturbed with a standard deviation of 0.07~MeV, the value of
\geehadtot\ fluctuates up and down by 0.2\% of itself, and the
$\chi^2$ increases by only 5--30 units.  \label{pag:notjitter} We
therefore assign a 0.2\% systematic uncertainty in \geehadtot\ and
\gee\ due to beam energy measurement errors.

\chapter{Lineshape Fitting}
\label{chp:fitting}

We have reached the core of the analysis, the fits of the \ups\
lineshapes.  All previous studies either deliver data to these fits or
define the curve that the data are fit to.  In this Chapter, we will
review the fit function and all of its parameters, present the fit
results, and discuss hadron-level interference as a fitting issue.

\section{The Fit Function}
\label{sec:fitfunction}

\subsection{Hadronic Peak}

The central feature of the fit function is a Breit-Wigner curve
representing the hadronic cross-section (Equation
\ref{eqn:breitwigner}), convoluted with beam energy spread and
initial-state radiation (ISR).  The beam energy spread is modeled by a
Gaussian with unit area, and the ISR distribution is calculated to
fourth order in perturbative QED by Kuraev and Fadin \cite{kf}
(Equation (28)).

Most of the floating parameters of the fit modify this contribution.
The area of the Breit-Wigner is the most important floating parameter,
as this is how we determine \geehadtot\ and \gee\ (Equation
\ref{eqn:gee}).  To let each 48-hour scan's \ups\ mass float
independently (Section~\ref{pag:massfloats} on
page~\pageref{pag:massfloats}), we fix the mass in the fit function
and allow the \ecm\ measurements in the data to be shifted
scan-by-scan.  This way, when we plot the lineshape fits, a single
curve represents the fit to all data.  In Chapter~\ref{chp:hardware},
we discussed the variability of beam energy spread
(Section~\ref{pag:beamenergyspread} on
page~\pageref{pag:beamenergyspread}).  To allow scans to have
different beam energy spreads (Gaussian widths), we use different fit
curves.  We do not transform the data in analogy with the mass shifts
because the ISR tail spoils the linearity of the transformation.

The full width $\Gamma$ is a parameter in the Breit-Wigner
distribution, but we do not allow this parameter to float.  Our fitted
\geehadtot\ is sensitive to $\Gamma$ only at the 0.03\% level, so we
fix each $\Gamma$ to its previously measured value \cite{pdg}.

The hadronic \ups\ efficiency is another multiplicative constant in
our fit function.  We multiply our fit function by efficiency rather
than dividing our data by efficiency because our data includes several
different components, each with a different efficiency.

We add interference with the continuum
(Section~\ref{sec:earlyinterference} on
page~\pageref{sec:earlyinterference}) to the Breit-Wigner before
convolution.  That is, the signal lineshape is
\begin{equation}
  \tilde{\sigma}_{\subs{res}+\subs{int}}(E_\subs{CM}) = \bigg(
  \sigma_\subs{res}(E') +
  \tilde{\sigma}_\subs{int}(E') \bigg) \otimes G \otimes ISR
\end{equation}
where $\sigma_\subs{res}$ and $\tilde{\sigma}_\subs{int}$ are defined
in Equation \ref{eqn:rescontinter} and $\otimes$ represents
convolution with the beam energy spread Gaussian ($G$) and the ISR
distribution ($ISR$).  This $\tilde{\sigma}_{\subs{res}+\subs{int}}$
does not represent a physical cross-section until we add the continuum
piece.  (It is negative for some values of \ecm.)  The interference
between resonant and continuum \qqbar\ decays is characterized by two
constants, $\phi_0$ and $\alpha_\subs{int}$ (Equation \ref{eqn:yint}),
both of which are known and do not need to float in the fit.

\subsection{Tau-Pair Peak}

The \tautau\ background term has the same form as the hadronic signal
term: it is the sum of a Breit-Wigner and $\tilde{\sigma}_\subs{int}^{\tau^+\tau^-}$,
convoluted by the same beam energy spread and ISR distribution.  This
term introduces no new floating parameters: the Breit-Wigner area for
\tautau\ is the Breit-Wigner area for hadronic \ups, multiplied by a
ratio of branching fractions and efficiencies
\begin{equation}
  \frac{\mbox{area}_{\tau^+\tau^-}}{\mbox{area}_\subs{had}} = \left.
  \frac{{\mathcal B}_{\tau^+\tau^-}}{{\mathcal B}_\subs{had}} \times
  \frac{\epsilon_{\tau^+\tau^-}}{\epsilon_\subs{had}} \right._{\mbox{.}}
\end{equation}

The value of \btt\ we use in the fit function is from \cite{jean}.  We
use this value to determine \geehadtot, but when we determine \gee, we
perform a separate fit with \btt\ = \bmm\ and \bmm\ from
\cite{istvan}.  This allows us to subtract \tautau\ backgrounds with
the same branching fraction that we use to multiply it back in, when
we convert \geehadtot\ to \gee\ (Equation \ref{eqn:geehadtot:to:gee}).

The magnitude of interference, $\alpha_\subs{int}^{\tau^+\tau^-}$, is also different
for the \tautau\ peak, because $\sigma(e^+e^- \to \tau^+\tau^-)$ and
$\Gamma(\Upsilon \to \tau^+\tau^-)$ are not equal to $\sigma(e^+e^-
\to q\bar{q})$ and $\Gamma(\Upsilon \to q\bar{q})$, respectively.

\subsection{Background Terms}

We add a single $1/s$ term to the fit function to represent continuum
\qqbar, radiative Bhabhas, \tautau, and any residual Bhabha or \mumu\
backgrounds.  The magnitude of this term floats in the fit, though it
is primarily influenced by our large off-resonance data point.

The resonance interferes with only part of the contniuum, so the
magnitude of resonance interference is not tied to the fitted value of
the $1/s$ term.  The \qqbar\ and \tautau\ cross-sections that enter
into $\alpha_\subs{int}$ and $\alpha_\subs{int}^{\tau^+\tau^-}$ are
either derived from other experiments (\cite{novor} for
$\alpha_\subs{int}$) or are calculated from QED (for
$\alpha_\subs{int}^{\tau^+\tau^-}$).

This dissociation of the continuum cross-section from the interference
term is artificial, since interference is only meaningful in the
presence of both resonance and continuum amplitudes.  It would be more
natural, for instance, to include the continuum cross-section in the
beam energy spread and ISR convolutions with the resonance and
interference terms, since the natural lineshape includes all three
terms before it is smeared by the physical beams.  However, the continuum
cross-section doesn't depend on \ecm\ sharply enough for this to
matter.

For \uss\ and \usss\ fits, we add the fitted lineshape(s) from \us\
and \uss\ as backgrounds, because the ISR tails from these resonances
introduce 0.4--0.6\% corrections.  These background components do not
float in the fit.  (We fit the three resonances separately.)

The $\log s$ correction motivated by two-photon fusion backgrounds is
tied to the overall continuum normalization through $f$ = 0.080 \PM\
0.005, the fraction of $\log s$ component to $1/s$ at 9~GeV,
determined by a fit from Section~\ref{pag:logs} on
page~\pageref{pag:logs}.  The total continuum cross-section is
therefore
\begin{equation}
  \sigma_\subs{cont}(E_\subs{CM}) = \sigma_\subs{cont}^0 \left(
  (1-f) \frac{(\mbox{9~GeV})^2}{{E_\subs{CM}}^2} +
  f \log\frac{{E_\subs{CM}}^2}{(\mbox{9~GeV})^2} \right)
\end{equation}
where $\sigma_\subs{cont}^0$ is the floating parameter.

We allow $\sigma_\subs{cont}^0$ to float independently in the \us,
\uss, and \usss\ fits.  Since it represents the cross-section at a
fixed \ecm, the three fits ought to return the same value.  Instead,
the \us, \uss, and \usss\ fits yield 9.355 \PM\ 0.010~nb, 9.318 \PM\
0.007~nb, and 9.315 \PM\ 0.011~nb, respectively.  The \uss\ and \usss\
values are consistent with each other, but the \us\ fit has a
$\sigma_\subs{cont}^0$ which is 3.0 standard deviations higher than
\uss.  However, this difference in $\sigma_\subs{cont}^0$ values is
only 0.04\% of $\sigma_\subs{cont}^0$ itself.  It is possible that our
ISR tail corrections are too large, an error which is absorbed into
$\sigma_\subs{cont}^0$, rather than biasing the fit for \geehadtot.

\subsection{Summary of all Floating Parameters}

The \us\ fit has a total of sixteen parameters: \geehadtot, $M_1,
\ldots M_{11}$ for its eleven 48-hour scans, $\Delta E_1$, $\Delta
E_2$, and $\Delta E_3$ for its three groups of scans with potentially
different beam energy spreads (see Table~\ref{tab:scansa}), and
$\sigma_\subs{cont}^0$ for the continuum background normalization at
9~GeV.

The \uss\ fit has a total of nine floating parameters: \geehadtot,
$M_1, \ldots, M_6$ for its six 48-hour scans, a single beam energy
spread $\Delta E$ (the CESR orbits were stable), and
$\sigma_\subs{cont}^0$.

The \usss\ fit has a total of sixteen floating parameters: \geehadtot,
$M_1, \ldots, M_7$ for its seven 48-hour scans, $\Delta E_1,
\ldots, \Delta E_7$, a beam energy spread for each scan, and
$\sigma_\subs{cont}^0$.

\section{Systematic Uncertainties from Constants in the Fit}
\label{sec:lineshapesyst}

Each constant in the fit function may introduce systematic error, but
in every case, this error is negligible.  Kuraev and Fadin estimate
that their ISR calculation has a 0.1\% uncertainty; varying the tail
normalization by this amount changes our \geehadtot\ result by 0.05\%.
As previously mentioned, the uncertainty in $\Gamma$ affects our result
by less than 0.03\%.  Uncertainty in the \qqbar\ interference
magnitude (from $R$) yields a 0.02\% uncertainty, and the \tautau\
interference magnitude (from \btt) yields 0.08\%.  Uncertainty in the
two-photon fraction $f$ yields 0.002\%.  The sum of all of these
uncertainties in quadrature is 0.10\%.

\label{pag:dontneedepsilon} We do not propagate the uncertainty in the
\tautau\ efficiency (first mentioned in
Section~\ref{pag:dontneedepsilon} on
page~\pageref{pag:dontneedepsilon}) because it is multiplied by \btt\
or \bmm\ (in the \geehadtot\ and \gee\ fits, respectively).  The
fractional uncertainty in the Monte Carlo modeling of \tautau\
efficiency is not significantly greater than 2\%, the smallest \btt\
or \bmm\ fractional uncertainty.

The hadronic efficiency also enters the fit function as a constant,
but we will treat this as a major systematic uncertainty in the next
Chapter.

\section{Fit Results}

\begin{sidewaysfigure}[p]
  \begin{center}
    \includegraphics[width=\linewidth]{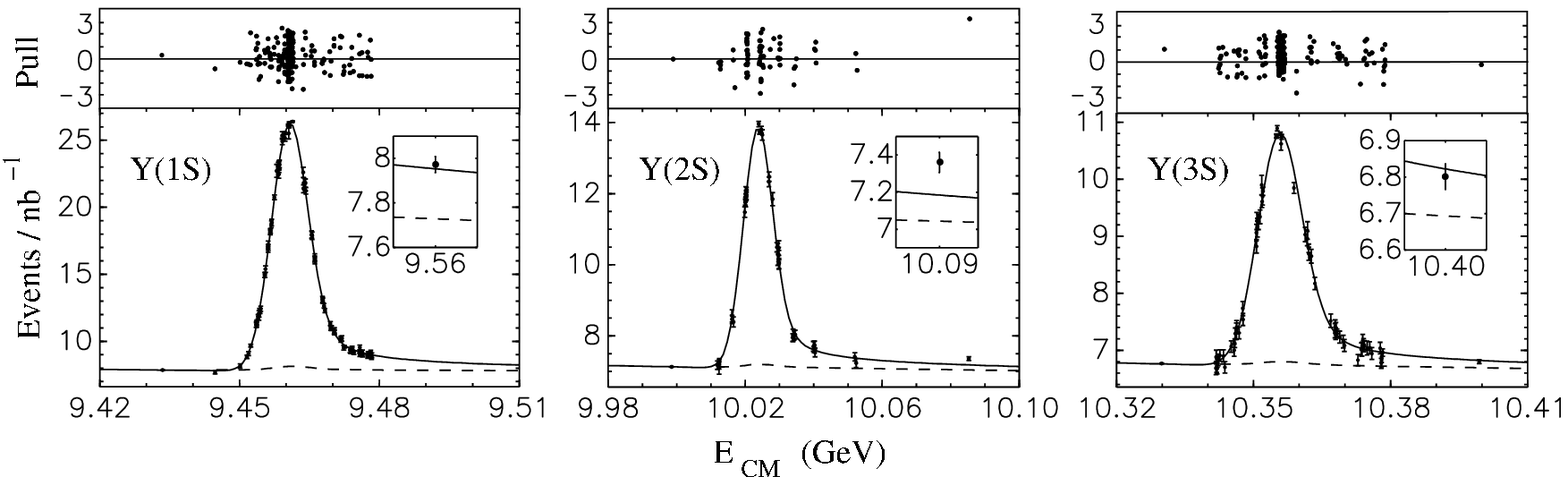}
  \end{center}
  \caption[Best-fit lineshapes for \us, \uss, and
  \usss]{\label{fitresults} Best-fit lineshapes (solid) to the \us,
  \uss, and \usss\ data (points), from which \geehadtot\ is
  determined.  Dashed curves are total background estimates, and the
  pull distributions above each plot indicate the statistical
  significance of all deviations from the best-fit line.  Insets for
  each resonance present a tail measurement 100~MeV, 60~MeV, and
  45~MeV above the \us, \uss, and \usss\ masses.  (Fit functions with
  multiple beam energy spreads are represented by their average beam
  energy spread.)}
\end{sidewaysfigure}

\begin{figure}[p]
  \begin{center}
    \includegraphics[width=0.9\linewidth]{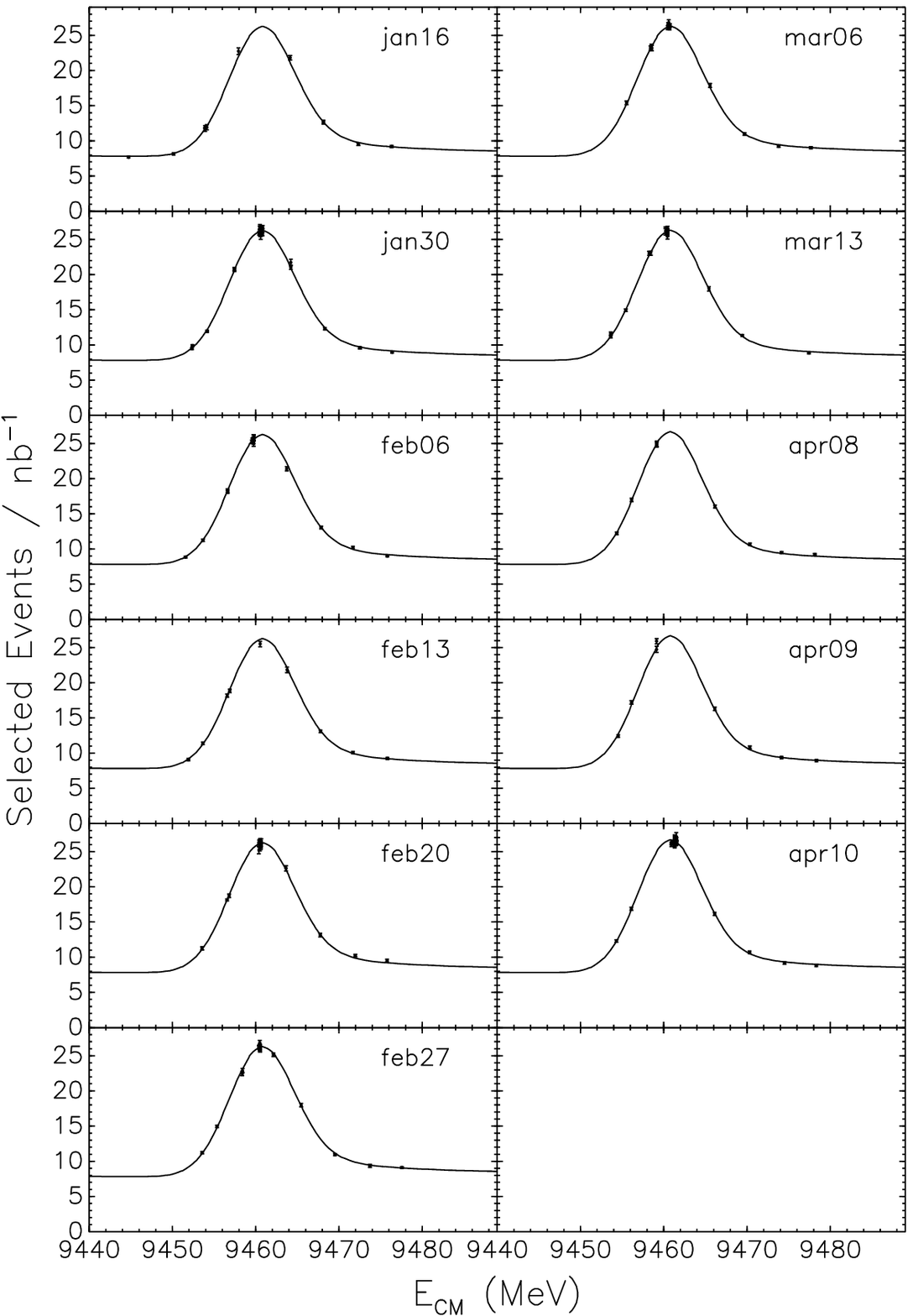}
  \end{center}
  \caption[Best-fit lineshape for each \us\ scan]{\label{allscanss} Best-fit \us\ lineshape,
  overlaid upon data from individual scans.}
\end{figure}

\begin{figure}[p]
  \begin{center}
    \includegraphics[width=0.9\linewidth]{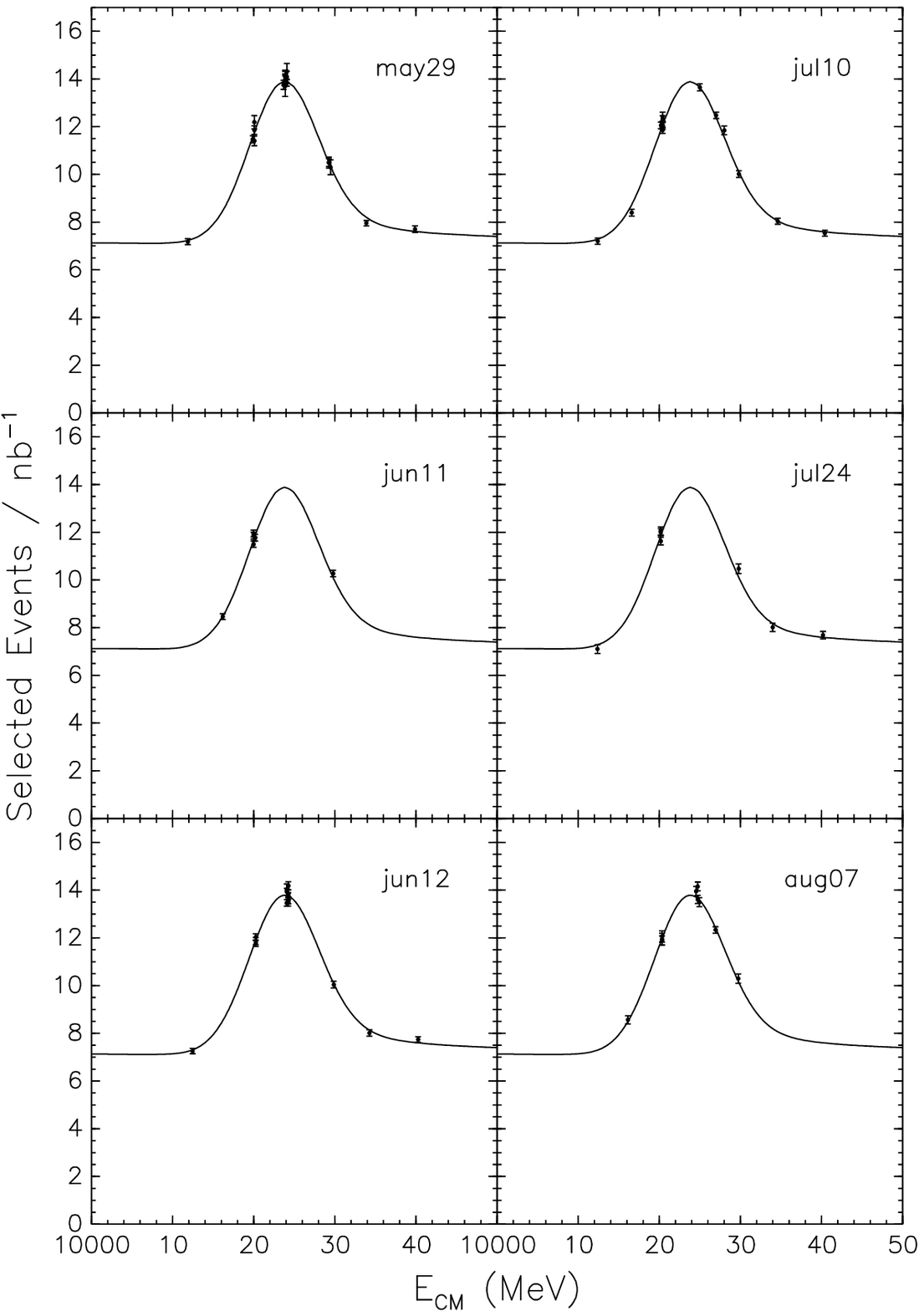}
  \end{center}
  \caption[Best-fit lineshape for each \uss\ scan]{\label{allscansss} Best-fit \uss\ lineshape,
  overlaid upon data from individual scans.}
\end{figure}

\begin{figure}[p]
  \begin{center}
    \includegraphics[width=0.9\linewidth]{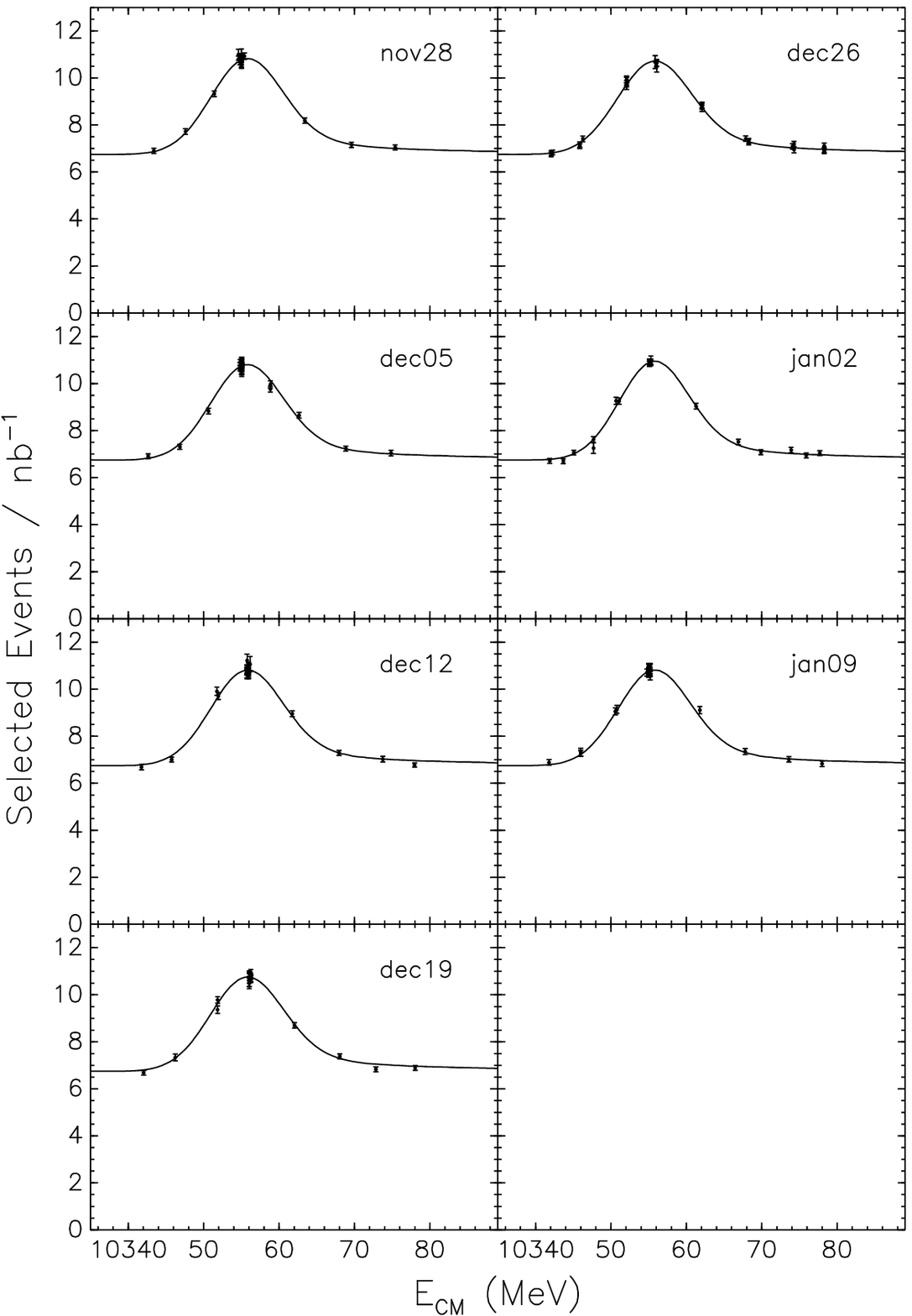}
  \end{center}
  \caption[Best-fit lineshape for each \usss\ scan]{\label{allscanssss} Best-fit \usss\ lineshape,
  overlaid upon data from individual scans.}
\end{figure}

\begin{figure}[p]
  \begin{center}
    \includegraphics[width=\linewidth]{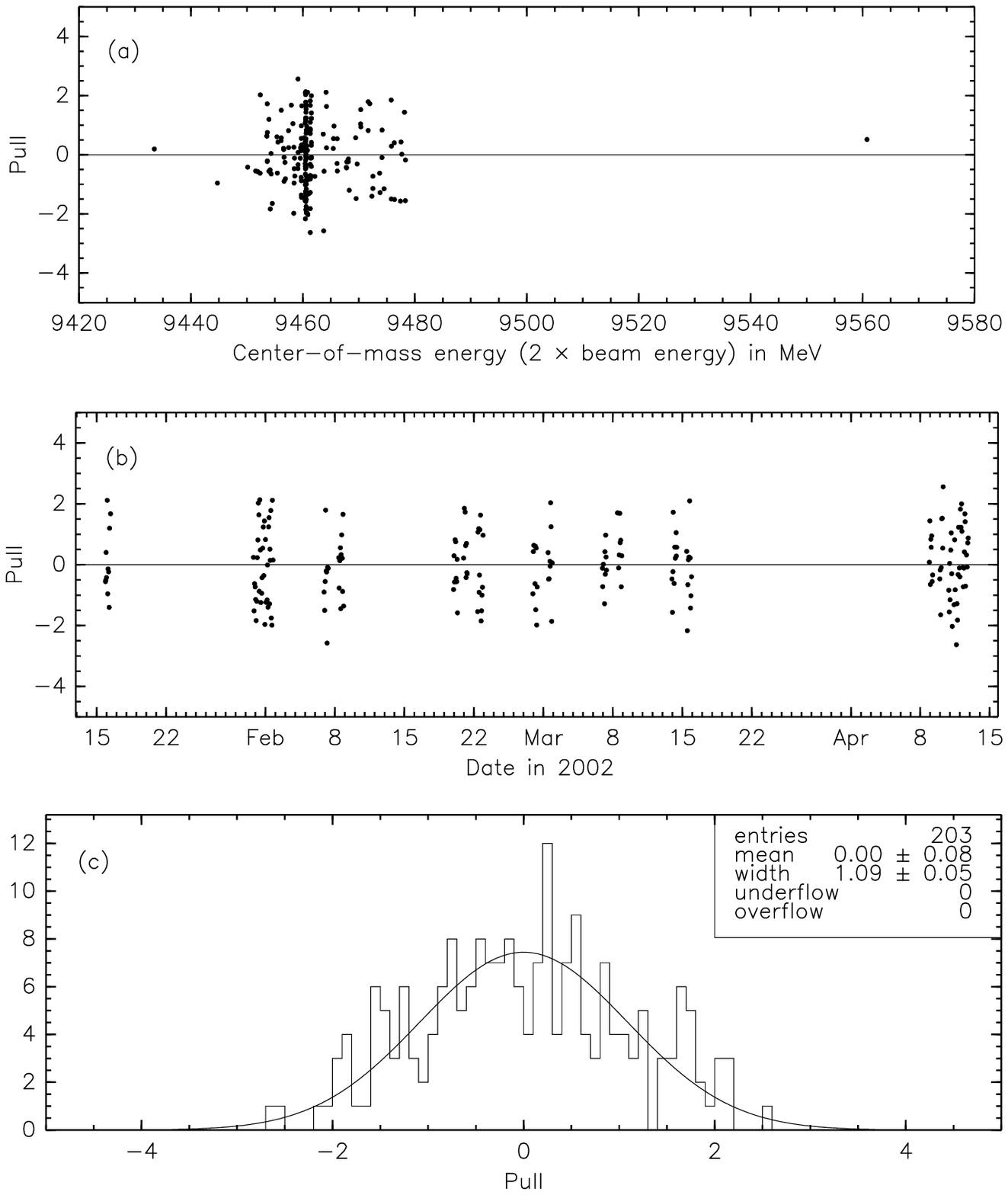}
  \end{center}
  \caption[Pull distributions for \us]{\label{pullsone} The pull distribution of \us\ lineshape
  fits (a) as a function of \ecm, (b) as a function of date, and (c),
  as a histogram, fitted to a Gaussian curve, which is almost two
  standard deviations wider than unity.}
\end{figure}

\begin{figure}[p]
  \begin{center}
    \includegraphics[width=\linewidth]{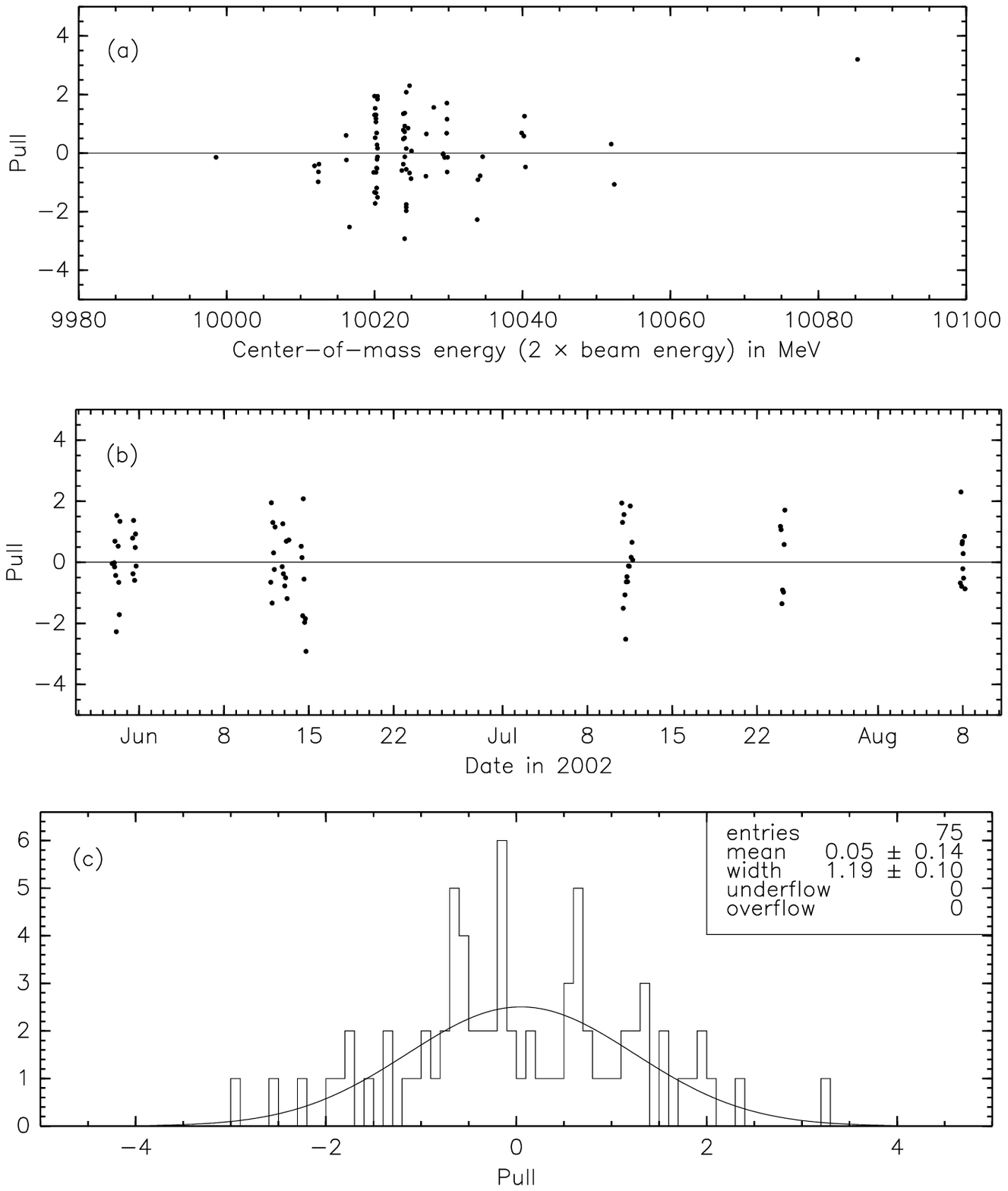}
  \end{center}
  \caption[Pull distributions for \uss]{\label{pullstwo} The pull distribution of \uss\ lineshape
  fits (a) as a function of \ecm, (b) as a function of date, and (c),
  as a histogram, fitted to a Gaussian curve, which is almost two
  standard deviations wider than unity.}
\end{figure}

\begin{figure}[p]
  \begin{center}
    \includegraphics[width=\linewidth]{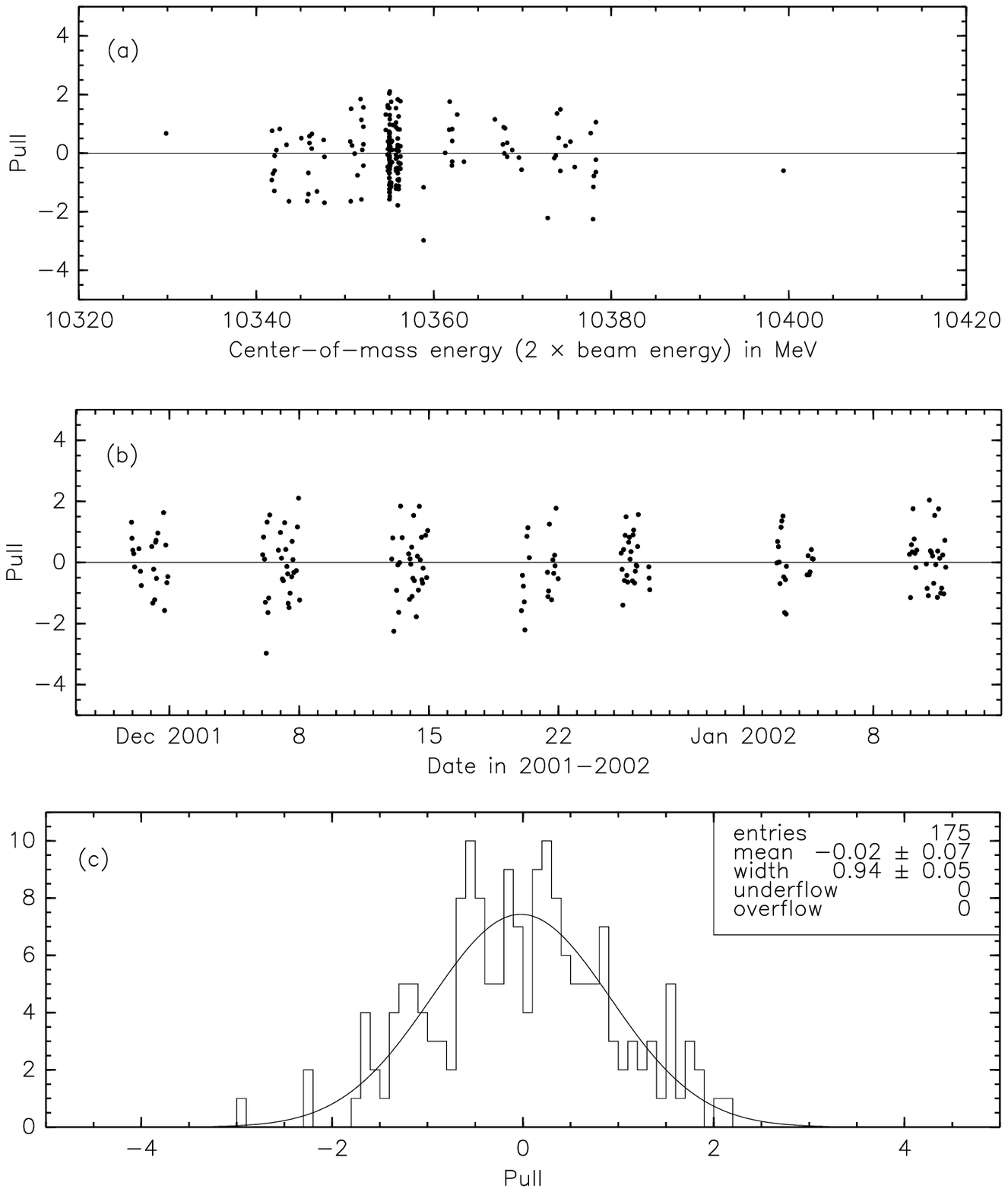}
  \end{center}
  \caption[Pull distributions for \usss]{\label{pullsthree} The pull distribution of \usss\
  lineshape fits (a) as a function of \ecm, (b) as a function of date,
  and (c), as a histogram, fitted to a Gaussian curve, which is
  statistically consistent with unity.}
\end{figure}

We used a {\tt C++} version (1.5.2) of {\tt MINUIT} from the {\tt
SEAL} package \cite{seal} to perform a $\chi^2$ fit of our
cross-section data to our fit function.  The resulting best-fit values
for the Breit-Wigner area used in determining \gee\ are presented in
Table~\ref{tab:bestfit}.  The best-fit functions are plotted in
Figure~\ref{fitresults}, and
Figures~\ref{allscanss}--\ref{allscanssss}, overlaying data.  At the
top of Figure~\ref{fitresults} and in Figures~\ref{pullsone},
\ref{pullstwo}, and \ref{pullsthree} are pull distributions.  Pull is
the ratio of residual to uncertainty, expressing the significance of
deviations from the best-fit line.  All data points are visible in
these plots; no point has a pull greater than 3.5.  Best-fit \ups\
masses and beam energy spreads are plotted in
Figures~\ref{beamenergydrift} and \ref{beamenergyspread},
respectively.

\begin{table}
  \caption[Best-fit Breit-Wigner area and $\chi^2$
  significance]{\label{tab:bestfit} Best-fit Breit-Wigner area and
  $\chi^2$ significance.}
  \begin{center}
    \renewcommand{\arraystretch}{1.35}
    \begin{tabular}{c c c c c}
      \hline\hline
      & \mbox{\hspace{-0.5 cm}} $\int \sigma(e^+e^- \to \Upsilon \to \mbox{hadronic}) \, dE_\subs{CM}$ \mbox{\hspace{-0.5 cm}} & reduced $\chi^2$ & confidence level \\\hline
      \us   & 324.12 \PM\ 0.92~MeV~nb & $\displaystyle \frac{240.4}{203-16}$ = 1.29 & 0.51\% \\
      \uss  & 133.65 \PM\ 0.94~MeV~nb & $\displaystyle \frac{107.2}{75-9}$   = 1.62 & 0.10\% \\
      \usss &  89.25 \PM\ 0.88~MeV~nb & $\displaystyle \frac{154.5}{175-16}$ = 0.97 & 58.6\% \\\hline\hline
    \end{tabular}
  \end{center}
\end{table}
    
This fit function describes our data well: no trends are evident in
pull versus energy, but the \us\ and especially the \uss\ pull
distributions are wider than would be expected from statistical
fluctuations.  We can also see this by noting that the $\chi^2$ per
degree of freedom ($N_\subs{dof}$) is improbably high
(Table~\ref{tab:bestfit}).  In the \uss\ data, the high-energy tail is
the largest deviation, contributing 9.1 units to the total $\chi^2$,
but this is only a fifth of the excess.  (Dropping this point changes
the \uss\ \geehadtot\ by 0.4\%.)  The simulations of fits with
jittering beam energy, described at the end of
Section~\ref{pag:notjitter} on page~\pageref{pag:notjitter}
demonstrate that uncertain beam energy measurements can increase the
fit $\chi^2$ by 5--30 units.  If the \us\ and \uss\ $\chi^2$ values
have both been artificially raised 30 units by jittering beam energy
measurements, their natural confidence levels would be 10--15\%, but
there is no conclusive evidence that beam energy jitter is the cause,
and assuming an increase of 30 units of $\chi^2$ is the most extreme
hypothesis.

The larger-than-expected deviations are symmetric around the best-fit
line for all \ecm\ and dates, and they do not favor a particular \ecm\
or date (Figures~\ref{pullsone}--\ref{pullsthree}), so we treat them
as though we had underestimated our statistical uncertainty
($\sigma_\subs{stat}$).  We add $\sigma_\subs{stat}
\sqrt{\chi^2/N_\subs{dof} - 1}$ to our systematic uncertainty in
\label{sec:chichiconsistency} quadrature, which has the same effect on
the total uncertainty as if we had multiplied $\sigma_\subs{stat}$ by
$\sqrt{\chi^2/N_\subs{dof}}$.

Note that our fit does not suggest that the beam energy spread
distribution is distorted from a pure Gaussian.  If this were the
case, we would observe trends in pull versus \ecm.  Therefore, our
assumption of a Gaussian beam energy spread is valid for our level of
precision (statistical plus the above-mentioned systematic
uncertainty, which is roughly $\pm 0.1$~nb).

\section{Hadron-Level Interference}
\label{sec:interference}

In our fit function, we assumed that continuum and \ups\ decays only
interfere at the parton level, and not at the hadron level.  That is,
we assume that $e^+e^- \to q\bar{q}$ can interfere with $e^+e^- \to
\Upsilon \to q\bar{q}$ but not $e^+e^- \to \Upsilon \to ggg$, even
though both \qqbar\ and $ggg$ hadronize into some of the same final
states.

\begin{figure}[p]
  \begin{center}
    \includegraphics[width=0.8\linewidth]{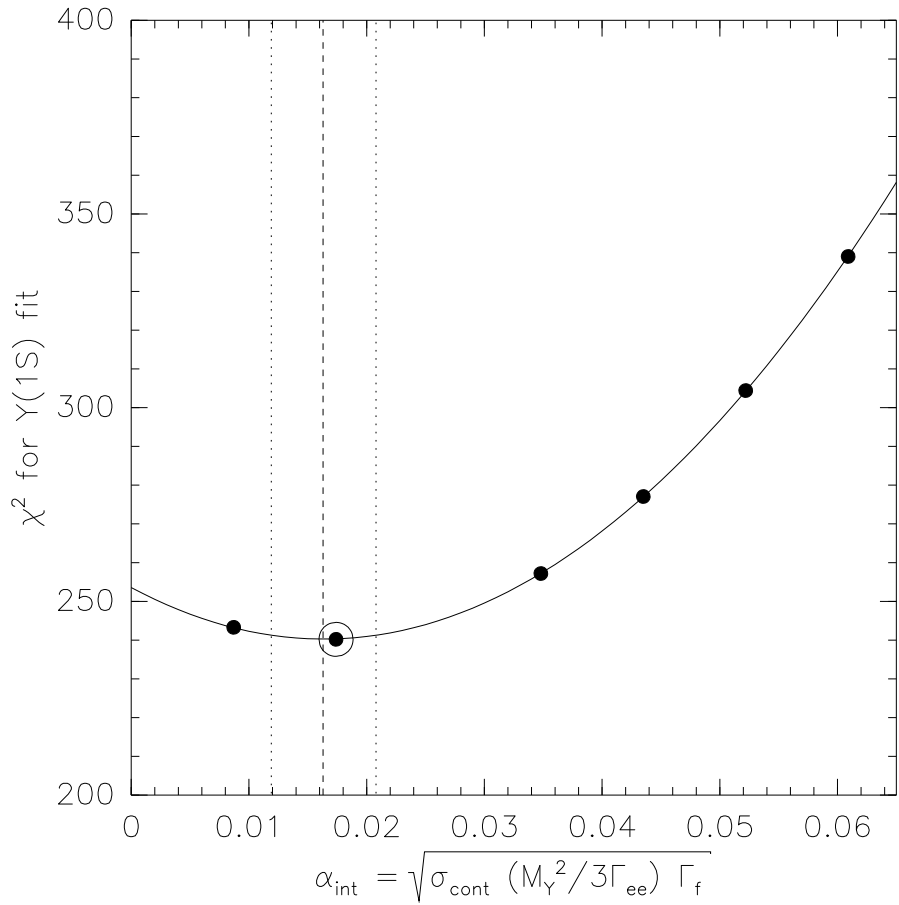}
  \end{center}
  \caption[Best-fit interference between resonance and continuum
  \qqbar]{\label{simpleintfit} A fit of the \us\ for the magnitude of
  hadronic interference ($\alpha_\subs{int}$), assuming $\phi_0 = 0$.
  The points are \us\ fits with trial $\alpha_\subs{int}$ values, the
  vertical axis is the $\chi^2$ of these fits, and the solid line is a
  parabola drawn through the points.  Dashed and dotted lines indicate
  the minimum and 68\% confidence level bands, and the circled point
  represents the \qqbar-only fit we used in this anaysis.}
\end{figure}

This \qqbar-only scheme optimizes the \us\ fit, assuming no phase
difference between resonance and continuum at $E_\subs{CM} \ll
M_\Upsilon$ ($\phi_0=0$).  We determined this by repeatedly fitting
our lineshape, assuming different interference magnitudes
$\alpha_\subs{int}$ (defined in Equation~\ref{eqn:yint}).  We plot the
$\chi^2$ of these fits in Figure~\ref{simpleintfit}.  The interference
magnitude $\alpha_\subs{int}$ is proportional to the square root of
the branching fraction of \ups\ to the interfering final state, which
is $\sqrt{{\mathcal B}_\subs{int}} = \sqrt{8.9\%}$ for \qqbar.  Our
fit is optimized by $\alpha_\subs{int} = 0.016 \pm 0.004$, or
${\mathcal B}_\subs{int} = 7.8 \pm 4.2$.  The $\chi^2$ of the \us\ fit
with no interference is 15 units larger than our \qqbar-only scheme,
which corresponds to 3.7 standard deviations.

However, if $e^+e^- \to \Upsilon \to ggg \to$ hadrons and $e^+e^- \to
q\bar{q} \to$ hadrons significantly interfere, the phase difference
$\phi_0$ need not be zero.  Strong and Electromagnetic \ups\ decays to
hadrons can differ in phase, an effect which has been observed in
exclusive decays of charmonium \cite{charmphase}
\cite{charmcharmphase}.  This Strong minus Electromagnetic ($ggg -
q\bar{q}$) phase difference is constant with respect to \ecm, which
gives the resonance minus continuum phase difference a constant offset
as it evolves through the resonance peak.  In this case, $\phi_0$ need
not be zero, and our fit is less well constrained.

The question of whether $ggg$ and \qqbar\ interfere inclusively (in
the sum over all hadronic final states) is fundamental and interesting
in its own right, concerning the quantum mechanics of the
hadronization process.  Unfortunately, our data cannot answer this
question for all phases, because if $\phi_0 \approx \pm \pi/2$, the
interference term $\tilde{\sigma}_\subs{int}$ has the same \ecm\
dependence as the resonance Breit-Wigner (Equation~\ref{eqn:yint}).
It has been argued \cite{yan} that inclusive $ggg$/\qqbar\ interference
is a small effect because every exclusive final state may have a
different $ggg$/\qqbar\ phase.  The sum of interference cross-terms
includes cancellations and therefore grows more slowly with the number
of exclusive final states than the direct cross-section terms.

Though improbable, it is possible that exclusive final states add
coherently, such that the inclusive $ggg$ and \qqbar\ cross-sections
significantly interfere with $ggg - q\bar{q}$ phase angle
$\phi_{ggg}$.  The magnitude of this interference cannot be maximal,
since some $ggg$ final states are distinguishable from \qqbar\ decays.
We define $f_{ggg}$ as the fraction of the $ggg$ amplitude which
interferes with \qqbar.  If $\phi_{ggg}=0$, our previous study applies
(Figure~\ref{simpleintfit}) and $f_{ggg} < 0.12$ at 95\% confidence
level.  We can also put limits on $f_{ggg}$ which are independent of
$\phi_{ggg}$ by considering the fact that $ggg$ and \qqbar\ have
different quantum numbers.  One-third of the \qqbar\ amplitude is
$c\bar{c}$, but $\Upsilon(1S) \to ggg \to c\bar{c}$ is negligible
\cite{watkins}.  Moreover, Strong decays preserve the \ups\ meson's
isospin of $I=0$, while Electromagnetic decays through a virtual
photon have $I=0$ or $1$.  At most half of the $u\bar{u}$ and
$d\bar{d}$ has $I_z=0$, and therefore overlap with $ggg$.  Adding
these fractions yields
\begin{equation}
  \begin{array}{c} c\bar{c} \\ \frac{2}{3} \end{array}
  \mbox{\hspace{-0.3 cm}}
  \begin{array}{c} \\ \times \end{array}
  \mbox{\hspace{-0.3 cm}}
  \begin{array}{c} \\ 0 \end{array}
  \begin{array}{c} \\ + \end{array}
  \begin{array}{c} s\bar{s} \\ \frac{1}{3} \end{array}
  \mbox{\hspace{-0.3 cm}}
  \begin{array}{c} \\ \times \end{array}
  \mbox{\hspace{-0.3 cm}}
  \begin{array}{c} \\ 1 \end{array}
  \begin{array}{c} \\ + \end{array}
  \left(
  \begin{array}{c} d\bar{d} \\ \frac{1}{3} \end{array}
  \begin{array}{c} \\ + \end{array}
  \begin{array}{c} u\bar{u} \\ \frac{2}{3} \end{array}
  \right)
  \mbox{\hspace{-0.3 cm}}
  \begin{array}{c} \\ \times \end{array}
  \mbox{\hspace{-0.3 cm}}
  \begin{array}{c} \\ 0.5 \end{array}
  \begin{array}{c} \\ = \end{array}
  \begin{array}{c} \\ 42\% \end{array}
\end{equation}
or $f_{ggg} \lesssim 0.4$.

\begin{figure}[p]
  \begin{center}
    \includegraphics[width=0.8\linewidth]{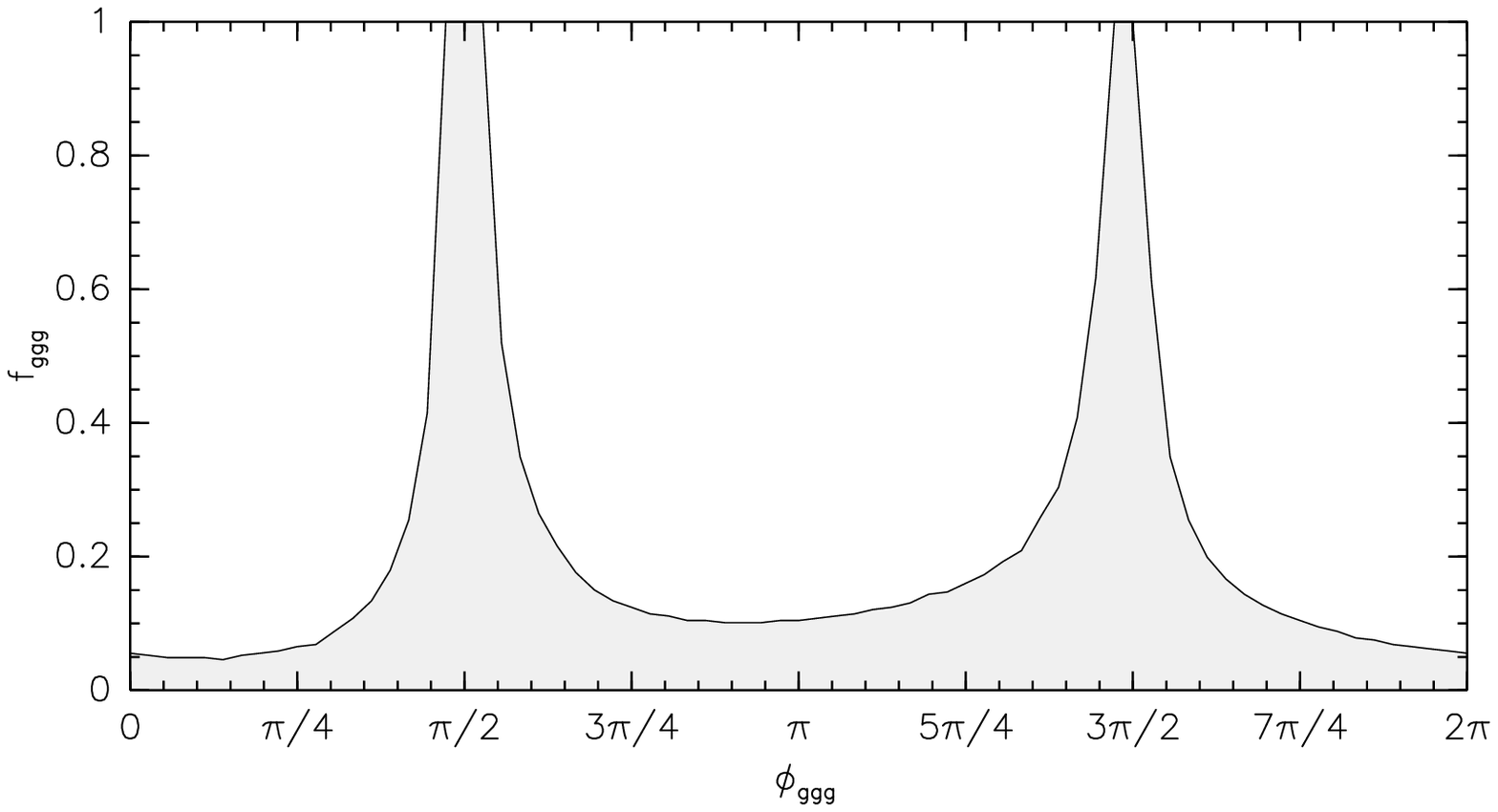}
  \end{center}
  \caption[Upper limits on interference between Strong and
  Electromagnetic decays to hadrons]{\label{intconstraint} The fraction of $ggg$ amplitude
  allowed to interfere with continuum \qqbar\ ($f_{ggg}$), according
  to the \us\ fit at 68\% confidence level.  We have no sensitivity to
  this parameter if the $ggg - q\bar{q}$ phase difference in \ups\
  decays ($\phi_{ggg}$) is near $\pm\pi/2$.}
\end{figure}

Our \us\ fits constrain $f_{ggg}$ as a function of $\phi_{ggg}$.
Assuming $\phi_{ggg}$ values between $0$ and $2\pi$, we fit for the
maximum $f_{ggg}$ consistent with the data at the 68\% confidence
level (the value which raises the fit $\chi^2$ by one unit), and plot
this allowed region in Figure~\ref{intconstraint}.  For most of the
$\phi_{ggg}$ range, our fit is more restrictive than the constraint we
derived from flavor and isospin considerations, though within 0.2~rad
of $\pm\pi/2$, our fit yields no information.  Outside of this range,
the minimum $\chi^2$ is attained within $0 < f_{ggg} < 0.1$.

\begin{figure}[p]
  \begin{center}
    \includegraphics[width=0.8\linewidth]{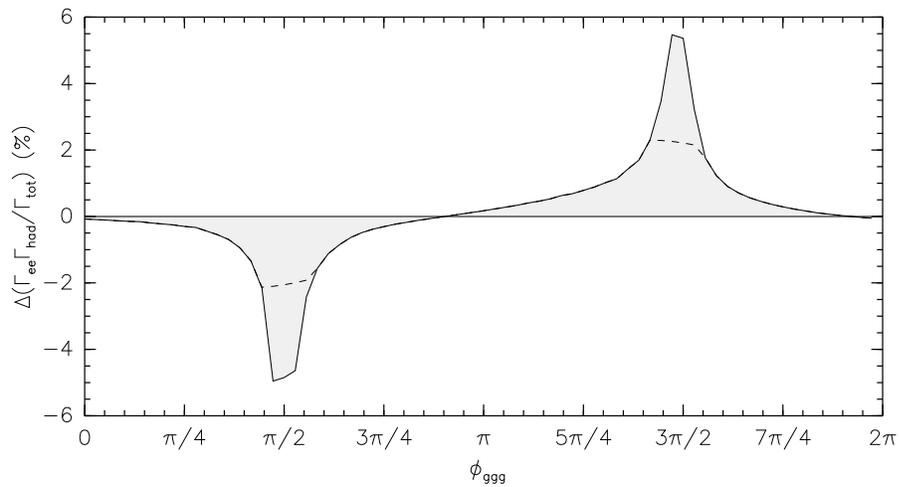}
  \end{center}
  \caption[Upper limits on \gee\ corrections due
  Strong/Electromagnetic interference]{\label{areaconstraint} The
  possible correction to \geehadtot\ implied by our ignorance of
  $f_{ggg}$ at 68\% confidence level (gray) as a function of
  $\phi_{ggg}$.  The dashed lines represent upper limits including
  isospin and flavor arguments.}
\end{figure}

Hadron-level interference also affects our \geehadtot\ result,
especially for $\phi_{ggg}$ near $\pm \pi/2$.  From the same fits, we
derive the following relationship between the correction that must be
applied to \geehadtot\ and $f_{ggg}$ and $\phi_{ggg}$.
\begin{equation}
  \Delta(\Gamma_{ee}\Gamma_\subs{had}/\Gamma_\subs{tot}) = -5.42\%
  \bigg(f_{ggg} \, \sin(\phi_{ggg} + 0.3) \bigg)
  + 0.14\% \bigg(f_{ggg} + 0.92 \, {f_{ggg}}^2 \bigg)
  \label{eqn:areashift}
\end{equation}
Applying the limits on $f_{ggg}$ shown in Figure~\ref{intconstraint}
to Equation~\ref{eqn:areashift}, we obtain the 68\% confidence level
upper limits on \us\ \geehadtot\ corrections shown in
Figure~\ref{areaconstraint}.  Corrections for the \uss\ are 70\% as
large as this and corrections for the \usss\ are 65\% as large.

Complete a priori ignorance of inclusive $ggg$/\qqbar\ interference
would imply a 2.0\% uncertainty in \geehadtot\ and \gee\ for the \us,
a 1.4\% uncertainty for the \uss, and a 1.3\% uncertainty for the
\usss.  However, the possibility that $f_{ggg}$ is as large as 0.4 is
not widely believed to be likely, nor is it clear that $\phi_{ggg}
\approx \pm\pi/2$

\chapter{Results and Conclusions}
\label{chp:results}

In this Chapter, we present final results for \geehadtot, for \gee,
for $\Gamma_{ee}(nS)/\Gamma_{ee}(mS)$, for $\Gamma$, and for
$|\psi(0,0,0)|^2$, all of which are derived from \geehadtot.  But
first, we review the systematic uncertainties in \geehadtot.

The hadronic \ups\ efficiency and integrated luminosity scale factor
together determine the multiplicative scale for cross-section, so
fractional uncertainties in these two factors add in quadrature to the
fractional uncertainty in \geehadtot.  Uncertainties in backgrounds
contribute negligibly to systematic uncertainty (their largest effect
is statistical).  At the end of Chapter~\ref{chp:beamenergy}, we
determined the effect of beam energy uncertainty on the lineshape fits
through simulations, and in Section~\ref{sec:lineshapesyst}, we
determined the uncertainty due to the parameterization of our fit
function.  Our data have two unexplained features: Bhabhas and
\gamgam\ events do not predict the same luminosity as a function of
\ecm, and the fit $\chi^2$ is significantly higher than the number of
degrees of freedom for the \us\ and \uss.  We have quantified both of
these as systematic uncertainties.  All uncertainties in \geehadtot\
and \gee\ are listed in Table~\ref{tab:systematics}, including
uncertainty from the correction for leptonic modes
(Equation~\ref{eqn:geehadtot:to:gee}), which applies to \gee\ and not
\geehadtot.

\begin{table}
  \caption[All uncertainties in \geehadtot\ and
  \gee]{\label{tab:systematics} All uncertainties in \geehadtot\ and
  \gee.  The ``correction for leptonic modes'' applies to \gee\ only,
  and ``common hadronic efficiency'' and ``overall luminosity scale''
  are common to all three resonances.}
  \begin{center}
    \begin{tabular}{l c c c}
      \hline\hline Contribution to \gee & \hspace{0 cm}\us\hspace{0 cm} & \hspace{0 cm}\uss\hspace{0 cm} & \hspace{0 cm}\usss\hspace{0 cm} \\\hline
      Common hadronic efficiency (Section \ref{sec:usefficiency})                                & $^{+0.4}_{-0.6}$\% & $^{+0.4}_{-0.6}$\% & $^{+0.4}_{-0.6}$\% \\
      \uss, \usss\ efficiency corrections (Section \ref{sec:ussusssefficiency})                  & 0      & 0.15\% & 0.13\% \\
      Overall luminosity scale (Section \ref{sec:luminosity})                                    & 1.3\%  & 1.3\%  & 1.3\%  \\
      Bhabha/$\gamma\gamma$ inconsistency (Section \ref{sec:luminosityconsistency})              & 0.4\%  & 0.4\%  & 0.4\%  \\
      Beam energy measurement drift (Chapter \ref{chp:beamenergy})                               & 0.2\%  & 0.2\%  & 0.2\%  \\
      Fit function shape (Subsection \ref{sec:varyslowly} and Section \ref{sec:fitfunction})     & 0.1\%  & 0.1\%  & 0.1\%  \\
      $\chi^2$ inconsistency (Section \ref{sec:chichiconsistency})                               & 0.2\%  & 0.6\%  & 0      \\
      Correction for leptonic modes (Equation \ref{eqn:geehadtot:to:gee})                        & 0.2\%  & 0.2\%  & 0.3\%  \\\hline
      Total systematic uncertainty                                                               & 1.5\%  & 1.6\%  & 1.5\%  \\
      Statistical uncertainty (Table~\ref{tab:bestfit})                                          & 0.3\%  & 0.7\%  & 1.0\%  \\\hline
      Total                                                                                      & 1.5\%  & 1.8\%  & 1.8\%  \\\hline\hline
    \end{tabular}
  \end{center}
\end{table}

\begin{sidewaysfigure}[p]
  \begin{center}
    \includegraphics[width=0.7\linewidth]{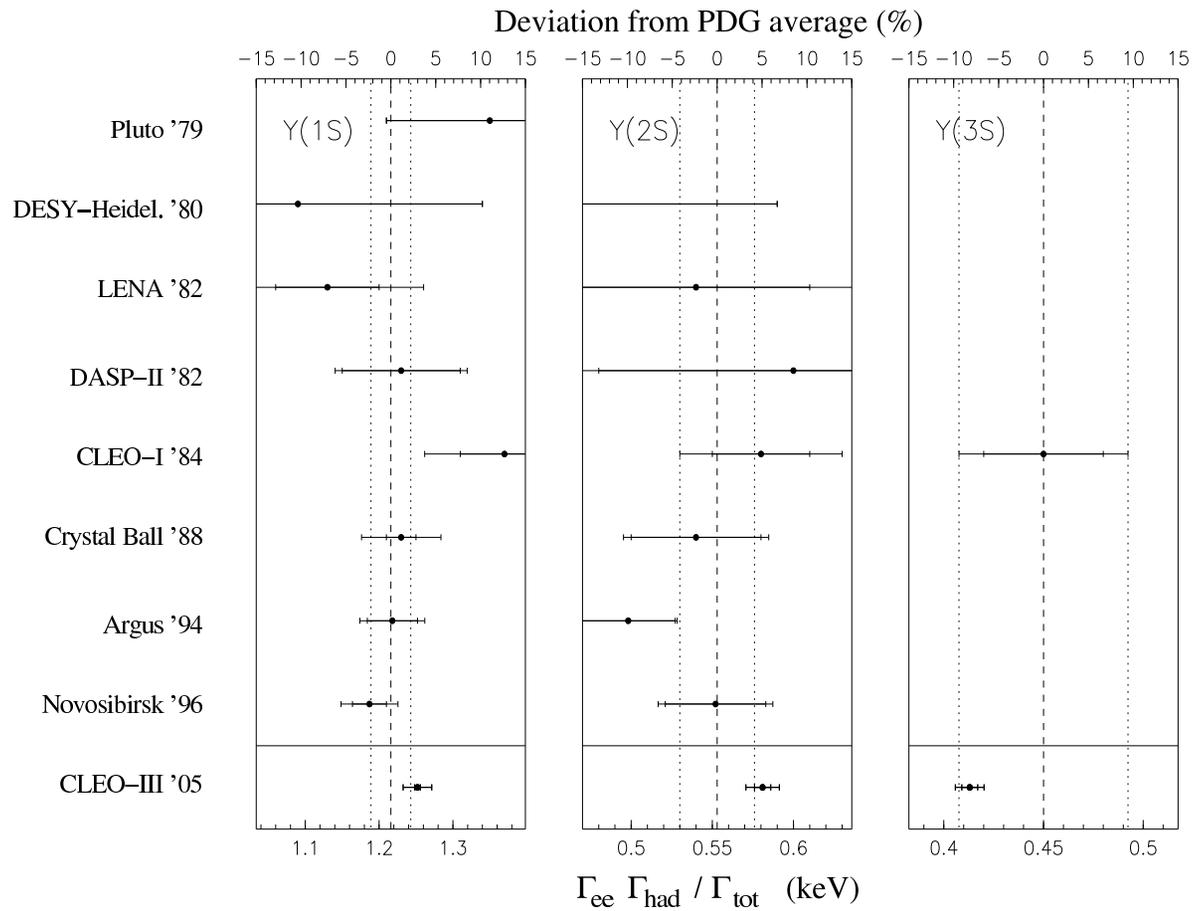}
  \end{center}
  \caption[Comparison of our \geehadtot\ with previous
  measurements]{\label{historyplot} Comparison of our \geehadtot\ with
  previous measurements.  The dashed and dotted lines are world
  averages and uncertainties, excluding our new
  measurements~\cite{pdg}.}
\end{sidewaysfigure}

Our values of \geehadtot, quoted with statistical uncertainties
first and systematic uncertainties second, are
\begin{eqnarray}
  (\Gamma_{ee}\Gamma_\subs{had}/\Gamma_\subs{tot})(1S) &=& 1.252 \pm 0.004 \pm 0.019 \mbox{ keV,} \\
  (\Gamma_{ee}\Gamma_\subs{had}/\Gamma_\subs{tot})(2S) &=& 0.581 \pm 0.004 \pm 0.009 \mbox{ keV, and} \\
  (\Gamma_{ee}\Gamma_\subs{had}/\Gamma_\subs{tot})(3S) &=& 0.413 \pm 0.004 \pm 0.006 \mbox{ keV.}
\end{eqnarray}
Correcting for leptonic modes with a factor of $(1-3{\mathcal B}_{\mu\mu})$, we obtain \gee:
\begin{eqnarray}
  (\Gamma_{ee})(1S) &=& 1.354 \pm 0.004 \pm 0.020 \mbox{ keV,} \\
  (\Gamma_{ee})(2S) &=& 0.619 \pm 0.004 \pm 0.010 \mbox{ keV, and} \\
  (\Gamma_{ee})(3S) &=& 0.446 \pm 0.004 \pm 0.007 \mbox{ keV.}
\end{eqnarray}
The total uncertainties for each resonance is less than 2\%.
Figure~\ref{historyplot} compares our \geehadtot\ to all previous
measurements: we find it to be consistent with, but more precise than,
the world average.

As a consequence of the measurement technique, our \gee\ values
represent the rate of $\Upsilon \to e^+e^-$ decays with no photons in
the final state, not even very soft photons.  This is because
final-state radiation in \gee\ ($\Upsilon \to \gamma e^+e^-$)
corresponds to initial-state radiation in the measured cross-section
($e^+e^- \to \gamma \Upsilon$), which we excluded with our fitting
technique.  Our result does include Electromagnetic vacuum
polarization, a few-percent effect in which the virtual photon
connecting \ee\ to \ups\ is interrupted by fermion loops, because this
affects the decay process and the production process equally, and we
never made any correction to remove it.

This analysis represents a substantial gain in precision, but it is
also more general than previous analyses in that we do not assume an
\ups\ decay model to determine our efficiency.  In previous analyses,
efficiency was determined by selecting events from a Monte Carlo
simulation, so potential errors due to modeling hadronization, unknown
decay modes, or the detector simulation were not represented by their
quoted efficiency uncertainties.  We relax this assumption for \us\
decays and only assume that our Monte Carlo correctly scales from \us\ to
\uss\ and \usss.  In addition, we obtain an upper limit on all-neutral
\us\ decays.  The \us\ branching fraction to events that generate zero
tracks with $|\cos\theta| < 0.93$ and $p_\perp > 150$~MeV is less than
1.01\% at 90\% confidence level.  See Section~\ref{pag:invisible} on
page~\pageref{pag:invisible} for more details.

This is the first analysis with sufficient precision to observe
interference between the \us\ resonance and the continuum.  This
effect was observed for the $J/\psi$ and $\psi'$ mesons in the \mumu\
channel within a year of their discovery \cite{spear} \cite{spearb}
because these charmonium resonances have much larger cross-sections
relative to continuum.  We therefore had the first opportunity to
explore the nature of this interference in the \ups\ meson.  We found
evidence of and corrected for interference in the \qqbar\ final state
(Figure~\ref{simpleintfit}) and observed interference in the \mumu\
final state (Figure~\ref{newmumu}), but not in inclusive hadrons.

These measurements provide useful constraints on the $b\bar{b}$
potential or checks on the validity of the approximation
\cite{oldpot}.  The degrees of freedom in these calculations are the
\ups\ masses and widths.  Our \gee\ measurements compliment the
Novosibirsk measurements of the \ups\ masses \cite{novomass}, forming
a set of six experimental inputs with better than 2\% precision each.
The \usss\ measurement is particularly valuable, as it is the second
ever performed (Figure~\ref{historyplot}), increasing the world
knowledge by a factor of five.  Also, the \usss\ wavefunction probes
more of the non-Coulomb ``confinement'' potential than the \us\ or
\uss.  The $\Upsilon(4S)$ wavefunction reaches further, but it is more
complicated to extract \gee\ due to its vicinity to the $B^0\bar{B^0}$
threshold \cite{argus}.  The full width of the $\Upsilon(4S)$ is a
function of \ecm\ and is non-negligible ($\Gamma \approx 20$ at the
$\Upsilon(4S)$ mass), making it necessary to model a QCD process to
express this function.

Finally, the \ups\ di-electron widths are valuable as a test of
Lattice QCD.  The ratios $\Gamma_{ee}(nS)/\Gamma_{ee}(mS)$ are
especially useful, since the renormalization factor for the virtual
photon current, $Z_\subs{match}^\subs{vector}$ cancels, which allows
us to test this aspect of the calculation in isolation.  Conversely,
the ratios test the rest of the calculation, particularly the NRQCD
treatment of $b$ quarks and the staggered-quark formalism for virtual
light quarks, in isolation and with higher precision than what can be
attained with the absolute \gee\ calculations.  These aspects are
shared with the calculation of $f_B$, which is used to extract \vtd\
from $B^0$-$\bar{B^0}$ mixing.

\begin{figure}[p]
  \begin{center}
    \includegraphics[width=\linewidth]{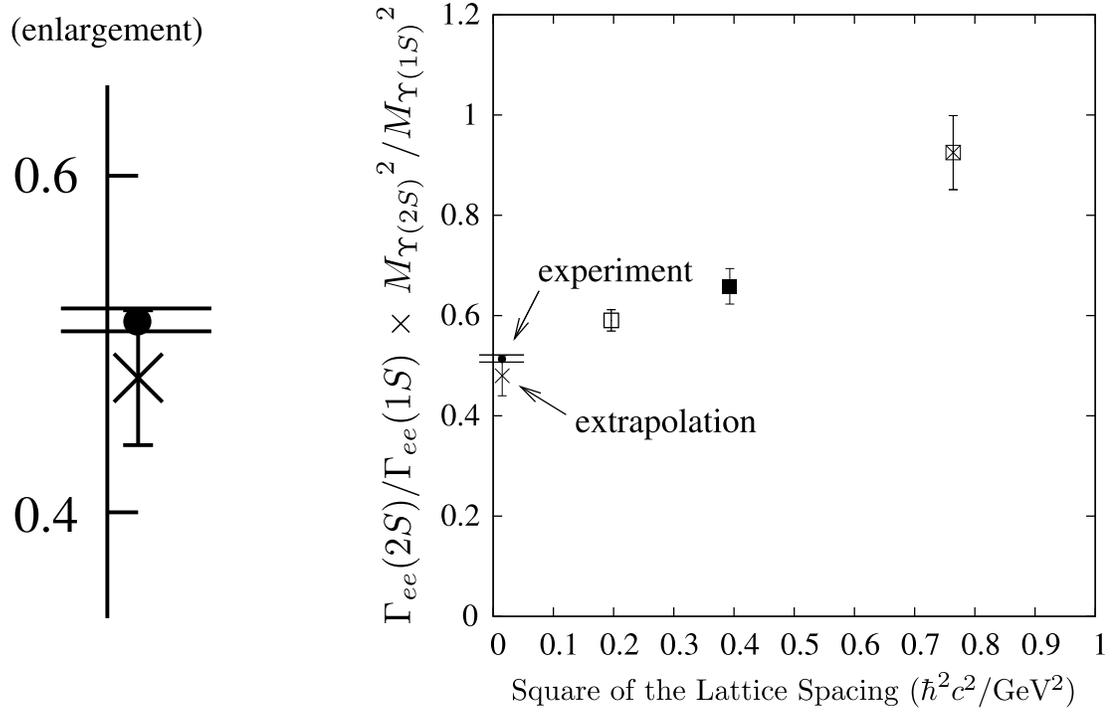}
  \end{center}
  \caption[Comparison of our $\Gamma_{ee}(2S)/\Gamma_{ee}(1S)$ with
  the HPQCD calculation]{\label{latticespacingagain} HPQCD
  calculations of $\Gamma_{ee}(2S)/\Gamma_{ee}(1S)$ times the ratio of
  masses squared as a function of lattice grid size squared.  The
  extrapolation to zero lattice size and our new measurement are
  overlaid.}
\end{figure}

Our experimental values of $\Gamma_{ee}(nS)/\Gamma_{ee}(mS)$ are
\begin{eqnarray}
  \Gamma_{ee}(2S)/\Gamma_{ee}(1S) &=& 0.457 \pm 0.004 \pm 0.004 \mbox{,} \label{eqn:ssovers} \\
  \Gamma_{ee}(3S)/\Gamma_{ee}(1S) &=& 0.329 \pm 0.003 \pm 0.003 \mbox{, and} \\
  \Gamma_{ee}(3S)/\Gamma_{ee}(2S) &=& 0.720 \pm 0.009 \pm 0.007 \mbox{,}
\end{eqnarray}
where we have canceled systematic uncertainties which are shared among
the pairs of resonances.  These shared systematics are the common
hadronic efficiency and the overall luminosity scale in
Table~\ref{tab:systematics}.  We overlay our
$\Gamma_{ee}(2S)/\Gamma_{ee}(1S)$ experimental result on the HPQCD
calculation (Equation~\ref{eqn:latticeresult}) in
Figure~\ref{latticespacingagain}.

We now use our \gee\ measurements to determine basic parameters of the
\ups\ mesons: their total decay rates (lifetimes) and wavefunctions at
the origin (sizes).  The only known experimental access to these
parameters is through \gee.  Again assuming \bee\ = \bmm, we find
\begin{eqnarray}
  \Gamma(1S) &=& 54.4 \pm 0.2 \pm 0.8 \pm 1.6 \mbox{ keV,} \\
  \Gamma(2S) &=& 30.5 \pm 0.2 \pm 0.5 \pm 1.3 \mbox{ keV, and} \\
  \Gamma(3S) &=& 18.6 \pm 0.2 \pm 0.3 \pm 0.9 \mbox{ keV.}
\end{eqnarray}
The first two uncertainties are statistical and systematic, and the
third uncertainty is propagated from \bmm.  The \uss\ and \usss\
values are lower than the averages quoted in \cite{pdg} because the
2005 measurements of \bmm\ \cite{istvan} are higher than previous
measurements and were not included in \cite{pdg}.  The above widths
correspond to the following lifetimes:
\begin{eqnarray}
  \tau(1S) &=& 12.1 \pm 0.0 \pm 0.2 \pm 0.4 \mbox{ zs,} \\
  \tau(2S) &=& 21.6 \pm 0.1 \pm 0.4 \pm 0.9 \mbox{ zs, and} \\
  \tau(3S) &=& 35.4 \pm 0.4 \pm 0.6 \pm 1.7 \mbox{ zs,}
\end{eqnarray}
where 1~zs (zeptosecond) is $10^{-21}$~s.
To the degree that the \ups\ system is non-relativistic, and therefore
Equation~\ref{eqn:waveatorigin} holds,
the value of the $b\bar{b}$ wavefunction at the origin is
\begin{eqnarray}
  |\psi(0,0,0)|^2(1S) &=& 53.0 \pm 0.2 \pm 0.8 \mbox{ fm$^{-3}$,} \\
  |\psi(0,0,0)|^2(2S) &=& 27.2 \pm 0.2 \pm 0.4 \mbox{ fm$^{-3}$, and} \\
  |\psi(0,0,0)|^2(3S) &=& 20.9 \pm 0.2 \pm 0.3 \mbox{ fm$^{-3}$.}
\end{eqnarray}
The inverse cube roots of these values (0.27~fm, 0.33~fm, and 0.36~fm)
characterize the physical sizes of the \ups\ mesons, though the form
of the potential is needed to express them as the RMS of the
wavefunction.  These bottomonium $|\psi(0,0,0)|^2$ values, with ratios
of 100:51:39, fall less rapidly from $1S$ to $3S$ than positronium, in
which $|\psi(0,0,0)|^2$ have ratios of 100:35:19.  We see in this
the effect of the linear part of the Strong force potential: the
\usss\ has more probability density at the origin than it would if its
potential were purely Coulombic.  In the sense that \gee\ is sensitive
to the physical size and shape of the \ups\ meson, we have truly used
a particle accelerator as a ``giant microscope'' to observe a
structured object a quadrillion times smaller than a meter.

\end{document}